\documentclass[a4paper,11pt]{article}
\usepackage{jheppub}
\usepackage{amsmath,amssymb,amsfonts,bm,bbm}
\usepackage{graphics,graphicx,epsfig,epstopdf}
\usepackage[active]{srcltx}
\usepackage{slashed}
\usepackage{textcomp}
\usepackage{xcolor}
\usepackage{hyperref}
\usepackage[ruled,linesnumbered]{algorithm2e}
\usepackage{tikz}
\usepackage{subfigure}
\input epsf
\graphicspath{{figures/}{fig/}}

\def \be {\begin{equation}}
\def \ee {\end{equation}}
\def \ba {\begin{array}}
\def \ea {\end{array}}
\def \bea{\begin{eqnarray}}
\def \eea{\end{eqnarray}}
\newcommand{\bean}{\begin{eqnarray*}}
\newcommand{\eean}{\end{eqnarray*}}
\newcommand{\nn}{\nonumber \\
}

\def\W #1{\widetilde{#1}}
\def\WH #1{\widehat{#1}}

\def\d{\partial}

\def\a{{\alpha}}
\def\b{{\beta}}

\def\eref#1{(\ref{#1})}

\def\Label#1{\label{#1}%
    \smash{\hbox to0pt{\raise1ex\hbox{\tiny[#1]}\hss}}}

\definecolor{ceil}{rgb}{0.57,0.63,0.81}

\newcommand{\mi}{\mathrm{i}}
\newcommand{\md}{\mathrm{d}}

\begin{document}

\title{An Algorithm for the Symbolic Reduction of Multi-loop Feynman Integrals via Generating Functions}
\author[a,b,c,d]{Bo Feng}
\author[e]{Xiang Li}
\author[g]{Yuanche Liu}
\author[e,f]{Yan-Qing Ma}
\author[g,d,f]{Yang Zhang}

\affiliation[a]{State Key Laboratory of Nuclear Physics and Technology, Institute of Quantum Matter, South China Normal University, Guangzhou 510006, China}
\affiliation[b]{Guangdong Basic Research Center of Excellence for Structure and Fundamental Interactions of Matter, Guangdong Provincial Key Laboratory of Nuclear Science, Guangzhou 510006, China}
\affiliation[c]{Beijing Computational Science Research Center, Beijing 100084, China}
\affiliation[d]{Peng Huanwu Center for Fundamental Theory, Hefei, Anhui, 230026, China}
\affiliation[e]{School of Physics, Peking University, Beijing 100871, China}
\affiliation[f]{Center for High Energy Physics, Peking University, Beijing 100871, China}
\affiliation[g]{Interdisciplinary Center for Theoretical Study, University of Science and Technology of China, Hefei, Anhui 230026, China}
\abstract{We develop a generating-function formulation for the symbolic reduction of multi-loop Feynman integrals. In this framework, integration-by-parts identities are rewritten as differential equations for sector-wise generating functions, so the reduction problem can be studied in a non-commutative algebra of differential operators rather than only through relations among individual integrals. This viewpoint leads to an iterative algorithm that generates candidate equations, extracts symbolic reduction rules, updates the active rule set, and tests completeness on the lattice of integral indices. We illustrate the method with the sunset topology, planar and non-planar massless double-box topologies, representative subsectors, and a degenerate example in which the top sector contains no master integral. Together, these examples show how symbolic reduction rules, descendant equations, and completeness criteria can be organized within a single algebraic framework.}
\preprint{USTC-ICTS/PCFT-26-25}
\maketitle

\section{Introduction}

Multi-loop Feynman integrals with many external legs and several kinematic scales remain a central bottleneck in state-of-the-art perturbative calculations. On the side of master-integral evaluation, the last decade has seen striking progress from differential-equation methods, especially in canonical form, together with increasingly automated basis transformations and complementary numerical strategies. These developments now support genuinely high-multiplicity systems: automated transformations to canonical bases are available, auxiliary-mass-flow methods provide an efficient route to numerical evaluation, the complete reduction of two-loop five-light-parton amplitudes has been achieved, all two-loop five-point one-mass master integrals and six-point planar massless are known analytically, and analytic results have recently reached the three-loop five-point level \cite{Henn:2013pwa,Meyer:2016zeb,Meyer:2017joq,Liu:2022chg,Guan:2019bcx,Abreu:2023rco, Abreu:2024fei, Henn:2025xrc, Liu:2024ont, Chicherin:2025mvc}. For the kind of frontier calculations we have in mind, however, these advances can only be fully exploited if the accompanying reduction problem can be handled with comparable efficiency and transparency.

To reach physical amplitudes, one must first reduce the large space of integrals generated by Feynman rules to a finite basis of master integrals. Historically, the modern framework for this step is the integration-by-parts (IBP) method introduced by Tkachov and Chetyrkin, while Laporta transformed IBP identities into a systematic elimination algorithm suitable for computer implementation \cite{Tkachov:1981wb,Chetyrkin:1981qh,Laporta:2000dsw}. Around this core framework, several complementary strands were developed, including dimensional shifts, generalized recurrence relations and difference-equation methods \cite{Tarasov:1996br,Laporta:2000dsw,Lee:2009dh}. In this sense, the history of reduction is not merely the history of solving larger linear systems; it is also the history of learning how to encode and organize relations among integrals in a form adapted to high-loop, multi-leg and multi-scale problems. Even today, for sufficiently complicated families, reduction is often the dominant algebraic bottleneck in the whole calculation \cite{Marquard:2021spf,Smirnov:2025dfy}.

The practical success of IBP reduction is inseparable from the development of public codes. Early automated implementations such as \texttt{AIR}, \texttt{FIRE} and \texttt{Reduze} established the viability of large-scale Laporta-style reductions, while \texttt{LiteRed} emphasized the heuristic search for symbolic sector reduction rules together with symmetry detection, differential equations and dimensional recurrences \cite{Anastasiou:2004vj,Smirnov:2008iw,Studerus:2009ye,Lee:2012cn,Lee:2013mka}. A modified algorithm of \texttt{LiteRed} appears in a recent program \texttt{SPIDER} \cite{Dlapa:2026oyq}. Later generations of software, including \texttt{Reduze}~2, \texttt{FIRE} and \texttt{Kira}, strengthened the field through distributed reduction, modular arithmetic, finite-field methods and improved equation management \cite{vonManteuffel:2012np,Smirnov:2019qkx,Smirnov:2023yhb,Maierhoefer:2017hyi,Klappert:2020nbg}. More recently, \texttt{Blade} and \texttt{NeatIBP}  have pushed in the direction of  block-triangular decompositions and trimmed IBP systems and \cite{Guan:2024byi,Wu:2023upw,Wu:2025aeg}. A concise recent snapshot of this software-centered landscape from the viewpoint of symbolic reduction is given in Ref.~\cite{Feng:2025leo}, while broader historical overviews can be found in Refs.~\cite{Marquard:2021spf,Smirnov:2025dfy}. At the same time, all these developments also highlight the basic limitation of the traditional paradigm: in hard multi-loop multi-leg problems, one is still often confronting extremely large sparse systems whose solution is expensive in both time and memory.

For this reason, recent work has increasingly focused on reformulating the reduction problem itself rather than only accelerating Laporta elimination. One line of development constructs shorter or better-adapted IBP systems using generalized generators, unitarity cuts, algebraic geometry, module intersections, small-size IBP systems and transverse integration identities \cite{vonManteuffel:2014ixa,Larsen:2015ped,Boehm:2020zig,Wu:2023upw,Chestnov:2025jam}. A second line seeks symbolic rules that remain valid on the lattice of indices, leading to syzygy-constrained symbolic reduction rules, diagonalized or triangularized IBP equations, seedless lowering operators and covariant-differentiation based reductions \cite{Smith:2025xes,Liu:2025udl,delaCruz:2026mas,vonGersdorff:2026zco}. A third line explores data-driven strategies, where machine-learning, reinforcement-learning and related heuristic approaches are used to optimize equation generation, equation selection or even online reduction itself \cite{vonHippel:2025okr,Zeng:2025xbh,Shih:2026jfe}. Recently, optimized IBP reduction is invented based on the fundamental filtration structure of Feynman integrals \cite{-collaboration:2026aup,Bree:2025tug,e-collaboration:2025frv}. Alternative non-IBP reduction frameworks based on intersection theory \cite{Mastrolia:2018uzb,Frellesvig:2019kgj,Frellesvig:2019uqt,Mizera:2019vvs,Weinzierl:2020xyy,Frellesvig:2020qot, Fontana:2023amt,Jiang:2023topdown,Crisanti:2024onv,
Lu:2024dsb} are also becoming increasingly practical for multi-leg problems \cite{Huang:2026xnq}. Taken together, these developments make clear that the modern question is not only how to solve the reduction equations faster, but also how to represent them in a form where their hidden structure becomes manifest.

Among these new developments for reduction, generating function method has  its own characteristic. 
at both conceptual and algorithmic levels. Instead of treating each lattice point of indices separately, one packages an entire sector into a single object, so that shifts of propagator powers or numerator degrees become differential operators, and IBP identities become differential equations for the generating function. Symbolic recurrence relations then arise naturally at the operator level. This makes the dependence on indices explicit, suggests completeness tests that do not rely on a manually chosen seed set, and opens the possibility of globally valid lowering relations. From this viewpoint, the generating-function program sits naturally at the intersection of the older recurrence-relation tradition and the modern search for seed-independent symbolic reductions \cite{Tarasov:1996br,Laporta:2000dsw,Lee:2009dh}.

The recent history of this approach is already rich enough to be instructive. The early use of generating functions appeared in \cite{Ablinger:2014yaa, Kosower:2018obg}, where specific task of reduction has been carried out. Generating functions for one-loop tensor-reduction
first appeared in \cite{Feng:2022gft}, after which explicit closed-form constructions were developed and the method was extended to arbitrary one-loop reduction \cite{Hu:2023egt,Li:2024sag} and towards multi-loop case\cite{Chen:2025gqu} recently. In parallel, generating functions were proposed as a way to explore the linear space of Feynman integrals and to derive reduction rules from differential equations satisfied by those generating functions \cite{Guan:2023lsg}. The next step was to connect this language directly to IBP relations in a systematic one-loop framework \cite{Hu:2025gibp}, then to Baikov-space generating functions for multi-loop reduction coefficients \cite{Hu:2025rrt}, and finally to a general proposal for complete symbolic reduction of multi-loop integrals via generating functions \cite{Feng:2025leo}. Closely related developments, such as syzygy-constrained symbolic rules and seedless constructions of lowering operators, reinforce the same lesson: reduction may be organized most efficiently at the level of operator identities rather than by solving large seeded systems case by case \cite{Smith:2025xes,delaCruz:2026mas}.

Our goal is to develop precisely such a formulation for sector-wise generating functions. The starting point is to package all integrals in a given sector into a single generating function. In that language, IBP identities become differential equations for the generating function, while symbolic reduction rules arise from relations among differential operators. The emphasis therefore shifts from isolated recurrence relations at specific lattice points to operator identities that can be generated, simplified, compared and organized systematically.

This perspective leads naturally to an iterative algorithm. Starting from the differential equations induced by IBP identities, we derive reduction rules for selected operator structures by solving small linear systems, simplify newly generated equations with the rules already obtained, and test whether the remaining irreducible lattice points match the expected master-integral content. The resulting workflow separates the problem into four tightly linked ingredients: the generating-function formalism, the operator algebra underlying the reduction rules, the iterative procedure that constructs and updates those rules, and a lattice-based completeness criterion.

The paper is organized as follows. In Section~\ref{sec:formalism}, we introduce the generating-function description of an integral family and the operator-algebra language used throughout the paper. In Section~\ref{sec:algorithm}, we present the iterative algorithm for generating and solving differential equations and for checking completeness. The remaining sections develop a sequence of examples. We begin with the sunset topology, which serves as the main pedagogical case, and then turn to planar and non-planar double-box topologies, representative subsectors, and finally a degenerate topology in which the top sector has no master integrals. Taken together, these examples show how the same framework handles top sectors, subsectors and sector elimination within a unified formulation.


\section{The Generating Function Formalism}
\label{sec:formalism}

In this section, we establish the basic language used throughout the paper. Rather than viewing integral reduction as a collection of linear relations among individual Feynman integrals, we encode the same information in generating functions that satisfy differential equations. Once this reformulation is in place, the reduction problem can be studied directly in terms of differential operators and their non-commutative algebra.

The central idea is to package all integrals in a fixed sector into a single generating function. Integration-by-parts identities then become differential equations for that generating function, and symbolic reduction rules arise from operator identities extracted from those equations. This perspective is useful for two reasons. First, it makes the structural content of the reduction problem more transparent. Second, it provides the algebraic framework required by the iterative algorithm developed in the next section.

\subsection{Generating Functions for Feynman Integrals}
\label{sec:gf_definition}

An $L$-loop Feynman integral family with $E$ independent external momenta can be generically represented by
\begin{equation}
    I(\vec{\nu}) = \int \prod_{i=1}^{L} \frac{\mathrm{d}^{D}\ell_i}{i\pi^{D/2}}
    \frac{\mathcal{D}_{K+1}^{-\nu_{K+1}}\cdots \mathcal{D}_N^{-\nu_N}}{\mathcal{D}_1^{\nu_{1}}\cdots \mathcal{D}_K^{\nu_{K}}},
    \label{eq:integral_family}
\end{equation}
where $D = 4-2\epsilon$ is the spacetime dimension, $\ell_i$ are the loop momenta, and $\vec{\nu} = (\nu_1, \dots, \nu_N)$ is a vector of integer indices. The denominators $\mathcal{D}_1,\dots,\mathcal{D}_K$ are the inverse propagators, which can have positive or negative powers. The $\mathcal{D}_{K+1},\dots,\mathcal{D}_N$ are the irreducible scalar products (ISPs) required for a complete basis of scalar denominators, and their powers $\nu_{i}$ are non-positive integers. The total number of denominators is $N = L(L+1)/2 + LE$.

An integral family is composed of numerous sectors, each defined by a subset of the denominators $\{\mathcal{D}_i\}$ being present. We encode this structure using a sector-defining binary vector $\vec{\mu} = (\mu_1, \dots, \mu_N)$ where $\mu_i=1$ if $\mathcal{D}_i$ is a propagator in this sector and $\mu_i=0$ otherwise\footnote{For ISPs that are present in the numerator across all sectors, their corresponding $\mu_i$ are always zero.}. For each sector $\vec{\mu}$, we define a generating function $G_{\vec{\mu}}$ in terms of a set of formal variables $\vec{\eta} = (\eta_1, \dots, \eta_N)$:
\begin{equation}
    G_{\vec{\mu}}(\vec{\eta}) = \int \prod_{i=1}^{L} \frac{\mathrm{d}^{D}\ell_i}{i\pi^{D/2}} 
    e^{\sum_{j=1}^N (1-\mu_j)\eta_j s_0^{-1}\mathcal{D}_j}
    \frac{1}{\prod_{i=1}^N (\mathcal{D}_i - s_0\eta_i)^{\mu_i}}.
    \label{eq:generating_function}
\end{equation}
Here, $s_0$ is an arbitrary momentum-scale constant to render the variables $\eta_i$ dimensionless.

The connection to the individual Feynman integrals $I(\vec{\nu})$ is revealed by Taylor expanding $G_{\vec{\mu}}(\vec{\eta})$ around $\vec{\eta}=\vec{0}$. As an illustrative example, consider a sector in which the first $K$ denominators are propagators ($\mu_i=1$ for $i=1,\dots,K$), while the remaining terms appear in the exponential ($\mu_j=0$ for $j=K+1,\dots,N$). The expansion yields:
\begin{align}
    G_{\vec{\mu}}(\vec{\eta}) &= \int \prod_{i=1}^{L} \frac{\mathrm{d}^{D}\ell_i}{i\pi^{D/2}} 
    \sum_{\vec{n}\geq 0} \left( \prod_{j=K+1}^N \frac{1}{n_j!} \right) 
    s_0^{\sum_{i=1}^K n_i - \sum_{j=K+1}^N n_j} \nonumber \\
    & \quad \times \frac{\mathcal{D}_{K+1}^{n_{K+1}}\cdots \mathcal{D}_N^{n_N}}{\mathcal{D}_1^{n_1+1}\cdots \mathcal{D}_K^{n_K+1}} 
    \vec{\eta}^{\vec{n}} \equiv \sum_{\vec{n}\geq 0} F_{\vec{\mu}, \vec{n}} \vec{\eta}^{\vec{n}},
    \label{eq:gf_expansion}
\end{align}
For later convenience, we introduce the following notation for an $N$-dimensional vector:
\bea \vec{a}\geq 0 & \Longrightarrow & \forall i\in [1,...,N],~~ \vec{a}_i\geq 0 \nn
\vec{a}\not\geq 0 & \Longrightarrow & \exists i\in [1,...,N], ~~~{\rm such~that}~~~~ \vec{a}_i<0~~~~\label{a-geq-def}\eea
%
By comparing the terms, we establish a direct mapping between the expansion coefficients $F_{\vec{\mu}, \vec{n}}$ and the integrals $I(\vec{\nu})$:
\begin{equation}
    F_{\vec{\mu}, \vec{n}} = \left(\prod_{j=K+1}^N \frac{1}{n_j!}\right) 
    s_0^{\sum_{i=1}^K n_i - \sum_{j=K+1}^N n_j} I(n_1+1, \dots, n_K+1, -n_{K+1}, \dots, -n_N).
    \label{eq:coeff_integral_map}
\end{equation}
This mapping is the cornerstone of our formalism. It identifies Feynman integrals with the expansion coefficients of the generating function, labeled by lattice points $\vec{n}$. We define the {\bf degree} of the coefficients $F_{\vec{n}}$ (or, equivalently, of the corresponding Feynman integrals $I$) as $\sum_{i=1}^N n_i$. The reduction problem for Feynman integrals is therefore translated into the problem of finding relations among these coefficients, which are induced by the differential equations (DEs) satisfied by the generating function.



It is worth noting that although there are potentially $2^N$ different generating functions corresponding to different choices of $\vec{\mu}$, some sectors vanish identically, so the corresponding generating functions can be set to zero from the outset. In addition, if global symmetries relate different sectors, it is sufficient to compute only one representative generating function and obtain the others by symmetry. Taking these observations into account, the number of generating functions that must actually be computed is far smaller than $2^N$ in practice.

\subsection{From IBP Identities to Differential Equations}
\label{sec:ibp_to_de}

The well-known IBP identities arise from the fact that the integral of a total derivative over loop momenta vanishes, i.e., schematically $\int \mathrm{d}^D\ell \, \dots \, \frac{\partial}{\partial \ell^\mu} (\dots) = 0$.
When applied to the integrand of Eq.~\eqref{eq:generating_function}, these identities become a system of linear partial differential equations for $G_{\vec{\mu}}(\vec{\eta})$. The building blocks of these equations are \textbf{differential operators}, namely polynomials in the variables $\eta_i$ and the partial derivatives $\partial_i \equiv \partial/\partial\eta_i$. Any such operator can be expressed as a linear combination of operators in a \textbf{standard form}, $\vec{\eta}^{\vec{a}} \vec{\partial}^{\vec{b}} \equiv \prod_i \eta_i^{a_i} \partial_i^{b_i}$. For example, the operator $\partial_1 \eta_1 \partial_1$ can be rewritten in standard form as $\eta_1 \partial_1^2 + \partial_1$.
In standard form, an operator is uniquely specified by a pair of vectors $(\vec{a},\vec{b})$, which we call the {\bf finer index} of the operator.

The general differential equations (DEs) for the generating function of a given sector $\vec{\mu}$ that we encounter in this paper have the form
\begin{equation}
	\sum_t c_t \WH{O}_t \, G_{\vec{\mu}}(\vec{\eta}) = B.
	\label{eq:generic_de}
\end{equation}
where the $\WH{O}_t$ are differential operators in standard form, and the coefficients $c_t$ are polynomials in the spacetime dimension $D$ and the kinematic invariants. The term $B$, or the boundary term, receives contributions from generating functions of subsectors (i.e., sectors with fewer propagators), possibly with additional operator insertions.
When performing the reduction for a given sector, we assume the reductions for all its subsectors are already known, thus treating $B$ as a known source term.

Applying the operator $\WH{O}_t = \vec{\eta}^{\vec{a}_t} \vec{\partial}^{\vec{b}_t}$ to the series expansion $G_{\vec{\mu}}(\vec{\eta}) = \sum_{\vec{n}} F_{\vec{n}} \vec{\eta}^{\vec{n}}$ yields:
\begin{equation}
	c_t(\vec{\eta}) \vec{\eta}^{\vec{a}_t} \vec{\partial}^{\vec{b}_t} \sum_{\vec{n}} F_{\vec{n}} \vec{\eta}^{\vec{n}} 
	\quad \longrightarrow \quad
	\sum_{\vec{n}} c_t(\vec{n}) \left( \prod_{j=1}^N \frac{(n_j - a_{t,j} + b_{t,j})!}{(n_j-a_{t,j})!} \right) F_{\vec{n} - \vec{a}_t + \vec{b}_t} \vec{\eta}^{\vec{n}}.
\end{equation}
where, for simplicity, we adopt the convention\footnote{The meaning of the notation $\vec{n}\not\geq 0$ can be found in \eref{a-geq-def}. }
\bea F_{\vec{n}}=0,~~~~~\vec{n}\not\geq 0~~~~\Label{lesson-1-1-1b}\eea
Thus the equation \eref{eq:generic_de} gives
\bea \sum_t c_t \left(\prod_{j=1}^N\prod_{p=1}^{(\vec{b}_t)_j}(n_j-(\vec{a}_t)_j+p)\right) F_{\vec{n}+\vec{b}_t-\vec{a}_t}= \W B~~~~\Label{lesson-1-1-1a}\eea
This translates the differential equation \eqref{eq:generic_de} into a \textbf{recurrence relation}.
From \eref{lesson-1-1-1a}, one sees that the role of an operator in the recurrence relation is largely determined by the combination $\vec{o} \equiv \vec{b} - \vec{a}$, which we call the {\bf index} of the operator. Similarly, we define the {\bf degree} of an operator as $|\vec{o}|\equiv \sum_{i=1}^N (\vec{o})_i$, which is related to the degree of lattice points through \eref{lesson-1-1-1a}. We say that a differential equation has degree $M$ when the highest degree among all operators appearing in that equation is $M$.

The mapping between \eref{eq:generic_de} and \eref{lesson-1-1-1a} indicates that the reduction of Feynman integrals---that is, expressing coefficients at higher-degree lattice points in \eref{eq:gf_expansion} in terms of lower-degree lattice points---becomes the problem of expressing a higher-degree operator as a linear combination of lower-degree operators using the DEs. We refer to such expressions as {\bf symbolic reduction rules}, since they generate recurrence relations that reduce the Feynman integrals.
The primary objective of this work is to develop a systematic algorithm to find the complete symbolic reduction rules.

Before going on, we want to emphasize that we are not trying to solve the differential equation \eref{eq:generic_de} for the generating function, i.e., expressing the generating function as an analytic function of the external data and $\eta$'s. Doing so will be extremely difficult. As pointed out, our aim is to solve some highest degree operators from a set of differential equations, which is just to solve some linear equations. The number of linear equations is much less than the one in the IBP system. 


\subsection{The Operator Algebra}
\label{sec:operator_algebra}

The discussion so far shows that the reduction problem is governed by a non-commutative algebra of differential operators. To use this structure algorithmically, we must identify which properties of the algebra remain stable under simplification and which ones control the shifts on the coefficient lattice. We therefore collect the operator facts that will be used repeatedly in the construction of reduction rules.


\subsubsection{The Algebra of Zero-Index Operators}
\label{sec:zero_index_ops}
We first consider a special class of operators, namely those with finer index $(\vec{a},\vec{a})$. Since their index is $\vec{o}=\vec{0}$, we call them {\bf zero-index operators}. In what follows, the phrase ``a zero-index operator'' may refer either to a single zero-index operator or to a sum of zero-index operators. These operators possess several crucial properties that simplify the algebra substantially.
\begin{itemize}
	\item[\textbf{(A)}] \textbf{Commutativity:} Any two zero-index operators commute, i.e., $[\WH{O}_{0,a}, \WH{O}_{0,b}] = 0$. This can be proven by showing that all operators of the form $\eta^k \partial^k$ can be expressed as polynomials in the fundamental operator $\WH{\eta} \equiv \eta \partial$. Since all such operators are functions of a single operator $\WH{\eta}$, they must commute with each other.
	
	\item[\textbf{(B)}] \textbf{Braiding Property:} For any operator $\WH{O}$ and any zero-index operator $\WH{O}_1$, there exists another zero-index operator $\WH{O}_2$ such that:
	\begin{equation}
		\WH{O} \WH{O}_1 = \WH{O}_2 \WH{O}.
		\label{eq:braiding}
	\end{equation}
In essence, a zero-index operator can be ``braided,'' or moved through any other operator, changing its form but not its zero-index nature. This allows us to collect all zero-index operators on one side of an expression. A similar property holds when moving an operator from right to left.
    Relation \eref{eq:braiding} will be useful for later simplification and its explicit construction will be given in the next subsubsection. 
\end{itemize}

\subsubsection{Operators with the same index}
\label{sec:same-index}

From \eref{lesson-1-1-1a}, we see that, in a recurrence relation, all contributions with the same operator index can be collected into a single term. Translating this observation into the language of operators, the sum of operators with the same index can be written compactly as
\begin{equation}
    \WH{O} = \WH{O}_0 \WH{O}^F_{\vec{o}},~~~~\label{gen-O}
\end{equation}
where $\WH{O}^F_{\vec{o}}$ is the \textbf{representative operator} for index $\vec{o}$, and $\WH{O}_0$ is a \textbf{zero-index operator}. The representative operator is defined as:
\begin{equation}
    \WH{O}^F_{\vec{o}} \equiv \vec{\eta}^{\vec{o}_{-}} \vec{\partial}^{\vec{o}_{+}},
    \label{eq:representative_op}
\end{equation}
where $\vec{o}_{+}$ contains the positive components of $\vec{o}$ (with the other components set to zero) and $\vec{o}_{-}$ contains the absolute values of the negative components, such that $\vec{o} = \vec{o}_{+} - \vec{o}_{-}$. Zero-index operators are linear combinations of operators of the form $\vec{\eta}^{\vec{a}}\vec{\partial}^{\vec{a}}$. The identity operator is a special zero-index operator with $\vec{a}=\vec{0}$. For instance, the combination of same-index operators $\WH{O} = \partial_1 + \eta_2\partial_1\partial_2$ can be decomposed into a zero-index operator $\WH{O}_0 = (1+\eta_2\partial_2)$ and a representative operator $\WH{O}_{\vec{o}}^F = \partial_1$. This decomposition is a powerful tool for handling combinations of operators with the same index.

Using the decomposition  \eref{gen-O} and \eref{eq:representative_op}, we can  give the explicit construction for the  relation between zero-index operators $\WH O_1, \WH O_2$ in \eref{eq:braiding}. First we compute   the commutative relation  $\left[\WH {O}^F_{\vec{o}}, \WH O_1\right]$. Matching the operator index, it must be the form  $\left[\WH {O}^F_{\vec{o}}, \WH O_1\right]=\WH O_{a}\WH {O}^F_{\vec{o}} $ where $\WH O_a$ is a zero-index operator. Multiplying $\WH O_0$ at both sides, we get 
		$\WH O \WH O_1-\WH O_1\WH O=\WH O_{a} \WH O$, thus we have $\WH O_2= \WH O_1+\WH O_{a}$.

\subsubsection{Hierarchy and Descendant Operators}
\label{sec:descendants}

The system of differential equations exhibits a hierarchy involving operators of different degrees. Many operators are not genuinely new; instead, they can be generated from simpler ones. This motivates the following notion: an operator $\WH Q_2$ is called a {\bf descendant} of an operator $\WH Q_1$ if there exists another operator (or a combination of operators) $\WH O_1$ such that
\bea \WH O_0\WH Q_2= \WH O_\delta \WH Q_1~~~~\label{Index-zero-des-1-1}\eea
where $\WH O_0$ is a zero-index operator (or a sum of such operators). Under \eref{Index-zero-des-1-1}, we also say that $\WH Q_1$ is the mother of $\WH Q_2$. Note that since $\WH O_\delta$ may be the identity operator, an operator can be the descendant (or mother) of itself.

A necessary and sufficient condition for $\WH{Q}_2 = \vec{\eta}^{\vec{a}_q} \vec{\partial}^{\vec{b}_q}$ to be a descendant of $\WH{Q}_1 = \vec{\eta}^{\vec{a}_o} \vec{\partial}^{\vec{b}_o}$ is that 
\bea \vec{a}_q - \vec{a}_o \ge 0,~~~\&~~~~\vec{b}_q - \vec{b}_o \ge 0~~~~\label{des-condition}\eea
Thus we can define $\WH O_\delta=\eta^{\vec{a}_q - \vec{a}_o}\vec{\d}^{\vec{b}_q - \vec{b}_o}$ and $\WH O_0$ can be solved easily. { An important point is that the $\WH O_0$ will contain the identity operator.}
This structure is fundamental to the algorithm because it allows newly generated differential equations to be simplified by referring back to a controlled set of previously solved ``mother'' rules.

For the convenience of later discussions, here we present the construction of $\WH O_\delta$ and $\WH O_0$ for two particular cases:
\begin{itemize}
	\item {\bf Case A:} For this case  $\WH{Q}_1 =  \vec{\partial}^{\vec{b}_o}$. Then it is easy to see that 
	\bea \WH O_\delta=\eta^{\vec{a}_q}\vec{\d}^{\vec{b}_q - \vec{b}_o},~~~~~~\WH O_0=1~~~~\label{des-A}\eea

	\item {\bf Case B:} For this case  $\WH{Q}_1 = \WH O_{0;m} \vec{\partial}^{\vec{b}_o}$ and $\WH{Q}_2 =\WH O_{0;d} \vec{\eta}^{\vec{a}_q} \vec{\partial}^{\vec{b}_q}$, where both $\WH O_{0;m}, \WH O_{0;d}$ are zero-index operators.
	Both  $\WH{Q}_1, \WH{Q}_2$ are, in fact, the combinations of operators with the same index and are written into the form \eref{gen-O}. 
	For this more general case, we can try with  
	\bean \WH O'_\delta=\eta^{\vec{a}_q}\vec{\d}^{\vec{b}_q - \vec{b}_o},\eean
	But now
	\bean & & \WH O'_\delta \WH{Q}_1 =\eta^{\vec{a}_q}\vec{\d}^{\vec{b}_q - \vec{b}_o} \WH O_{0;m} \vec{\partial}^{\vec{b}_o}=\WH O'_{0;m} \eta^{\vec{a}_q}\vec{\d}^{\vec{b}_q - \vec{b}_o} \vec{\partial}^{\vec{b}_o} = \WH O'_{0;m} \vec{\eta}^{\vec{a}_q} \vec{\partial}^{\vec{b}_q}\eean
	where we have used the result \eref{eq:braiding}. Using this computation, we see that now we have
	\bea \WH O'_\delta=\WH O_{0;d}\eta^{\vec{a}_q}\vec{\d}^{\vec{b}_q - \vec{b}_o},~~~~\WH O_0=\WH O'_{0;m},~~ \label{des-B}\eea

\end{itemize}



\subsection{The Derivation of Reduction Rules}
\label{sec:reduction_rules}

With the operator algebra in place, we can now state precisely what we mean by a {\bf symbolic reduction rule}. Such a rule isolates a highest-degree operator, or a highest-degree operator class, and rewrites it by solving linear equations in terms of lower-degree operators together with possible boundary contributions. It is therefore the operator-level analogue of a recurrence relation for the coefficients of the generating function.

For the recurrence relation \eref{lesson-1-1-1a} to be useful for all choices $\vec{n}\geq 0$ (i.e., the LHS is not zero), the be-solved operator, which will be called a \textbf{good operator}, must satisfy following two conditions:
\begin{enumerate}
	\item The target operator's index $\vec{o} = \vec{b}-\vec{a}$ must be non-negative, i.e., $\vec{o} \ge 0$. 
	
	\item If we have solved a sum of operators with the same highest degree, as we have mentioned in the previous subsection, the sum can be expressed as $\WH{O}_0 \WH{O}^F_{\vec{o}}$. Then we require that 
	 $\WH{O}_0$ must contain the identity operator with a non-zero coefficient. 
\end{enumerate}
If an equation does not contain any good operator, it will not be used to produce a reduction rule. 

With this clarification in place, we classify the differential equations, and the rules they yield, according to their operator content:
\begin{itemize}
    \item \textbf{T1-type (Type 1):} The equation contains a single highest-degree operator index. This is the simplest case, allowing for a direct solution for that operator index.
    \begin{itemize}
        \item \textbf{T1A-type:} The equation involves only one operator at the highest degree. It yields a rule of the form:
              \begin{equation}
                  \WH{O}_H = \sum_i c_i \WH{O}_i + B, \quad |\vec{o}_H| > |\vec{o}_i|~\forall i.
                  \label{eq:RRule-T1A}
              \end{equation}
        \item \textbf{T1B-type:} The equation involves multiple operators, all sharing the same highest-degree index $\vec{o}$. They can be bundled to yield a rule for the representative operator $\WH{O}^F_{\vec{o}}$:
              \begin{equation}
                  \WH{O}_{0} \WH{O}^F_{\vec{o}} = \sum_i c_i \WH{O}_i + B, \quad |\vec{o}|> |\vec{o}_i|~\forall i.
                \label{eq:RRule-T1B}
              \end{equation}
         where $\WH{O}_{0}$  is a zero-index operator. 
              
    \end{itemize}
    
    The T1-type reduction rules are perfect in the sense that after applying it to reduce a Feynman integrals, the degree will always be reduced at least one. 
    
    \item \textbf{T2-type (Type 2):} The equation contains multiple operator indices of the same maximal degree. This represents a more complex scenario where a choice must be made about which operator index to solve for. E.g., an equation might relate $\partial_1^2$ and $\partial_2^2$. Once a target index (say, $\vec{o}_1$) is chosen, the resulting rule is further classified:
    \begin{itemize}
        \item \textbf{T2A-type:} The chosen target index $\vec{o}_1$ corresponds to a single operator in the equation:
        \begin{equation}
            \WH{O}_{\vec{o}_1} = \sum_j b_j \WH{O}_{\vec{o}_j}+\sum_i c_i \WH{O}_i + B, \quad |\vec{o}_1|= |\vec{o}_j| > |\vec{o}_i|~\forall j,i.
                  \label{eq:RRule-T2A}
        \end{equation}
        \item \textbf{T2B-type:} The chosen target index $\vec{o}_1$ corresponds to multiple operators in the equation:
        \begin{equation}
            \WH{O}_{0} \WH{O}^F_{\vec{o}_1} = \sum_j b_j \WH{O}_{\vec{o}_j}+\sum_i c_i \WH{O}_i + B, \quad |\vec{o}_1|= |\vec{o}_j| > |\vec{o}_i|~\forall j,i.
                  \label{eq:RRule-T2B}
        \end{equation}
    \end{itemize}
    
    The T2-type reduction rules are less direct because they do not completely lower the degree in a single step. One must therefore choose the solving strategy with care: the algorithm needs a global ordering of operator indices so that higher-priority operators are expressed in terms of lower-priority ones of the same degree. Once this ordering is fixed, the remaining operators can later be converted into T1-type reductions, and the degree is eventually lowered in a controlled way.  
    
\end{itemize}


These conditions ensure that our algorithm systematically reduces any integral with sufficiently high powers to simpler ones. Notably, in sectors with no master integrals (degenerate sectors), the algorithm will naturally produce a reduction rule for the identity operator itself (i.e., $\vec{b}=\vec{0}$), proving that all integrals in that sector reduce to zero or to subsector contributions.

\subsubsection{Descendant operators of reduction rules}\label{des-red}

An important application of the descendant concept arises when a highest-degree operator in a newly generated equation is itself a descendant of an operator that has already been solved by one of the reduction rules above. In that situation, the previously solved rule can be used to eliminate the descendant operator from the equation. We call this manipulation {\bf simplification}; its algorithmic implementation will be discussed in Section~\ref{sec:simplification}.

Consider an equation of the form $0=\WH Q+\cdots$, where $\WH Q$ is a highest-degree operator. If $\WH Q$ is a descendant of a TA-type reduction rule, for example \eref{eq:RRule-T1A}, then by \eref{des-A} we immediately obtain
\bea \WH Q & = & \WH O_\delta \WH{O}_H = \sum_i c_i \WH O_\delta \WH{O}_i +\WH O_\delta B~~~~~\label{des-apply-TA}\eea
Substituting this back into the original equation replaces the highest-degree operator $\WH Q$ by lower-degree operators. If instead $\WH Q$ is a descendant of a TB-type reduction rule, for example \eref{eq:RRule-T1B}, then by \eref{des-B} we have immediately
\bea \WH O_0\WH Q & = & \WH O_\delta \WH{O}_{0} \WH{O}^F_{\vec{o}} = \sum_i c_i  \WH O_\delta \WH{O}_i + \WH O_\delta B~~~~~\label{des-apply-TB}\eea
To eliminate $\WH Q$ from the equation, one first acts with $\WH O_0$ on both sides and then arrives at 
 $0=\sum_i c_i  \WH O_\delta \WH{O}_i + \WH O_\delta B+\WH O_0(\cdots)$, where the highest-degree operator has again been replaced by lower-degree structures. The same logic applies to descendants of the T2A-type and T2B-type reduction rules.

\subsection{Operators with index $\vec{o}\not \geq 0$}
\label{sec:neg-index}

When solving differential equations, especially in subsectors, one frequently encounters highest-degree operators whose index satisfies $\vec{o}\not\geq 0$ (see \eref{a-geq-def}). Such operators are not suitable targets for reduction rules, but they still affect the structure of the equations in which they appear. The practical way to control them is to use the corresponding relations only on restricted regions of the lattice where the problematic terms vanish automatically. {  For example, with the operator  $\eta_2 {\d^2\over \d\eta_5^2}$ appearing in the DE, the term $(n_5+2)(n_5+1) F_{\vec{n}+2\vec{e}_5-\vec{e}_2}$ will appear in the corresponding  recurrence relation. Thus when applying the recurrence relation to reduce the lattice point $\vec{n}=(n_1,0,n_3,...)$, the contribution of $\eta_2 {\d^2\over \d\eta_5^2}$ is zero because $F_{\vec{n}\not\geq 0}=0$}. 

When designing the solving order for a T2-type equation, operators with index $\vec{o}\not \geq 0$ must therefore be handled with care. To avoid an infinite loop, the reduction rule extracted from such an equation is best applied only on restricted lattice regions (for example, $\vec{n}=(n_1,0,n_3,\dots)$ in the example above), unless the rule is being used precisely to reduce the problematic index itself. In practice, this means that other reduction rules must first shrink the irreducible set to the appropriate boundary, after which the partially restricted rule can be applied safely and effectively.


\section{The Algorithm}
\label{sec:algorithm}

Having established the generating-function formalism, we now present the algorithmic core of the paper. The purpose of the algorithm is to derive a complete set of symbolic reduction rules for a given integral family by working directly with the differential equations satisfied by the generating function. Conceptually, the algorithm converts those equations into a controlled set of operator identities that act as recurrence relations on the coefficient lattice.

At a high level, the workflow is iterative. Starting from the fundamental IBP-induced differential equations, we extract whichever reduction rules can already be solved, use them to simplify newly generated equations, and then determine which regions of the lattice remain irreducible. If the resulting rule set is incomplete, the surviving irreducible structure indicates what kind of descendant equations should be generated next. Equation generation, rule extraction, and completeness checking are therefore tightly coupled.

\begin{figure}[ht]
	\centering
	\includegraphics[width = 1.0\textwidth]{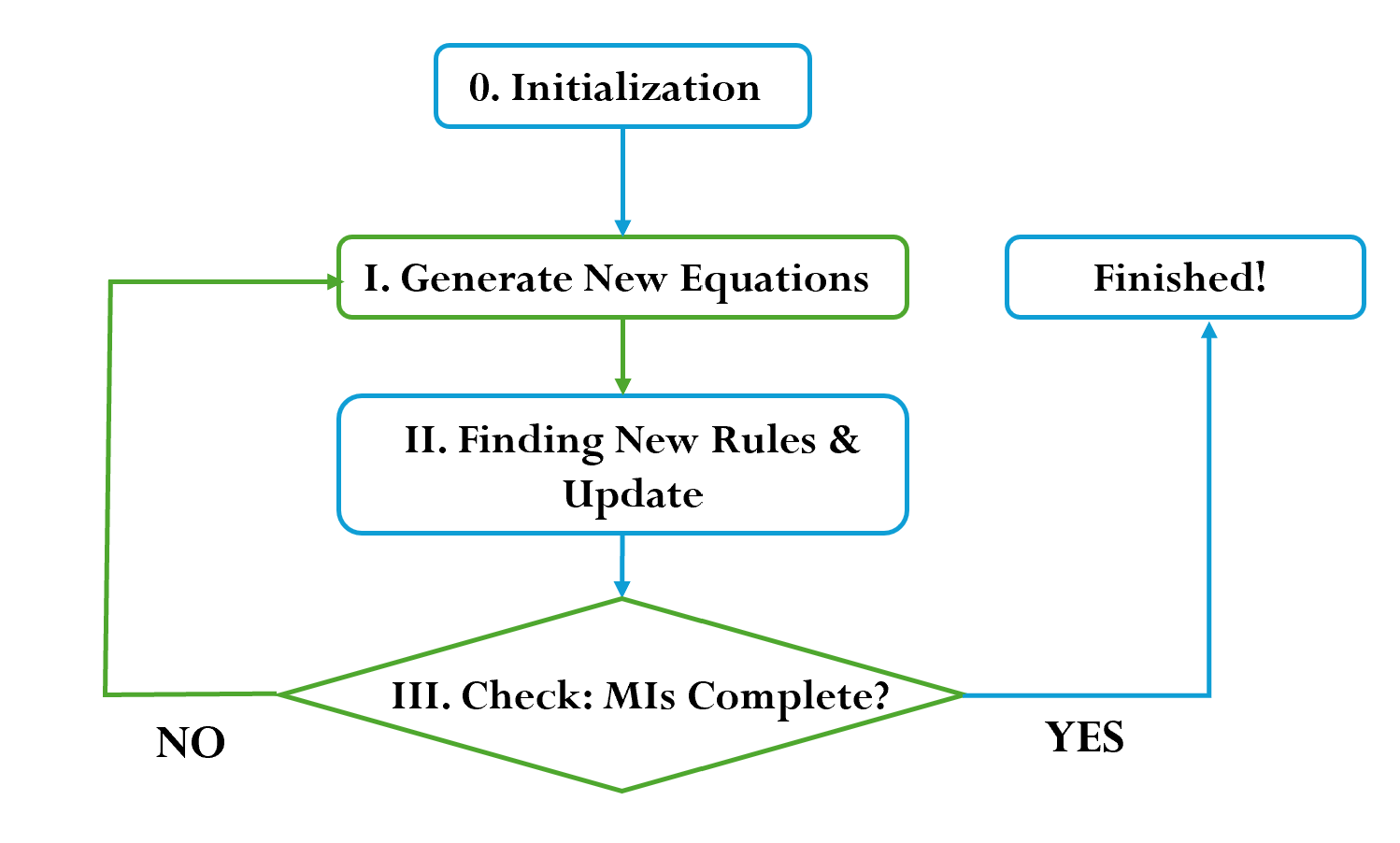}
	\caption{The iterative structure of the reduction algorithm.}
	\label{fig:alg_diagram}
\end{figure}

This section is organized as follows. In Section~\ref{sec:lattice_coverage}, we define the geometric interpretation of reduction rules and the completeness criterion in terms of reducible ($\mathcal{S}$) and irreducible ($\mathcal{U}$) lattice sets. We then describe the iterative procedure, which consists of three main modules:
\begin{itemize}
    \item \textbf{Module I (Section~\ref{sec:generation}):} Generating a system of differential equations.
    \item \textbf{Module II (Section~\ref{sec:solving}):} Solving these equations to extract valid reduction rules.
    \item \textbf{Module III (Section~\ref{sec:checking}):} Checking the completeness of the derived rules against the lattice criteria.
\end{itemize}
The overall workflow is illustrated in Figure~\ref{fig:alg_diagram}. If the completeness check in Module III fails, its output is fed back into Module I to guide the next round of equation generation. The algorithm is therefore not a one-pass elimination procedure; it is a structured loop in which the remaining irreducible lattice points determine what information must be generated next.

\subsection{Reduction Rules and Lattice Coverage}
\label{sec:lattice_coverage}
Before detailing the iterative steps, we first describe the geometric meaning of a reduction rule on the coefficient lattice. A symbolic reduction rule associated with an operator of index $\vec{o}$ determines exactly which lattice points can be moved to points of lower complexity. This viewpoint is useful because completeness is ultimately a statement about lattice coverage, not merely about the existence of formal operator identities.

Mathematically, the set of lattice points reducible by a rule with index $\vec{o}$ is defined as:
\begin{equation}
    \mathcal{S}_{\vec{o}} = \{ \vec{m} \in \mathbb{Z}^N \mid \vec{m} \ge 0 \quad \& \quad \vec{m} - \vec{o}\geq 0 \},
    \label{eq:red-set}
\end{equation}
where the inequality is understood component-wise. Conversely, the lattice points that cannot be reduced by this specific rule form the complementary set:
\begin{equation}
    \mathcal{U}_{\vec{o}} = \{ \vec{m} \in \mathbb{Z}^N \mid \vec{m} \ge 0 \quad \& \quad \vec{m} - \vec{o}\not\geq 0 \}.
    \label{eq:unred-set}
\end{equation}
The goal of our algorithm is to find a collection of $k$ rules $\{ \vec{o}_1, \dots, \vec{o}_k \}$ such that the intersection of their irreducible sets,
\begin{equation}
    \mathcal{U}_{total}^{(k)} \equiv \bigcap_{i=1}^k \mathcal{U}_{\vec{o}_i},
    \label{eq:total-unred}
\end{equation}
is minimized. Specifically, when the number of points in $\mathcal{U}_{total}$ equals the number of Master Integrals (MI), the reduction is complete.

\subsection{Module I: Equation Generation}
\label{sec:generation}

The first module of each iteration is to generate the differential equations that will feed the solver. These equations come from two sources: the fundamental IBP identities and the descendant equations\footnote{We use the term ``descendant'' because these equations are obtained by acting with operators on previously known equations, in direct analogy with descendant operators.} generated from rules found in previous iterations. The role of this module is exploratory: it produces candidate operator structures that may lead to new reduction rules.

\subsubsection{Initial Seed Equations from IBP}
\label{sec:initial_ibp}

The algorithm begins with a set of seed equations derived directly from the application of IBP identities to the generating function $G(\vec{\eta})$, which gives
\begin{equation}
    0= \int\prod_{i=1}^{L}\frac{\md^{D}\ell_i}{\mi\pi^{D/2}} {\d\over \d\ell_a}\left\{ R
{e^{\sum_{j=K+1}^N \eta_j s_0^{-1}\mathcal{D}_j}\over \prod_{i=1}^K(\mathcal{D}_i-s_0\eta_i) }\right\}, \label{eq:initial_ibp}
\end{equation}
where $R=\ell_i, K_j$. These IBP identities can be transformed to the differential equations of $\eta_i$ by mapping:
\begin{align}
    {s_0\over (\mathcal{D}_i-s_0\eta_i)^2}&\to {\d\over \d\eta_i}{1\over (\mathcal{D}_i-s_0\eta_i)}  \quad i\leq K,\nn
    {\mathcal{D}_i\over s_0}&\to {\d\over \d\eta_i}\quad i > K.
\end{align}
The explicit forms of these IBP identities are as follows:
\begin{itemize}
\item For the case $R=\ell_b$ we will have 
    \begin{align}
        0 = & D\delta_{ab}G_{\vec{\mu}}(\eta) %
	+\sum_{j=K+1}^N\left( \sum_{t=K+1}^N \a_{ab,j}^t \eta_j{\d\over \d\eta_t}G_{\vec{\mu}}(\eta)+ \sum_{t=1}^{K} \a_{ab,j}^t{\eta_j\over s_0}G_{\vec{\mu}-\vec{e}_t}(\eta)|_{\eta_t=0}\right. \nn & +\left. 
	{\eta_j}\left({\b_{ab,j}\over s_0}+\sum_{t=1}^{K}\a_{ab,j}^t \eta_t\right)G_{\vec{\mu}}(\eta)\right)
	- \sum_{i=1}^K\left(\a_{ab,i}^i\left( 1+\eta_i{\d\over \d\eta_i}\right) G_{\vec{\mu}}(\eta)\right.\nn & \left.
	+\left({\b_{ab,i}\over s_0}+ \sum_{t=1, t\neq i}^K \a_{ab,i}^t \eta_t \right){\d\over \d\eta_i} G_{\vec{\mu}}(\eta)
	+ \sum_{t=K+1}^N \a_{ab,i}^t {\d\over \d\eta_t }{\d\over \d\eta_i} G_{\vec{\mu}}(\eta)\right. \nn & 
	\left.  + \sum_{t=1, t\neq i}^K \a_{ab,i}^t{\d\over \d\eta_i}G_{\vec{\mu}-\vec{e}_t}(\eta)|_{\eta_t=0}\right),
    \label{eq:IBP1}
    \end{align}
    where $\a_{ab,k}$ and $\b_{ab,k}$ are from
    \begin{equation}
        \ell_b \cdot {\d\mathcal{D}_k\over \d \ell_a }=\sum_{t=1}^N \a_{ab,k}^t \mathcal{D}_t +\b_{ab,k},~~~~~~k=1,...,N.
    \end{equation}
\item For the case $R=K_b$ we will have
    \begin{align}
        0 = & 
	\sum_{j=K+1}^N\left( \sum_{t=K+1}^N \W\a_{ab,j}^t \eta_j{\d\over \d\eta_t}G_{\vec{\mu}}(\eta)+ \sum_{t=1}^{K} \W\a_{ab,j}^t{\eta_j\over s_0}G_{\vec{\mu}-\vec{e}_t}(\eta)|_{\eta_t=0}\right. \nn & +\left. 
	{\eta_j}\left({\W\b_{ab,j}\over s_0}+\sum_{t=1}^{K}\W\a_{ab,j}^t \eta_t\right)G_{\vec{\mu}}(\eta)\right)
	- \sum_{i=1}^K\left(\W\a_{ab,i}^i\left( 1+\eta_i{\d\over \d\eta_i}\right) G_{\vec{\mu}}(\eta)\right.\nn & \left.
	+\left({\W\b_{ab,i}\over s_0}+ \sum_{t=1, t\neq i}^K \W\a_{ab,i}^t \eta_t \right){\d\over \d\eta_i} G_{\vec{\mu}}(\eta)
	+ \sum_{t=K+1}^N \W\a_{ab,i}^t {\d\over \d\eta_t }{\d\over \d\eta_i} G_{\vec{\mu}}(\eta)\right. \nn & 
	\left.  + \sum_{t=1, t\neq i}^K \W\a_{ab,i}^t{\d\over \d\eta_i}G_{\vec{\mu}-\vec{e}_t}(\eta)|_{\eta_t=0}\right),
	\label{eq:IBP2}
    \end{align}
    where $\W\a_{ab,k}$ and $\W\b_{ab,k}$ are from
    \begin{equation}
        K_b \cdot {\d\mathcal{D}_k\over \d \ell_a }=\sum_{t=1}^N \W\a_{ab,k}^t \mathcal{D}_t +\W \b_{ab,k},~~~~~~k=1,...,N.
    \end{equation}
\end{itemize}
From this derivation, we can make several important observations about the initial seed equations:
\begin{enumerate}
    \item They are at most degree-two differential equations.
    \item The degree-two operators are of the form $\partial_i \partial_j$, where one index corresponds to a propagator and the other to an ISP.
    \item The degree-one operators $\partial_i$ only involve propagator indices.
    \item The boundary terms involve generating functions of subsectors evaluated at $\eta_k=0$, signifying a simpler topology.
\end{enumerate}
These properties are related to the form \eref{eq:initial_ibp} of generating functions.  When generalized to other situations with different generating functions, they should change accordingly. 

\subsubsection{Generating Descendant Equations}
\label{sec:descendant_eqns}

After the initial round, subsequent iterations generate new, more complex equations. This is achieved by differentiating the equations or reduction rules obtained from the previous solving module. 

We act with derivative operators $\partial_i \equiv \partial/\partial\eta_i$ on the set of equations identified as ``solvable'' from the previous round (further details in Section~\ref{sec:solving}). In principle, one could act with all possible $\partial_i$ on all available equations. However, this is computationally expensive and often redundant. We employ a selection strategy to prune the generation process:
\begin{itemize}
    \item \textbf{Criterion 1 (Efficiency):} If a reduction rule for a specific operator, e.g., $\partial_a$, has already been found, it implies that any power of the index $n_a$ can be reduced. Therefore, we can exclude acting with $\partial_a$ to generate new equations, as this direction is already under control.
    
    \item \textbf{Criterion 2 (Hierarchy):} We can prioritize generating equations from a specific subset of the existing ones, for example, only from T1-type rules, or only from degree-one rules. The choice of this criterion may evolve in later rounds of the algorithm.
    
    \item \textbf{Criterion 3 (Cutoff):} To manage the computational complexity, we introduce a cutoff degree $M_0$ and solve equations only up to degree $M \leq M_0$ after full simplification (see Section~\ref{sec:simplification}). We also do not solve equations with degree $M<0$. As discussed in Section~\ref{sec:reduction_rules}, the target of a reduction rule should have operator index $\vec{o}\geq 0$. For $M<0$, every highest-degree operator necessarily satisfies $\vec{o}\not\geq 0$, so such equations are used only as sources for further equation generation rather than as direct reduction rules.
    
    For equations with degree $M=0$, the situation is slightly more subtle, since the only good operator index is $\vec{o}=\vec{0}$, i.e., the zero-index operator. If the equation contains the identity operator as the highest-degree operator, we can solve it and conclude that the lattice point $\vec{n}=\vec{0}$ reduces to subsectors, which means that the sector contains no master integral. By contrast, if the sector is known to contain nontrivial master integrals, then no good operator is available and the equation should not be solved. We will encounter both situations in the examples below.
    
     The IBP relations naturally set the initial value to $M_0=2$. In most practical examples, $M_0=2$ is sufficient; however, the algorithm allows $M_0$ to be increased if necessary.
\end{itemize}
From our experience, Criterion 1 is highly effective. If the algorithm stalls (i.e., fails to find a complete set of rules), these restrictions should be relaxed to include a broader range of source equations.

Each newly-generated equation must be simplified using the full set of currently known reduction rules. This is arguably the most crucial step of the algorithm, as it reveals the truly novel operator structures that need to be solved. If an equation simplifies to $0=0$, it is considered a \textbf{trivial descendant} and is discarded, as it provides no new information. Otherwise, it is a \textbf{non-trivial descendant} and is added to the pool of equations to be solved in Module II.

\subsubsection{The Simplification Procedure}
\label{sec:simplification}

Once an equation has been generated, whether by $\partial_i$ action or by linear combination, it must be simplified with the full set of currently known reduction rules. This step is essential because it removes operator content that is already under control and leaves behind only the genuinely new structures that still need to be solved.

The simplification strategy relies on two ordered lists of reduction rules: $\mathcal{R}_{A}$ (TA-type rules) and $\mathcal{R}_{B}$ (TB-type rules) (see, for example, \eref{eq:RRule-T1A},  \eref{eq:RRule-T1B}, \eref{eq:RRule-T2A} and \eref{eq:RRule-T2B}) in the subsection \ref{sec:reduction_rules}). We prioritize simpler rules to keep expressions manageable. Thus, we perform the simplification in two distinct rounds:
\begin{itemize}
    \item \textbf{First Round:} Simplify using only $\mathcal{R}_{A}$.
    \item \textbf{Second Round:} Simplify the result of the first round using $\mathcal{R}_{B}$.
\end{itemize}
Within each list, simpler reduction rules found in earlier iterations and T1-type rules are given higher priority.

\begin{figure}
	\centering
	\includegraphics[width=1.0\linewidth]{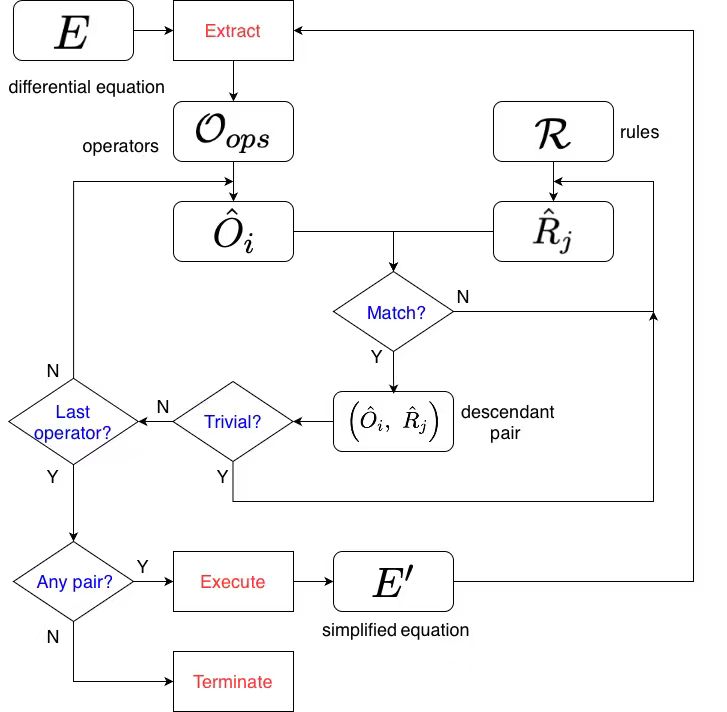}
	\caption{Flowchart of the simplification.}
	\label{fig:simplification_flow}
\end{figure}

The logic for a single round of simplification (given a rule list $\mathcal{R}$) follows the workflow depicted in Figure~\ref{fig:simplification_flow}.
\begin{itemize}
    \item \textbf{Step 1: Extract and Sort Highest Degree Operators.}
    Given an equation $E$, we identify all  highest degree differential operators and sort them to optimize the matching process. The sorting logic uses the finer index $(\vec{a}, \vec{b})$. An operator with a smaller degree of $\vec{b}$ comes first. For operators with the same degree, we adopt a convention, such as prioritizing those involving ISPs over propagators. Let this sorted list of highest-degree operators be $\mathcal{O}_{ops} = \{ \WH{O}_1, \WH{O}_2, \dots \}$.
    
    \item \textbf{Step 2: Selection and Matching.}
    We iterate through each operator $\WH{O}_i$ in $\mathcal{O}_{ops}$ and attempt to match it against the rules in $\mathcal{R}$ based on \ref{sec:descendants}. For a specific operator $\WH{O}_i$ and a candidate reduction rule for operator $\WH{R}_j$, we check if $\WH{O}_i$ is a \textbf{descendant} of the operator $\WH{R}_j$ according to the criterion \eref{des-condition}:
   	\begin{enumerate}
    \item If it is not satisfied, we continue checking other rules. 
    \item If satisfied, $\WH{O}_i$ is a descendant of $\WH{R}_j$. 
    Before finalizing this match, we verify if substituting this expression into equation $E$ yields a trivial identity ($0=0$)\footnote{ In the practice, this information is trivially obtained if we remember the origin of this equation, i.e., acting which differential operator on which reduction rule. }:
    	\begin{enumerate}
    	\item  If it is trivial, we discard this match and continue checking other rules. 
    	\item If it is non-trivial, we record the pair $(\WH{O}_i, \WH{R}_j)$ as a valid simplification (which will be carried out later  according to the prescription given in \eref{des-apply-TA} or \eref{des-apply-TB}) and immediately stop searching for $\WH{O}_i$.
		\end{enumerate}
    \end{enumerate}

    \item \textbf{Step 3: Execute the simplification.}
    After checking all operators in $\mathcal{O}_{ops}$, we determine if any valid simplification pairs were found.
    \begin{enumerate}
        \item If \textbf{No pairs are found}: The current set of operators cannot be reduced further by list $\mathcal{R}$. The process terminates for this round.
        
        \item If \textbf{Pairs are found}: We execute the substitutions for all recorded pairs according to the prescription given in \eref{des-apply-TA} or \eref{des-apply-TB}. For T1-type reduction rules, the substitutions can be carried out simultaneously. For T2-type reduction rules, it is safer to apply the substitutions sequentially and to check after each step whether the pair $(\WH{O}_i, \WH{R}_j)$ still appears in the updated equation.
        
        After done all substitutions, we produce a modified equation $E'$. 
        
        
    \end{enumerate}

    \item \textbf{Step 4: Verification and Iteration}
        To ensure the robustness of the result, we perform a final check on $E'$. We verify if the modified equation still contains any highest-degree operators that are candidates for reduction.
        \begin{enumerate}
             \item If such operators remain, we update the current equation $E \leftarrow E'$ and return to Step 1 to initiate another pass of extraction and matching.
             \item If no such operators remain, the process terminates for this round.   We must emphasize that we should put the resulted equation 
             into the category of lower degree, which will be simplified later. 
        \end{enumerate}
\end{itemize}

In practice, the same high-degree descendant operator $\WH{O}_i$ often appears across multiple different generated equations. To enhance computational efficiency, we implement a caching mechanism that records the mapping between a descendant operator and its corresponding "mother" rule, i.e., $\WH{O}_i \to \WH{R}_j$. When the algorithm encounters an operator $\WH{O}_i$ that has been successfully matched in a previous simplification instance, it retrieves the substitution directly from the cache, bypassing the redundant search and derivative construction in Step 2.

For the second-round simplification using rules from $\mathcal{R}_{B}$, we need some additional notation beyond the basic logic. As noted in Section~\ref{sec:reduction_rules}, TB-type rules may require acting with a zero-index operator $\WH{O}_{0}$ on the entire equation $E$ to isolate the target term, transforming $E \to \WH{O}_{0} E$ (see \eref{des-apply-TB}). While rigorous, this increases complexity. In practice, we often skip this global multiplication during the generation phase and reserve TB-type simplifications for the solving module, where they are applied strictly as needed.

Upon completion of the simplification rounds, the resulting system of equations is naturally filtered. Equations that have been reduced to mathematical identities ($0=0$) are discarded, as they provide no new constraints. The remaining set of non-trivial differential equations serves as the clean input for the subsequent \textbf{Module II}.

\subsection{Module II: Equation Solving and Rule Extraction}
\label{sec:solving}

The primary input for this module is the simplified system of differential equations produced by Module I. The objective is to decouple that system and extract explicit reduction rules. Whenever a good high-degree operator can be solved, the corresponding rule enlarges the reducible region $\mathcal{S}_{\vec{o}}$ of the lattice. In this sense, Module II is the stage at which algebraic information is converted into geometric progress.

To handle the complexity of the system systematically, we organize the equations by their total differential degree. The algorithm proceeds hierarchically, resolving the highest-degree sector (typically degree-2) before propagating the results to lower-degree sectors.

For a fixed degree $M$, the solving procedure is structured into two distinct steps: \textbf{Linearization} (Sec.~\ref{sec:linearization}), which classifies equations via Gaussian elimination, and \textbf{Operator Selection} (Sec.~\ref{sec:op_selection}), which implements the logical criteria for determining reduction targets. Finally, an \textbf{Update Mechanism} (Sec.~\ref{sec:update_mech}) ensures that new reduction rules are propagated globally. { These three manipulations should be repeated for each set of equations divided using the degree and type.} 

\subsubsection{Step 1: Linearization and Classification}
\label{sec:linearization}

Let $\mathcal{E}_M$ denote the subset of equations with maximum degree $M$. We aim to isolate the linearly independent constraints on the operators of degree $M$. Each equation in $\mathcal{E}_M$ can be decomposed as:
\begin{equation}
    E_i = \sum_{a} c_{i, a} \WH{O}_{0,i,a} \WH{O}_{\vec{o}_a}^{(M)} + \WH{R}_i^{(<M)} = 0 \,,
    \label{eq:linearized_form}
\end{equation}
where $\WH{O}_{\vec{o}_a}^{(M)}$ represent unique operators with a given operator index $\vec{o}_a$ of degree $M$ (i.e., $\WH{O}^{F}_{\vec{o}_a}$ in \eref{gen-O} in the previous section), $c_{i,a}$ are rational coefficients and $\WH{O}_{0,i,a}$ are zero-index operators, and $\WH{R}^{(<M)}$ encapsulates all terms of degree strictly less than $M$.. The combinations $c_{i, a} \WH{O}_{0,i,a}$ are treated as scalars since zero-index operators commute with each others.

We construct a coefficient matrix $C$ with entries $c_{i,a}\WH{O}_{0,i,a}$ and perform Gaussian elimination\footnote{With these zero-index operators, when carrying out Gaussian elimination, we allow only three types of manipulations: addition, subtraction, and multiplication.} to bring it to row-reduced form. This linear algebra step follows naturally from the structure of the input and classifies the resulting rows into three categories:

\begin{enumerate}
    \item \textbf{Null Rows (Seeds for Lower Degrees):}
    If a row results in zero coefficients for all degree-$M$ operators, the equation reduces to the form $\WH{R}'^{(<M)} = 0$. Such equations describe constraints entirely within lower-degree sectors. They are removed from the current degree-$M$ processing, and injected into the corresponding lower-degree subsets only after fully simplified by known reduction rules following \ref{sec:simplification}.

    \item \textbf{Type-1 Rows (Direct Reductions):}
    A row containing exactly one non-zero coefficient for a degree-$M$ operator $\WH{O}_k$ allows for immediate reduction:
    \begin{equation}
        c_k \WH{O}_k + \WH{R}' =0 \,.~~~~\label{comb-1-1}
    \end{equation}
    This unambiguously establishes a reduction rule for $\WH{O}_k$. We refer to these as \textbf{T1-type} equations.

    \item \textbf{Type-2 Rows (Coupled Relations):}
    A row with two or more non-zero coefficients defines a coupled constraint:
    \begin{equation}
        c_1 \WH{O}_{p_1} + c_2 \WH{O}_{p_2} + \dots + \WH{R}' = 0 \,.~~~~\label{comb-1-2}
    \end{equation}
    Solving such \textbf{T2-type} equations requires a choice: we must select one operator $\WH{O}_{p_i}$ to be the reduction target (LHS), expressing it in terms of the others. This selection is non-trivial and determines the efficiency of the reduction on the lattice $\mathcal{S}_{\vec{o}}$, as detailed in the following subsection.
\end{enumerate}

Crucially, the linear combinations \eref{comb-1-1} and \eref{comb-1-2} act as {\bf ``seeds''} that generate non-trivial relations in subsequent steps, as discussed in Section~\ref{sec:descendant_eqns}.

\subsubsection{Step 2: Solving and Operator Selection Strategy}
\label{sec:op_selection}
For T1-type equations, solving is straightforward: they directly yield T1A-type or T1B-type reduction rules. The T2-type equations are more delicate. In that case, we must define a global priority ordering that determines which operator should be solved for and placed on the left-hand side. The stability and convergence of the algorithm depend strongly on this choice.

The solving stage is carried out iteratively, always prioritizing the simplest reductions first. We begin by resolving every available T1-type equation. The resulting rules are then used to {\bf update} the remaining T2-type equations (as described in Section~\ref{sec:update_mech}), and some T2-type equations may simplify into new T1-type equations. We repeat this loop until no additional T1-type equations remain. Only then do we choose targets inside the surviving T2-type equations, using the following ordered criteria:

\begin{itemize}
    \item \textbf{Crit. 1: Validity (Negative-index avoidance).} Operators with index $\vec{o}\not\geq 0$ should not be chosen as solved targets. If an equation contains only such operators at highest degree, it should be retained as a source for further equation generation rather than turned directly into a reduction rule.

    \item \textbf{Crit. 2: Lattice reduction (the ``good'' operator).}
    We prioritize solving for an operator $\WH{O}_i$ if it lowers a specific lattice index $n_a$ in a way that is not neutralized by the competing operators in the same equation.
    
    Technically, let $\vec{v}_i$ and $\vec{v}_j$ be the lattice shifts associated with operators $\WH{O}_i$ and $\WH{O}_j$ (i.e., the operator indices). We define $\vec{v}_{ij} \equiv \vec{v}_i - \vec{v}_j$. If there exists a component $a$ such that $(\vec{v}_{ij})_a > 0$ for every competing operator $j$ in the equation, then solving for $\WH{O}_i$ guarantees a net reduction in the index $n_a$. In geometric terms, the recurrence moves lattice points toward the boundary of $\mathcal{S}_{\vec{v}_i}$.
    
    We emphasize that a single equation may contain several such good operators, each reducing a different lattice direction. The later criteria are used only when this ambiguity is not already resolved by the desired reduction flow.

    \item \textbf{Crit. 3: Boundary-aware hierarchy.}
    In some cases, an equation contains an operator $\WH{O}_{\text{bad}}$ whose index has a negative component in the $t$-th direction. Such an operator vanishes automatically on the boundary $n_t=0$. This observation can be used strategically: first employ other equations to force the irreducible set onto that boundary, and only then reuse the present equation to solve an additional reduction rule after the problematic term has become ineffective.
    
    A closely related possibility is that the same equation also contains a good operator that directly reduces $n_t$, for example $\partial/\partial\eta_t$. In that situation, the equation can still be used safely, because the presence of $\WH{O}_{\text{bad}}$ does not obstruct the reduction of the very index that controls its boundary behavior.


    \item \textbf{Crit. 4: Empirical hierarchy for variables.}
    If the preceding criteria are still insufficient to select a unique target, we fall back on an empirical ordering that has proven effective in practice. For degree-two operators, the suggested precedence is:
    \begin{equation}
        \partial_i \partial_a \succ \partial_a \partial_b \succ \partial_a^2 \succ \partial_i \partial_j \succ \partial_i^2 \,,
    \end{equation}
    where indices $a,b$ refer to ISP variables and $i,j$ refer to propagators. For degree-1, we prefer $\partial_a \succ \partial_i$.
\end{itemize}


\subsection{ Global Update Mechanism}
\label{sec:update_mech}

Once a new {\bf TA-type} reduction rule has been derived, it must be propagated through the entire system by the simplification procedure (see Section~\ref{sec:simplification}). This global update keeps the iterative process coherent: it removes solved operators from all remaining equations and often exposes new T1-type structures that were not visible before the update.

During the update, we treat the unsolved equations and the library of previously established rules on the same footing. The procedure is as follows:

\begin{enumerate}
    \item \textbf{Unification:}
    We first convert all previously solved reduction rules back into their equation forms ($\text{LHS} - \text{RHS} = 0$). These are then pooled together with the current set of unsolved  equations.

    \item \textbf{Global Substitution:}
    We apply the newly derived reduction rule(s) to this unified pool of equations to do the simplification procedure (see \ref{sec:simplification}). Every occurrence of the newly reduced operator is substituted by its corresponding expression. This step eliminates the solved operator from the entire system, ensuring it never appears as a variable again.
\end{enumerate}

After the global update, the system is effectively reset. We feed the updated pool of equations back into the solver and repeat the process until the set of reduction rules stops changing. At that point, the rules are maximally decoupled within the chosen strategy.

 We want to remark that for simplicity, the update could be performed only for TA-type reduction rules. For TB-type reduction rules, we use them in the solving step..

\subsection{Module III: Completeness Check}
\label{sec:checking}
Having obtained a candidate set of reduction rules, we must determine whether it is complete. This is done by analyzing the total irreducible set $\mathcal{U}_{total}$ defined in Eq.~\eqref{eq:total-unred}. The completeness check therefore plays a dual role: it is both a stopping criterion and a diagnostic tool. If the rule set is incomplete, the geometry of $\mathcal{U}_{total}$ indicates what kind of equations should be generated next.
We denote the number of lattice points in $\mathcal{U}_{total}$ by $|\mathcal{U}_{total}|$. Two situations must be distinguished:
\subsubsection{Case 1: Known Number of Master Integrals.}
If the number of master integrals, $N_{MI}$, is known beforehand (for example from geometric or algebraic methods), the stopping condition is simply:
\begin{equation}
|\mathcal{U}_{total}| = N_{MI}.
\end{equation}
If $|\mathcal{U}_{total}| > N_{MI}$, then the rules are incomplete and the next iteration must generate additional equations.

\subsubsection{Case 2: Unknown Number of Master Integrals.}
A useful feature of the algorithm is that it can also determine $N_{MI}$. If we arrive at a {\bf finite} set $\mathcal{U}_{total}$ without prior knowledge of the master-integral count, we still need to check whether that set is already complete. To do so, we perform a consistency test: we reduce selected points outside $\mathcal{U}_{total}$ along different reduction paths, that is, with different combinations of rules.
\begin{itemize}
\item If all paths yield identical expressions, the reduction is consistent and $N_{MI} = |\mathcal{U}_{total}|$.
\item If different paths yield conflicting expressions, then hidden relations (IBP identities) still exist among the points in $\mathcal{U}_{total}$. Those relations must be added to the system, and $\mathcal{U}_{total}$ must be reduced further until consistency is restored.
\end{itemize}

\section{The top sector of the sunset diagram}
\label{sec:sunset-top}

Having established the general algorithm in Section~\ref{sec:algorithm}, we now turn to explicit examples. We begin with the top sector of the sunset diagram because it is the most economical setting in which all of the main ingredients already appear. Although the topology is simple, it still contains the essential interplay between generating functions, operator identities, descendant equations, and lattice-based completeness. For that reason, we treat the sunset example in more detail than the later case studies and use it as the main pedagogical thread of the paper.

We begin by defining the kinematics and the integral family. The sunset topology is characterized by the following inverse propagators:
\begin{align}
\mathcal{D}_1 = \ell_1^2 - m_1^2 , \quad
\mathcal{D}_2 = \ell_2^2 - m_2^2 , \quad
\mathcal{D}_3 = (\ell_1 + \ell_2 - K)^2 - m_3^2 ,
\label{sunset-props}
\end{align}
where $K$ is the external momentum. To complete the basis of scalar products involving loop momenta, we introduce two auxiliary propagators (irreducible scalar products) which appear in the numerators:
\begin{equation}
\mathcal{D}_4 = \ell_1 \cdot K , \quad
\mathcal{D}_5 = \ell_2 \cdot K .
\label{sunset-isps}
\end{equation}
In our generating-function formalism, a generic sector is defined by the subset of propagators appearing in the denominator. For the sunset family, since there are three physical propagators, there are $2^3=8$ potential generating functions. However, because the generating functions are defined as formal sums of Feynman integrals, the non-vanishing functions relevant to this topology are restricted to the top sector and its subsectors:

\begin{equation}
G_{111}, \quad G_{110}, \quad G_{101}, \quad G_{011}.
\label{sunset-sectors}
\end{equation}
Explicitly, using the definition in Eq.~\eqref{eq:generating_function}, these generating functions are constructed as follows:
\begin{align}
G_{111} &= \int \frac{\mathrm{d}^{D}\ell_1 \mathrm{d}^{D}\ell_2}{(\mathrm{i}\pi^{D/2})^2} \, \frac{ \exp\left( s_0^{-1} \sum_{j=4}^5 \eta_j \mathcal{D}_j \right) }{ \prod_{i=1}^3 (\mathcal{D}_i - \eta_i s_0) },\\
G_{110} &= \int \frac{\mathrm{d}^{D}\ell_1 \mathrm{d}^{D}\ell_2}{(\mathrm{i}\pi^{D/2})^2} \, \frac{ \exp\left( s_0^{-1} \sum_{j=3}^5 \eta_j \mathcal{D}j \right) }{ \prod_{i=1}^2 (\mathcal{D}_i - \eta_i s_0) } , \\
G_{011} &= \int \frac{\mathrm{d}^{D}\ell_1 \mathrm{d}^{D}\ell_2}{(\mathrm{i}\pi^{D/2})^2} \, \frac{ \exp\left( s_0^{-1} \sum_{j \in {1,4,5}} \eta_j \mathcal{D}j \right) }{ \prod_{i=2}^3 (\mathcal{D}_i - \eta_i s_0) } , \\
G_{101} &= \int \frac{\mathrm{d}^{D}\ell_1 \mathrm{d}^{D}\ell_2}{(\mathrm{i}\pi^{D/2})^2} \, \frac{ \exp\left( s_0^{-1} \sum_{j \in {2,4,5}} \eta_j \mathcal{D}j \right) }{ \prod_{i \in {1,3}} (\mathcal{D}_i - \eta_i s_0) }.
\end{align}
In the remainder of this section, we focus on the top-sector generating function $G_{111}$. Our aim is not only to derive the corresponding symbolic reduction rules, but also to show explicitly how the iterative workflow operates in practice: which equations are generated, how they are simplified, what remains irreducible after each round, and how the reduction closes in different kinematic regimes.

\subsection{The first iteration of the computation}
\label{sec:sunset-round1}

We start from the fundamental IBP relations and use them to obtain the first layer of reduction rules. This first iteration is important because it already exposes the basic geometry of the reduction problem: it identifies which directions in the lattice are immediately controlled and which subsets remain untouched and must be addressed in later rounds.

\subsubsection*{Generating and organizing the IBP relations}

Applying the general formulas in Eqs.~\eqref{eq:IBP1} and \eqref{eq:IBP2} to the sunset topology yields six fundamental differential equations. We organize them according to the degree of the operators acting on the generating function. Since the present discussion is devoted to the top sector, we suppress boundary terms involving subsector generating functions in order to keep the operator structure transparent.

The first relation is given by:
\begin{align}
    0 = & 2 {\d\over \d \eta_5} {\d\over \d \eta_3} G_{111} -{2m_1^2 \over s_0} {\d \over \d \eta_1} G_{111}-{f_{+-+}\over s_0}{\d\over \d \eta_3} G_{111} +(D-3) G_{111}\nn
    & + \eta_4{\d\over \d \eta_4} G_{111}-2 \eta_1 {\d \over \d \eta_1} G_{111}-(\eta_3+\eta_1-\eta_2){\d\over \d \eta_3} G_{111}, \label{sunset-2-1-4}
\end{align}
where we have introduced the shorthand notation for kinematic combinations: \begin{equation} f_{abc} = a m_1^2 + b m_2^2 + c m_3^2 - K^2, \quad \text{with } a,b,c = \pm1. \label{f-abc-def} \end{equation}
The remaining five differential equations are listed below:
\begin{align}
0 = & - 2 {\d \over \d \eta_4}{\d \over \d \eta_1} G_{111}- 2 {\d \over \d \eta_5}{\d \over \d \eta_1} G_{111}- 2 {\d\over \d \eta_4} {\d\over \d \eta_3} G_{111} -{ f_{--+}\over s_0}{\d \over \d \eta_1} G_{111} -{f_{-++}\over s_0}{\d\over \d \eta_3} G_{111} \nn
& + \eta_4{\d\over \d \eta_5}G_{111}-(\eta_3-\eta_1-\eta_2){\d \over \d \eta_1} G_{111}-(\eta_3+\eta_2-\eta_1){\d\over \d \eta_3} G_{111}, ~~~~\label{sunset-2-2-4} \end{align}
\begin{align} 0 = & - 2 {\d \over \d \eta_4}{\d \over \d \eta_1} G_{111} - 2 {\d \over \d \eta_4}{\d \over \d \eta_3} G_{111} - 2 {\d \over \d \eta_5}{\d \over \d \eta_3} G_{111} + {2K^2 \over s_0}{\d \over \d \eta_3} G_{111}+{K^2\over s_0}\eta_4 G_{111},
~~~~~~~~\label{sunset-2-3-4} \end{align}
%
%
\begin{align} 0 = & - 2 {\d \over \d \eta_4}{\d \over \d \eta_2} G_{111}- 2 {\d \over \d \eta_5}{\d \over \d \eta_2} G_{111}- 2 {\d\over \d \eta_5} {\d\over \d \eta_3} G_{111}  -{ f_{--+}\over s_0}{\d \over \d \eta_2} G_{111} -{f_{+-+}\over s_0}{\d\over \d \eta_3} G_{111} \nn
& + \eta_5{\d\over \d \eta_4}G_{111}-(\eta_3-\eta_1-\eta_2){\d \over \d \eta_2} G_{111}-(\eta_3+\eta_1-\eta_2){\d\over \d \eta_3} G_{111}  ~~~~\label{sunset-2-4-4} \end{align}
%
%
\begin{align} 0 = & 2 {\d\over \d \eta_4} {\d\over \d \eta_3} G_{111}    +{-2m_2^2\over s_0}  {\d \over \d \eta_2} G_{111}-{f_{-++}\over s_0} {\d\over \d \eta_3} G_{111} +(D-3) G_{111} \nn
& + \eta_5{\d\over \d \eta_5} G_{111}-2 \eta_2 {\d \over \d \eta_2} G_{111}-(\eta_3-\eta_1+\eta_2){\d\over \d \eta_3} G_{111} ~~~~\label{sunset-2-5-4} \end{align}
%
%
\begin{align} 0 = &- 2 {\d \over \d \eta_5}{\d \over \d \eta_2} G_{111}- 2 {\d \over \d \eta_4}{\d \over \d \eta_3} G_{111}- 2 {\d \over \d \eta_5}{\d \over \d \eta_3}G_{111} + {2K^2\over s_0} {\d \over \d \eta_3} G_{111} + {K^2\over s_0}  \eta_5G_{111}  ~~~~~~~~\label{sunset-2-6-4} \end{align}

\subsubsection*{Solving the system of equations}

With the differential system in hand, we next isolate its highest-degree part and solve for the associated reduction rules. In the present case, all six equations have degree-two leading operators, so the first task is to analyze the corresponding coefficient matrix:
\begin{align}
    \left( \begin{array}{c| c |  c |c |c |c |c} \partial_{1}\partial_{4} & \partial_{2}\partial_{4} & \partial_{3}\partial_{4} & \partial_{1}\partial_{5} & \partial_{2}\partial_{5} & \partial_{3}\partial_{5}  & \text{Source Eq.}\\  \hline
	 &  &  &  &  & 2 & {\eref{sunset-2-1-4}  }  \\  \hline 
	-2 &  & -2  & -2 &  &   & {\eref{sunset-2-2-4}  } \\  \hline 
	-2 &  & -2 &  &  & -2  & {\eref{sunset-2-3-4}  }  \\  \hline 
	 & -2 &  &  & -2 & -2  & {\eref{sunset-2-4-4}  }   \\  \hline 
	 &  & 2 &  &  &  & {\eref{sunset-2-5-4}  }  \\   \hline 
	 &  & -2 &  & -2 & -2 & {\eref{sunset-2-6-4}  } 
\end{array}\right),~~~~~~~\label{sunset-B1-1-1}
\end{align}
where $\partial_i \equiv \frac{\partial}{\partial \eta_i}$. Applying Gaussian elimination, we transform the matrix into a form where each row contains a leading term, corresponding to a specific highest-degree operator:
\begin{align}
    \left( \begin{array}{c | c | c |c |c |c |c} \partial_{1}\partial_{4} & \partial_{2}\partial_{4} & \partial_{3}\partial_{4} & \partial_{1}\partial_{5} & \partial_{2}\partial_{5} & \partial_{3}\partial_{5} & \text{Linear Comb.}\\     \hline
    -2 &  &  &  &  &   & {\eref{sunset-2-3-4}  }+{\eref{sunset-2-1-4}  } + {\eref{sunset-2-5-4}  } \\  \hline 
    & -2 &  &  &  &   & {\eref{sunset-2-4-4}  } -{\eref{sunset-2-6-4}  } -{\eref{sunset-2-5-4}  } \\  \hline 
    &  & 2 &  &  &  & {\eref{sunset-2-5-4}  }  \\   \hline 
	 &  &   & -2 &  &   & {\eref{sunset-2-2-4}  } -{\eref{sunset-2-1-4}  }- {\eref{sunset-2-3-4}  }\\  \hline 
	&  &  &  & -2 &  & {\eref{sunset-2-6-4}  } + {\eref{sunset-2-1-4}  }+ {\eref{sunset-2-5-4}  }\\ \hline 
    &  &  &  &  & 2 & {\eref{sunset-2-1-4}  }
\end{array}\right),~~~~~~~\label{sunset-B1-1-2}
\end{align}
Each row in \eref{sunset-B1-1-2} is of T1A type, so each one directly yields a degree-two reduction rule in terms of lower-degree operators. We therefore obtain the following six primary reduction rules:

\begin{align} 2 {\d\over \d \eta_5} {\d\over \d \eta_3} G_{111}= & -\frac{2m_1^2}{s_0} {\d \over \d \eta_1} G_{111}-\frac{f_{+-+}}{s_0}{\d\over \d \eta_3} G_{111} +(D-3) G_{111}\nn
& + \eta_4{\d\over \d \eta_4} G_{111}-2 \eta_1 {\d \over \d \eta_1} G_{111}-(\eta_3+\eta_1-\eta_2){\d\over \d \eta_3} G_{111}, ~~~~~~~\label{sunset-B1-2-1}\end{align}
%
%
\begin{align}  2 {\d \over \d \eta_5}{\d \over \d \eta_1} G_{111}= & -{ f_{+-+}\over s_0}{\d \over \d \eta_1} G_{111} -{2m_3^2\over s_0}{\d\over \d \eta_3} G_{111} + \eta_4{\d\over \d \eta_4} G_{111} + \eta_4{\d\over \d \eta_5}G_{111} \nn &+ (D-3) G_{111}-(\eta_3+\eta_1-\eta_2) {\d \over \d \eta_1} G_{111}-2 \eta_3{\d\over \d \eta_3} G_{111}\nn
&-\eta_4 \frac{K^2}{s_0} G_{111},~~~~~~~\label{sunset-B1-2-2}
\end{align}
%
%
\begin{align}  2 {\d \over \d \eta_4}{\d \over \d \eta_1}  G_{111}= & {2m_3^2\over s_0}{\d \over \d \eta_3} G_{111}+ {2m_1^2 \over s_0} {\d \over \d \eta_1} G_{111}+{2m_2^2\over s_0}{\d\over \d \eta_2} G_{111} -2(D-3) G_{111}\nn &  - \eta_4{\d\over \d \eta_4} G_{111} -\eta_5{\d\over \d \eta_5} G_{111}+2 \eta_1 {\d \over \d \eta_1} G_{111}+ 2 \eta_2 {\d \over \d \eta_2} G_{111}+2\eta_3{\d\over \d \eta_3} G_{111}\nn &  +  \eta_4 \frac{K^2}{s_0} G_{111},~~~~~~~\label{sunset-B1-2-3}
\end{align}
%
%
\begin{align} 2 {\d \over \d \eta_4}{\d \over \d \eta_2} G_{111}=&-{ f_{-++}\over s_0}{\d \over \d \eta_2} G_{111} -{2 m_3^2\over s_0}{\d\over \d \eta_3} G_{111} + \eta_5{\d\over \d \eta_4} G_{111} + \eta_5{\d\over \d \eta_5}G_{111}\nn &+ (D-3) G_{111}-(\eta_3-\eta_1+\eta_2) {\d \over \d \eta_2} G_{111}-2 \eta_3{\d\over \d \eta_3} G_{111} \nn & -\eta_5 \frac{K^2}{s_0}G_{111},~~~~~~~~\label{sunset-B1-2-4}
\end{align}
%
%
%
\begin{align} 2 {\d\over \d \eta_4} {\d\over \d \eta_3} G_{111}=&  -2m_2^2 s^{-1}_0 {\d \over \d \eta_2} G_{111}-f_{-++}s^{-1}_0{\d\over \d \eta_3} G_{111} +(D-3) G_{111}, \nn
&  + \eta_5{\d\over \d \eta_5} G_{111}-2 \eta_2 {\d \over \d \eta_2} G_{111}-(\eta_3-\eta_1+\eta_2){\d\over \d \eta_3} G_{111}, ~~~~~~~\label{sunset-B1-2-5}\end{align}
%
%
\begin{align}  2 {\d \over \d \eta_5}{\d \over \d \eta_2} G_{111}=&  {2m_3^2\over s_0}{\d \over \d \eta_3} G_{111}+ {2m_1^2 \over s_0} {\d \over \d \eta_1} G_{111}+{2m_2^2\over s_0}{\d\over \d \eta_2} G_{111} -2(D-3) G_{111},\nn
& - \eta_4{\d\over \d \eta_4} G_{111}  -\eta_5{\d\over \d \eta_5} G_{111}+2 \eta_1 {\d \over \d \eta_1} G_{111}+ 2 \eta_2 {\d \over \d \eta_2} G_{111}+2\eta_3{\d\over \d \eta_3} G_{111} \nn & + \eta_5 \frac{K^2}{s_0} G_{111}.~~~~~~~\label{sunset-B1-2-6}
\end{align}
The six highest-degree operators on the left-hand side follow a distinct pattern:
\begin{equation}
    \left\{\frac{\partial}{\partial \eta_1}, \frac{\partial}{\partial \eta_2}, \frac{\partial}{\partial \eta_3}\right\}  \times  \left\{\frac{\partial}{\partial \eta_4}, \frac{\partial}{\partial \eta_5}\right\} ,
\label{rule-pattern}
\end{equation}
which corresponds to the product of derivatives with respect to propagator variables ($\eta_1, \eta_2, \eta_3$) and derivatives with respect to ISP variables ($\eta_4, \eta_5$). This is the first concrete manifestation of the mixed propagator--ISP pattern already anticipated in the general formalism.

\subsubsection*{Completeness check}

We now interpret these rules on the coefficient lattice in order to determine what has, and has not, been reduced. Translating the differential equations into recurrence relations for the Taylor coefficients makes this interpretation explicit. For example, Eq.~\eqref{sunset-B1-2-1} implies:
\begin{align} 
&2 (n_5+1)(n_3+1) F_{\vec{n}+\vec{e}_3+\vec{e}_5}\nn
= &-2{m_1^2 \over s_0} (n_1+1)F_{\vec{n}+\vec{e}_1}-{f_{+-+}\over s_0}(n_3+1)F_{\vec{n}+\vec{e}_3}+(D-3 + n_4 -2 n_1 -n_3) F_{\vec{n}}\nn
&-(n_3+1) F_{\vec{n}+\vec{e}_3-\vec{e}_1} +(n_2+1) F_{\vec{n}+\vec{e}_3-\vec{e}_2}~~~~~~~\label{sunset-B1-2-1-rec}
\end{align}
where $\vec{e}_i$ denotes the unit vector in the $i$-th direction. The term on the left-hand side, $F_{\vec{n}+\vec{e}_3+\vec{e}_5}$, has the highest index sum. This relation allows us to reduce any integral with index $(n_1, n_2, n_3, n_4, n_5)$ provided that $n_3 \ge 1$ and $n_5 \ge 1$. Consequently, the integrals that cannot be reduced by this specific rule belong to the set $(n_1, n_2, 0, n_4, n_5) \cup (n_1, n_2, n_3, n_4, 0)$.

Combining the constraints from all six reduction rules derived above, we identify the lattice points that remain irreducible at this stage. Following the completeness check encoded in Eq.~\eqref{eq:total-unred}, we find the unreduced sets to be:
\begin{equation}
\mathcal{U}_1 = { (0, 0, 0, n_4, n_5) }, \quad
\mathcal{U}_2 = { (n_1, n_2, n_3, 0, 0) }.
\label{unreduced-set-1}
\end{equation}
Since both $\mathcal{U}_1$ and $\mathcal{U}_2$ still contain infinitely many lattice points, the first iteration is not complete. Nevertheless, it has already organized the unreduced region into two sharply defined pieces: one dominated by ISP powers and one dominated by propagator powers. This geometric split is precisely the information needed to guide the next iteration.

\subsection{Case study 1: Complete reduction in the massless case}
\label{sec:sunset-massless}

Before continuing, it is useful to separate the discussion into two kinematic branches. The degree structure of the descendant equations depends on the masses: when $m_i=0$, several nominally degree-two equations degenerate to degree-one equations, which changes both the available reduction rules and the final master-integral count. We therefore discuss the massless and massive cases separately. We begin with the massless case, where the mechanism of complete reduction is most transparent.

\subsubsection*{Generating the descendant equations}
At this stage no index has yet been reduced to a finite set, so the algorithm proceeds by generating descendant equations. We act with $\partial_i$ for $i=1,\dots,5$ on the six first-round rules in Eqs.~\eqref{sunset-B1-2-1}--\eqref{sunset-B1-2-6}, which initially produces $6\times 5=30$ candidate degree-three equations.

However, not all generated equations are independent or useful. We simplify them following the descendant logic and integrability conditions discussed in Section~\ref{sec:descendant_eqns}. We illustrate that simplification with two representative examples:

\begin{enumerate}
\item Consider the action of $\d_5$ on the reduction rule \eqref{sunset-B1-2-2}. This generates a degree-three operator $\partial_5 \partial_5 \partial_1$. This operator is clearly a descendant of the solved degree-two operator $\partial_5 \partial_1$. Upon substituting the existing rule for $\partial_5 \partial_1$, the equation reduces to exactly zero. This equation is trivially generated and provides no new information.

\item Consider the action of $\partial_4$ on the reduction rule \eqref{sunset-B1-2-2}. This generates the operator $\partial_4 \partial_5 \partial_1$. This operator has two parents: $\partial_5 \partial_1$ (from rule \eqref{sunset-B1-2-2}) and $\partial_4 \partial_1$ (from rule \eqref{sunset-B1-2-3}). Substituting rule \eqref{sunset-B1-2-2} yields zero. However, employing rule \eqref{sunset-B1-2-3} reduces the degree-three operator to lower degrees. This manipulation is equivalent to enforcing the integrability condition $[\partial_4, \partial_5] = 0$. After simplifying all resulting degree-two terms using the six fundamental rules, we arrive at a non-trivial degree-one equation:
\begin{align}
0= &   -\eta_4 {\d^2 \over \d \eta_4^2}G_{111}-\eta_5 {\d^2 \over \d \eta_5^2}G_{111} -2\eta_4 {\d\over \d \eta_4}{\d \over \d \eta_5} G_{111}+ {2m_1^2(m_2^2-m_3^2)\over s_0^2}{\d \over \d \eta_1}G_{111} \nn 
&  +{m_2^2 f_{++-}\over s_0^2}{\d \over \d \eta_2}G_{111}
	-{m_3^2 f_{+-+}\over s_0^2}{\d \over \d \eta_3}G_{111}-(D-2){\d \over \d \eta_4}G_{111}+(4-2D) {\d \over \d \eta_5}G_{111}       \nn
 &+ {2(m_2^2-m_3^2)\eta_1+2 m_1^2(\eta_2-\eta_3)\over  s_0}  {\d \over \d \eta_1}G_{111}- {m_2^2(\eta_1+\eta_2-\eta_3)+ f_{++-}\eta_2\over  s_0}{\d \over \d \eta_2}G_{111} \nn
	& -{f_{+-+}\eta_3+m_3^2 (\eta_1-\eta_2+\eta_3)\over s_0}{\d \over \d \eta_3}G_{111}+ {2K^2-f_{-+-}\over 2 s_0}\eta_4{\d \over \d \eta_4}G_{111}\nn
	&   +{2(K^2+m_1^2)\eta_4- f_{++-}\eta_5\over 2 s_0} {\d \over \d \eta_5}G_{111}+{(D-2)K^2-(D-3)(m_2^2-m_3^2)\over s_0} G_{111}    \nn
	&+ 2(\eta_2-\eta_3)\eta_1 {\d \over \d \eta_1}G_{111}+ (\eta_1+\eta_2-\eta_3)\eta_2{\d \over \d \eta_2}G_{111}-(\eta_1-\eta_2+\eta_3)\eta_3{\d \over \d \eta_3}G_{111} \nn
	& +{(\eta_1-\eta_2+\eta_3)\over 2}\eta_4{\d \over \d \eta_4}G_{111}+ (\eta_1\eta_4-{1\over 2} \eta_5(\eta_1+\eta_2-\eta_3))  {\d \over \d \eta_5}G_{111}\nn
	&  +({ K^2 f_{--+}\over 2 s_0^2}\eta_4+{K^2 m_2^2\over s_0^2}\eta_5-(D-3) (\eta_2-\eta_3))G_{111}    \nn
	&+({K^2\over s_0}\eta_2\eta_5-{K^2\over 2 s_0} \eta_4(\eta_1+\eta_2-\eta_3)) G_{111}   , 
\label{sunset-I145-1}
\end{align}
where operators ${\partial_i}$ appear as the highest degree operators, having the potential to give new reduction rule.
\end{enumerate}

Out of the 30 generated equations, 12 are trivially zero. Among the remaining 18, symmetry considerations (specifically the integrability conditions) reveal that only 9 are distinct. Filtering further, we identify 8 independent descendant equations. The first is Eq.~\eqref{sunset-I145-1} above, and the remaining 7 are listed below:

\begin{enumerate}

\item {\bf From $[\partial_4, \partial_5]$ acting on $\partial_2$: } we have 
\begin{align} 
0= &   -\eta_4 {\d^2 \over \d \eta_4^2}G_{111}-\eta_5 {\d^2 \over \d \eta_5^2}G_{111} -2\eta_5 {\d\over \d \eta_4}{\d \over \d \eta_5} G_{111}+ {2m_2^2(m_1^2-m_3^2)\over s_0^2}{\d \over \d \eta_2}G_{111}\nn &  +{m_1^2 f_{++-}\over s_0^2}{\d \over \d \eta_1}G_{111}-{m_3^2 f_{-++}\over s_0^2}{\d \over \d \eta_3}G_{111}+(4-2D){\d \over \d \eta_4}G_{111}+(2-D) {\d \over \d \eta_5}G_{111}     \nn
& + {f_{++-}\eta_1+m_1^2(\eta_1+\eta_2-\eta_3)\over  s_0}  {\d \over \d \eta_1}G_{111}- {2m_2^2(\eta_1-\eta_3)+2(m_1^2-m_3^2)\eta_2\over  s_0}{\d \over \d \eta_2}G_{111} \nn
& - {f_{-++}\eta_3+m_3^2 (\eta_2+\eta_3-\eta_1)\over s_0}{\d \over \d \eta_3}G_{111}+ {2(K^2+m_2^2)\eta_5-f_{++-}\eta_4\over 2 s_0}{\d \over \d \eta_4}G_{111}\nn
&  +{2K^2-f_{+--}\over 2 s_0}\eta_5 {\d \over \d \eta_5}G_{111}+{(D-2)K^2-(D-3)(m_1^2-m_3^2)\over s_0} G_{111}   \nn
& + (\eta_1+\eta_2-\eta_3)\eta_1 {\d \over \d \eta_1}G_{111}+ 2(\eta_1-\eta_3)\eta_2{\d \over \d \eta_2}G_{111}+(\eta_1-\eta_2-\eta_3)\eta_3{\d \over \d \eta_3}G_{111} \nn
& +(\eta_2\eta_5-{1\over 2}(\eta_1+\eta_2-\eta_3)\eta_4){\d \over \d \eta_4}G_{111}+ {1\over 2}(-\eta_1+\eta_2+\eta_3) \eta_5 {\d \over \d \eta_5}G_{111}\nn
&  +({ K^2 f_{--+}\over 2 s_0^2}\eta_5-(D-3)(\eta_1-\eta_3)+ {K^2 m_1^2\over s_0^2}\eta_4)G_{111}   \nn
& +    ({K^2\over s_0}\eta_1\eta_4-{K^2\over 2 s_0} \eta_5(\eta_1+\eta_2-\eta_3)) G_{111},   
~~~\label{sunset-I245-1}\end{align}

\item {\bf From $[\partial_4, \partial_5]$ acting on $\partial_3$: } we have 
\begin{align} 
0 = &    -\eta_4 {\d^2 \over \d \eta_4^2} G_{111}+\eta_5 {\d^2 \over \d \eta_5^2} G_{111}+ {m_1^2  f_{+-+}\over s_0^2}{\d \over \d \eta_1}G_{111}- {m_2^2  f_{-++}\over s_0^2}{\d \over \d \eta_2}G_{111}\nn &  + {2 m_3^2  (m_1^2-m_2^2)\over s_0^2}{\d \over \d \eta_3}G_{111} -(D-2) {\d \over \d \eta_4}G_{111}+(D-2) {\d \over \d \eta_5}G_{111}  \nn
& + {f_{+-+}\eta_1+m_1^2(\eta_1-\eta_2+\eta_3)\over s_0}  {\d \over \d \eta_1}G_{111}- {m_2^2 (-\eta_1+\eta_2+\eta_3)+f_{-++}\eta_2\over s_0}{\d \over \d \eta_2}G_{111} \nn
& + {2 \eta_3  (m_1^2-m_2^2)+2 m_3^2(\eta_1-\eta_2)\over s_0}{\d \over \d \eta_3}G_{111}-{f_{+-+}\eta_4\over s_0}{\d \over \d \eta_4}G_{111}+{f_{-++}\eta_5\over s_0}{\d \over \d \eta_5}G_{111}\nn
&  -(D-3){m_1^2-m_2^2\over s_0} G_{111}  +  (\eta_1-\eta_2+\eta_3)\eta_1 {\d \over \d \eta_1}G_{111}+ (\eta_1-\eta_2-\eta_3)\eta_2{\d \over \d \eta_2}G_{111}\nn
& +2(\eta_1-\eta_2)\eta_3{\d \over \d \eta_3}G_{111} -{(\eta_1-\eta_2+\eta_3)\eta_4\over 2}{\d \over \d \eta_4}G_{111}-{(\eta_1-\eta_2-\eta_3)\eta_5\over 2}{\d \over \d \eta_5}G_{111}\nn
&\nn
&  -(D-3)(\eta_1-\eta_2)G_{111}+{K^2\over s_0}((m_1^2+\eta_1)\eta_4-(m_2^2+\eta_2)\eta_5) G_{111},  
~~~\label{sunset-I345-1}\end{align}

\item 
{\bf From $[\partial_1, \partial_2]$ acting on $\partial_4$:}  We get
\begin{align} 
0 = & - {2 m_2^2\over s_0} {\d^2 \over \d \eta_2^2}G_{111}+{f_{+++} \over s_0 }{\d\over \d \eta_2}{\d \over \d \eta_1} G_{111}+{2m_3^2 \over s_0 }{\d\over \d \eta_3}{\d \over \d \eta_1} G_{111}+{2m_3^2 \over s_0 }{\d\over \d \eta_3}{\d \over \d \eta_2} G_{111}\nn
& +  (2D-7){\d \over \d \eta_2}G_{111}+\eta_5 {\d\over \d \eta_2}{\d \over \d \eta_5} G_{111}+\eta_4 {\d\over \d \eta_2}{\d \over \d \eta_4} G_{111}-2\eta_3 {\d\over \d \eta_2}{\d \over \d \eta_3} G_{111}  \nn & -2 \eta_2 {\d^2 \over \d \eta_2^2}G_{111}  + (D-3){\d \over \d \eta_1}G_{111}+\eta_5 {\d\over \d \eta_1}{\d \over \d \eta_5} G_{111}+\eta_5 {\d\over \d \eta_1}{\d \over \d \eta_4} G_{111}
 \nn & -2\eta_3 {\d\over \d \eta_1}{\d \over \d \eta_3} G_{111} -(\eta_1+\eta_2+\eta_3){\d\over \d \eta_2}{\d \over \d \eta_1} G_{111} \nn &-{K^2\over s_0}\eta_4 {\d\over \d \eta_2} G_{111}-{K^2\over s_0}\eta_5 {\d\over \d \eta_1} G_{111} 
~~~\label{sunset-I124-simpler}\end{align}

\item 
{\bf From $[\partial_1, \partial_3]$ acting on $\partial_4$:}  we get
\begin{align} 
0= &-  {2m_2^2 \over s_0 }{\d\over \d \eta_3}{\d \over \d \eta_2} G_{111}+{2m_2^2 \over s_0 }{\d\over \d \eta_2}{\d \over \d \eta_1} G_{111}+{f_{+++}\over s_0 }{\d\over \d \eta_3}{\d \over \d \eta_1} G_{111}+{2 m_3^2 \over s_0 }{\d^2 \over \d \eta_3^2} G_{111}\nn
& +\eta_5 {\d\over \d \eta_3}{\d \over \d \eta_5} G_{111} +\eta_4 {\d\over \d \eta_3}{\d \over \d \eta_4} G_{111}-2\eta_3 {\d^2 \over \d \eta_3^2} G_{111}-2\eta_2 {\d\over \d \eta_3}{\d \over \d \eta_2} G_{111} \nn & +(D-3){\d\over \d \eta_1} G_{111} +  (2D-7){\d\over \d \eta_3} G_{111} + \eta_5 {\d\over \d \eta_1}{\d \over \d \eta_5} G_{111}\nn
&-(\eta_1+\eta_2+\eta_3) {\d\over \d \eta_3}{\d \over \d \eta_1} G_{111}-2\eta_2 {\d\over \d \eta_2}{\d \over \d \eta_1} G_{111} -{K^2\over s_0}\eta_4 {\d\over \d \eta_3} G_{111}  ~~~\label{sunset-I134-simpler}
\end{align}

\item {\bf 
From $[\partial_2, \partial_3]$ acting on $\partial_4$:} we get 
\begin{align} 
0= &- {2 m_2^2 \over s_0 }{\d^2 \over \d \eta_2^2} G_{111}-{2 m_3^2 \over s_0 }{\d^2 \over \d \eta_3^2} G_{111}+   -(D-4) {\d\over \d \eta_3} G_{111}+ (D-4){\d\over \d \eta_2} G_{111} \nn & -\eta_5 {\d\over \d \eta_3}{\d \over \d \eta_5} G_{111} -\eta_5 {\d\over \d \eta_3}{\d \over \d \eta_4} G_{111}+ \eta_5 {\d\over \d \eta_2}{\d \over \d \eta_5} G_{111} \nn
&+ 2\eta_3 {\d^2 \over \d \eta_3^2} G_{111}-2\eta_2 {\d^2 \over \d \eta_3^2} G_{111}+  {K^2\over s_0}\eta_5 G_{111} ~~~\label{sunset-I234-simpler}
\end{align}

\item 
{\bf From $[\partial_1, \partial_2]$ acting on $\partial_5$:} we get
\begin{align} 
0= & -{-2 m_1^2\over s_0} {\d^2 \over \d \eta_1^2}G_{111}-{f_{+++} \over s_0 }{\d\over \d \eta_2}{\d \over \d \eta_1} G_{111}-{2m_3^2 \over s_0 }{\d\over \d \eta_3}{\d \over \d \eta_1} G_{111}-{2m_3^2 \over s_0 }{\d\over \d \eta_3}{\d \over \d \eta_2} G_{111}\nn
& - (D-3){\d \over \d \eta_2}G_{111}  + (7-2D){\d \over \d \eta_1}G_{111}-\eta_4 {\d\over \d \eta_2}{\d \over \d \eta_5} G_{111} -\eta_4 {\d\over \d \eta_2}{\d \over \d \eta_4} G_{111} \nn & +2\eta_3 {\d\over \d \eta_2}{\d \over \d \eta_3} G_{111} -\eta_5 {\d\over \d \eta_1}{\d \over \d \eta_5} G_{111} -\eta_4 {\d\over \d \eta_1}{\d \over \d \eta_4} G_{111}+2\eta_3 {\d\over \d \eta_1}{\d \over \d \eta_3} G_{111} \nn & +(\eta_1+\eta_2+\eta_3) {\d\over \d \eta_2}{\d \over \d \eta_1} G_{111}+2 \eta_1 {\d^2 \over \d \eta_1^2}G_{111}\nn& +{K^2\over s_0}\eta_4{\d\over \d \eta_2} G_{111}  + {K^2\over s_0}\eta_5{\d\over \d \eta_1} G_{111}~~~~~~~\label{sunset-I125-simpler}
\end{align}

\item 
{\bf From $[\partial_1, \partial_3]$ acting on $\partial_5$:} we get
\begin{align} 
0= &{2 m_1^2 \over s_0 }{\d^2 \over \d \eta_1^2} G_{111}-{2 m_3^2 \over s_0 }{\d^2 \over \d \eta_3^2} G_{111}+ (D-4) {\d\over \d \eta_3} G_{111}- (D-4){\d\over \d \eta_1} G_{111} \nn &  +\eta_4 {\d\over \d \eta_3}{\d \over \d \eta_5} G_{111}+\eta_4 {\d\over \d \eta_3}{\d \over \d \eta_4} G_{111} -\eta_4 {\d\over \d \eta_1}{\d \over \d \eta_4} G_{111}\nn &+2\eta_1 {\d^2 \over \d \eta_1^2} G_{111}-2\eta_3 {\d^2 \over \d \eta_3^2} G_{111} +  {K^2\over s_0}\eta_4{\d\over \d \eta_3} G_{111} ~~~\label{sunset-I135-simpler}
\end{align}

\item 
{\bf From $[\partial_2, \partial_3]$ acting on $\partial_5$:} we get
\begin{align}
0= &- -{f_{+++} \over s_0 }{\d\over \d \eta_3}{\d \over \d \eta_2} G_{111}-{2m_1^2 \over s_0 }{\d\over \d \eta_2}{\d \over \d \eta_1} G_{111}-{2m_1^2\over s_0 }{\d\over \d \eta_3}{\d \over \d \eta_1} G_{111}-{2 m_3^2 \over s_0 }{\d^2 \over \d \eta_3^2} G_{111}\nn
&+    (7-2D) {\d\over \d \eta_3} G_{111}- (D-3){\d\over \d \eta_2} G_{111} - \eta_5 {\d\over \d \eta_3}{\d \over \d \eta_5} G_{111} -\eta_4 {\d\over \d \eta_3}{\d \over \d \eta_4} G_{111}  \nn & +2 \eta_3 {\d^2 \over \d \eta_3^2} G_{111} -\eta_4 {\d\over \d \eta_2}{\d \over \d \eta_4} G_{111} +(\eta_1+\eta_2+\eta_3){\d\over \d \eta_3}{\d \over \d \eta_2} G_{111}\nn & +2\eta_1 {\d\over \d \eta_2}{\d \over \d \eta_1} G_{111} +2\eta_1 {\d\over \d \eta_3}{\d \over \d \eta_1} G_{111} +  {K^2\over s_0}\eta_5{\d\over \d \eta_3} G_{111}  ~~~\label{sunset-I235-simpler}\end{align}

\end{enumerate}

\subsubsection*{Solving the system in the massless case}

The new equations exhibit a crucial feature: the coefficients of their highest-degree operators depend on the masses. As a result, when $m_i=0$, several equations that are degree two in the generic massive case collapse to degree one. This change is not merely technical; it is directly tied to the smaller number of master integrals in the massless theory than in the generic massive case\footnote{For the massive case, especially when all $m_i$ are different, there are four master integrals. When the nonzero masses are equal, the number drops to two once global symmetry is used. Since the present discussion relies only on IBP relations, however, the analysis here still produces four master integrals even when the nonzero masses are equal.}. In the massless limit, only a single master integral remains.

We proceed by setting $m_i = 0$ in the equations derived above. Equations \eqref{sunset-I234-simpler} and \eqref{sunset-I135-simpler} degenerate to degree-one forms. Simplifying them with the rules from the previous round, we obtain:
\begin{align} 
0 = &     (D-4) {\d\over \d \eta_3} G_{111}- (D-4){\d\over \d \eta_2} G_{111}-2\eta_3 {\d^2 \over \d \eta_3^2} G_{111}+2\eta_2 {\d^2 \over \d \eta_2^2} G_{111} \nn
%
%
& -2\eta_5\eta_3 {\d\over \d \eta_3} G_{111}- 2\eta_5 \eta_1{\d\over \d \eta_1} G_{111}- 2\eta_5 \eta_2{\d\over \d \eta_2} G_{111} + \eta_4\eta_5 {\d\over \d \eta_4} G_{111} \nn 
& +  \eta_5^2{\d\over \d \eta_5} G_{111} +2(D-3)\eta_5 G_{111}
%
-   {K^2\over 2 s_0} \eta_5^2 G_{111}, ~~~\label{sunset-I234-m=0}
\end{align}
\begin{align}
0 = &   (D-4) {\d\over \d \eta_3} G_{111}- (D-4){\d\over \d \eta_1} G_{111}-2\eta_3 {\d^2 \over \d \eta_3^2} G_{111}+2\eta_1 {\d^2 \over \d \eta_1^2} G_{111} \nn
& -2\eta_4\eta_3 {\d\over \d \eta_3} G_{111}- 2\eta_4 \eta_1{\d\over \d \eta_1} G_{111}- 2\eta_4 \eta_2{\d\over \d \eta_2} G_{111}+ \eta_4 \eta_5{\d\over \d \eta_5} G_{111} \nn
& + \eta_4^2 {\d\over \d \eta_4} G_{111}+2(D-3)\eta_4 G_{111} 
%
-{K^2\over 2 s_0} \eta_4^2 G_{111}. ~~~\label{sunset-I135-m=0}
\end{align}

We now solve the system, starting with the remaining degree-two equations. The coefficient matrix for the operators $\partial_1\partial_2$, $\partial_1\partial_3$, and $\partial_2\partial_3$ is:
\begin{align} \left( \begin{array}{c| c |c |l} {\d_1} {\d_2} &  {\d_1} {\d_3} & {\d_2} {\d_3} &  \text{Source Eq.}\\  
\hline
	{K^2\over s_0} &   &   &  {\eref{sunset-I124-simpler}}\\  \hline 
	& {K^2\over s_0}    &   &  {\eref{sunset-I134-simpler}}\\  \hline 
	-{K^2\over s_0} &   &   &  {\eref{sunset-I125-simpler}}\\  \hline
	& & -{K^2\over s_0}    &  {\eref{sunset-I235-simpler}}
\end{array}\right).~~~~~\label{sunset-massless-B-1}
\end{align}
Thus, there is a combination giving degree-one equation. After simplification using the descendant reduction rules, the new degree-one equation ${\eref{sunset-I125-simpler}}+ {\eref{sunset-I124-simpler}}$ yields:
\begin{align} 0 = &   (D-4) {\d\over \d \eta_2} G_{111} -2\eta_2 {\d^2 \over \d \eta_2^2} G_{111}- (D-4) {\d\over \d \eta_1} G_{111} +2\eta_1 {\d^2 \over \d \eta_1^2} G_{111}\nn
& +  2(D-3)(\eta_4-\eta_5) G+(\eta_4-\eta_5)\eta_5 {\d\over \d \eta_5} G_{111}+(\eta_4-\eta_5)\eta_4 {\d\over \d \eta_4} G_{111}\nn 
&  - 2(\eta_4-\eta_5)\eta_3 {\d\over \d \eta_3} G_{111} - 2(\eta_4-\eta_5)\eta_2 {\d\over \d \eta_2} G_{111}- 2(\eta_4-\eta_5)\eta_1 {\d\over \d \eta_1} G_{111}\nn
&+  {K^2(-\eta_4^2+\eta_5^2)\over 2 s_0} G.~~~\label{sunset-I124plus125-m=0}
\end{align}

There remains three degree-two T1A-type equations, which give reduction rules for operators ${\d_1} {\d_2}$, ${\d_1} {\d_3}$ and ${\d_2} {\d_3}$ as
\begin{align}
&-{K^2 \over s_0 }{\d\over \d \eta_2}{\d \over \d \eta_1} G_{111}\nn
= &   -2 \eta_2 {\d^2 \over \d \eta_2^2}G_{111}-2\eta_3 {\d\over \d \eta_3}{\d \over \d \eta_1} G_{111}-2\eta_3 {\d\over \d \eta_3}{\d \over \d \eta_2} G_{111}-(\eta_1+\eta_2+\eta_3){\d\over \d \eta_2}{\d \over \d \eta_1} G_{111}\nn
& + (D-3){\d \over \d \eta_1}G_{111}+ (2D-7){\d \over \d \eta_2}G_{111} - {K^2\eta_5\over 2 s_0}  {\d \over \d \eta_1}G_{111}-{K^2\eta_4\over 2 s_0}{\d \over \d \eta_2}G_{111} \nn
%
& +{1\over 2}(3\eta_1+\eta_2-\eta_3)\eta_5 {\d \over \d \eta_1}G_{111}+(2 \eta_2\eta_5+{1\over 2} (\eta_1-\eta_2-\eta_3)\eta_4){\d \over \d \eta_2}G_{111}\nn
& +(\eta_5-\eta_4)\eta_3{\d \over \d \eta_3}G_{111}  +(\eta_4-\eta_5)\eta_5{\d \over \d \eta_5}G_{111}+{(D-3)\over 2} (\eta_4-3\eta_5)G_{111} \nn
&+   {K^2\over 2s_0}(\eta_5-\eta_4)\eta_5 G_{111},
~~~\label{sunset-I124-m=0}
\end{align}
\begin{align}
    &-{K^2\over s_0 }{\d\over \d \eta_3}{\d \over \d \eta_1} G_{111}\nn
= & -2\eta_3 {\d^2 \over \d \eta_3^2} G_{111}-2\eta_2{\d\over \d \eta_3}{\d \over \d \eta_2} G_{111} -2\eta_2 {\d\over \d \eta_2}{\d \over \d \eta_1} G_{111} -(\eta_1+\eta_2+\eta_3) {\d\over \d \eta_3}{\d \over \d \eta_1} G_{111}\nn 
&  +(D-3){\d\over \d \eta_1} G_{111}+(2D-7){\d\over \d \eta_3}G_{111}+  {K^2(\eta_5-\eta_4)\over 2 s_0} {\d\over \d \eta_3} G_{111}+{K^2\over 2s_0}\eta_5{\d\over \d \eta_1} G_{111} \nn
%
%
&+     {(\eta_1-\eta_2-\eta_3)(\eta_4-\eta_5)-2 \eta_3\eta_5\over 2} {\d\over \d \eta_3} G_{111} -{3\eta_1-\eta_2+\eta_3\over 2}\eta_5{\d\over \d \eta_1} G_{111}  \nn
&   -\eta_2\eta_4{\d\over \d \eta_2} G_{111}+ \eta_4 \eta_5{\d\over \d \eta_5} G_{111}+ \eta_4 \eta_5 {\d\over \d \eta_4} G_{111}+{(D-3)\over 2}(\eta_4+2\eta_5) G_{111} \nn
&  -{K^2\over 2 s_0} \eta_4\eta_5 G_{111}, ~~~\label{sunset-I134-m=0}
\end{align}

\begin{align}
& -{K^2\over s_0 }{\d\over \d \eta_3}{\d \over \d \eta_2} G_{111}\nn
%
= & -2\eta_3 {\d^2 \over \d \eta_3^2} G_{111} -2\eta_1 {\d\over \d \eta_2}{\d \over \d \eta_1} G_{111}- 2\eta_1 {\d\over \d \eta_3}{\d \over \d \eta_1} G_{111} -(\eta_1+\eta_2+\eta_3){\d\over \d \eta_3}{\d \over \d \eta_2} G_{111} \nn 
& + (D-3){\d\over \d \eta_2} G_{111}+ (2D-7) {\d\over \d \eta_3} G_{111} + {K^2(\eta_4-\eta_5)\over 2 s_0} {\d\over \d \eta_3} G_{111}+{K^2\over 2 s_0}\eta_4{\d\over \d \eta_2} G_{111},  \nn
%
%
& + {{(\eta_1-\eta_2-\eta_3)(\eta_4-\eta_5)-2 \eta_3\eta_4}\over 2} {\d\over \d \eta_3} G_{111} -{3\eta_2-\eta_1+\eta_3\over 2}\eta_4{\d\over \d \eta_2} G_{111}\nn
& -\eta_5 \eta_1{\d\over \d \eta_1} G_{111} + \eta_4 \eta_5{\d\over \d \eta_5} G_{111}+ \eta_4 \eta_5 {\d\over \d \eta_4} G_{111}  +{(D-3)\over 2} (2\eta_4+{\eta_5}) G_{111} \nn
&-{K^2\over 2 s_0} \eta_4\eta_5 G_{111},  ~~~\label{sunset-I235-m=0}
\end{align}
which allow us to reduce cross-terms between propagators.

Having found three new reduction rules, we would in principle update the six previously solved reduction rules together with the remaining degree-one equations according to Section~\ref{sec:update_mech}. In the present case, however, no further update is required.

We next turn to the six degree-one equations. Combining the degenerated equations with the results from the degree-two simplification, we construct the coefficient matrix for single derivatives as
\begin{align}
    \left( \begin{array}{c| c |c |c| c| l} {\d_1} & {\d_2} &   {\d_3} & {\d_4} & {\d_5} & \text{Source Eq.}  \\  \hline
    &   &   & -\WH O_{0;4,1}  &  \WH O_{0;5,2}  &  {\eref{sunset-I145-1}} \\  \hline
    &   &   & \WH O_{0;4,2}  &  -\WH O_{0;5,1}   &   {\eref{sunset-I245-1}} \\  \hline
    &   &   &  -\WH O_{0;4,1} &  \WH O_{0;5,1}  &  {\eref{sunset-I345-1}}\\  \hline
    & \WH O_{0;2,1}  & -\WH O_{0;3,1}  &   &    &  {\eref{sunset-I234-m=0}} \\  \hline
   -\WH O_{0;1,1} &   & \WH O_{0;3,1}  &   &    &  {\eref{sunset-I135-m=0}} \\  \hline
   -\WH O_{0;1,1} & \WH O_{0;2,1}  &   &   &    &  {\eref{sunset-I124plus125-m=0}} \\  
\end{array}\right),~~~~~\label{sunset-massless-B-3}
\end{align}
where for compactness we have defined the following differential operators:
\bea \WH O_{0;4,1} & = & (D-2)+ \eta_4 {\d\over \d\eta_4},~~~~\WH O_{0;4,2}  =  4-2D-\eta_4 {\d\over \d\eta_4}-2 \eta_5 {\d\over \d\eta_5}  \nn
\WH O_{0;5,1} & = & (D-2)+ \eta_5 {\d\over \d\eta_5},~~~~\WH O_{0;5,2}  =  4-2D-\eta_5 {\d\over \d\eta_5}-2 \eta_4 {\d\over \d\eta_4}\nn
\WH O_{0;i,1} & = & (D-4)- 2\eta_i {\d\over \d\eta_i},~~~~~~i=1,2,3.~~~~~\label{sunset-massless-B-3-O}\eea
Applying Gaussian elimination, we obtain the reduced matrix form:
\begin{align}
    \left( \begin{array}{c| c |c |c| c| l} {\d_1} & {\d_2} &   {\d_3} & {\d_4} & {\d_5} &  \text{Linear Combination} \\  \hline
	&   &   &  &    & \WH O_{0;I}{\eref{sunset-I145-1}}+  \WH O_{0;-}{\eref{sunset-I245-1}} + \WH O_{0;+} {\eref{sunset-I345-1}}) \\  \hline
	&   &   & \WH O_{0;I}  &    &   {\eref{sunset-I245-1}} + {\eref{sunset-I345-1}}\\  \hline
	&   &   &   &  \WH O_{0;I}  & {\eref{sunset-I145-1}} - {\eref{sunset-I345-1}}\\  \hline
	& \WH O_{0;2,1}  & -\WH O_{0;3,1}  &   &    &  {\eref{sunset-I234-m=0}} \\  \hline
	-\WH O_{0;1,1} &   & \WH O_{0;3,1}  &   &    &  {\eref{sunset-I135-m=0}} \\  \hline
	&   &   &   &    &  {\eref{sunset-I124plus125-m=0}}-{\eref{sunset-I234-m=0}}-{\eref{sunset-I135-m=0}} \\  
\end{array}\right)~~~~~\label{sunset-massless-B-5}
\end{align}
where
\begin{align}
    O_{0;I} &= \WH O_{0;4,2} - \WH O_{0;4,1} = \WH O_{0;5,2} - \WH O_{0;5,1},\\
    O_{0;\pm} &= \WH O_{0;4,1} \pm \WH O_{0;5,2},
\end{align}
We observe that the last row yields a trivial equation, confirming the redundancy anticipated from integrability conditions. The first row yields a degree-zero equation, but that equation contains no good operator to solve for and therefore will not be used further. We thus focus on the four independent equations represented by rows 2, 3, 4, and 5.

The second and third rows of the reduced matrix decouple from the propagator derivatives, allowing us to solve explicitly for $\partial/\partial \eta_4$ and $\partial/\partial \eta_5$, giving two T1B-type reduction rules:
\begin{align}
    & -\WH O_{0;I}{\d\over \d\eta_4} G_{111}\nn
=& {K^2(D-2)\over s_0} G_{111}  +{K^2\over s_0} \eta_5 {\d \over \d \eta_5} G_{111}+{K^2\over s_0} (\eta_4+\eta_5) {\d \over \d \eta_4} G_{111}  +{K^2\over s_0} \eta_3 {\d \over \d \eta_3} G_{111}  \nn 
& +{K^2\over s_0} \eta_2 {\d \over \d \eta_2} G_{111}  -{2K^2\over s_0} \eta_1 {\d \over \d \eta_1} G_{111}+(\eta_2+\eta_3-\eta_1)\eta_5 {\d \over \d \eta_5} G_{111}+(\eta_2\eta_5-\eta_1\eta_4){\d \over \d \eta_4} G_{111}\nn
& +(3\eta_1-3\eta_2-\eta_3)\eta_3 {\d \over \d \eta_3} G_{111} -(\eta_2-3\eta_1+3\eta_3)\eta_2 {\d \over \d \eta_2} G_{111} +2\eta_1^2 {\d \over \d \eta_1} G_{111} \nn 
&+((D-3)(\eta_2+\eta_3-2\eta_1)+{(K^2)^2\eta_5\over 2s_0^2}) G_{111}+ { K^2(4  \eta_1\eta_4-(\eta_1+3\eta_2-\eta_3)\eta_5)\over 2 s_0} G_{111},
~~~\label{sunset-I245plus345-m=0}
\end{align}
\begin{align}
    & -\WH O_{0;I}{\d\over \d\eta_5} G_{111}\nn
=& {K^2\eta_4\over  s_0}{\d \over \d \eta_4}G_{111} +{K^2(\eta_5+\eta_4)\over  s_0} {\d \over \d \eta_5}G_{111}+{K^2(D-2)\over s_0} G_{111}+{K^2\eta_3\over  s_0}{\d \over \d \eta_3}G_{111}\nn 
& - {2 K^2\eta_2\over  s_0}{\d \over \d \eta_2}G_{111}+ {K^2\eta_1\over  s_0}{\d \over \d \eta_1}G_{111}+(\eta_1-\eta_2+\eta_3)\eta_4{\d \over \d \eta_4} G_{111}+(\eta_1\eta_4-\eta_2\eta_5) {\d \over \d \eta_5} G_{111}\nn
& -(3\eta_1-3\eta_2+\eta_3)\eta_3 {\d \over \d \eta_3} G_{111} +2\eta_2^2 {\d \over \d \eta_2} G_{111} -(\eta_1-3\eta_2+3\eta_3)\eta_1 {\d \over \d \eta_1} G_{111}\nn 
& +  ((D-3)(\eta_1-2\eta_2+\eta_3)-{(K^2)^2\eta_4\over 2s_0^2}) G_{111} +{K^2 (4 \eta_2\eta_5- (3\eta_1+\eta_2-\eta_3)\eta_4)\over 2 s_0} G_{111}, 
~~~\label{sunset-I145minus345-m=0}
\end{align}
which guarantee perfect reduction relations for any integrals with nonzero indexes $(n_4, n_5)$ to $(0, 0)$.

The fourth and fifth rows of the matrix relate the derivatives of the propagator variables. Rearranging Eq.~\eqref{sunset-I234-m=0}, we can solve for the operator involving $\eta_1$ and $\eta_2$, giving two T2B-type reduction rules:
\begin{align}
    & (D-4){\d\over \d \eta_1} G_{111}- 2\eta_1 {\d^2 \over \d \eta_1^2} G_{111} = (D-4) {\d\over \d \eta_3} G_{111}-2\eta_3 {\d^2 \over \d \eta_3^2} G_{111}\nn
& - 2\eta_4 \eta_1{\d\over \d \eta_1} G_{111}- 2\eta_4 \eta_2{\d\over \d \eta_2} G_{111}-2\eta_4\eta_3 {\d\over \d \eta_3} G_{111}+ \eta_4^2 {\d\over \d \eta_4} G_{111}+ \eta_4 \eta_5{\d\over \d \eta_5} G_{111}\nn 
& +2(D-3)\eta_4 G_{111} -{K^2\over 2 s_0} \eta_4^2 G_{111},~~~\label{sunset-I135-m=0-n1}
\end{align}
\begin{align}
    & (D-4){\d\over \d \eta_2} G_{111}-2\eta_2 {\d^2 \over \d \eta_2^2} G_{111}
= (D-4) {\d\over \d \eta_3} G_{111}-2\eta_3 {\d^2 \over \d \eta_3^2} G_{111}\nn
%
%
&- 2\eta_5 \eta_1{\d\over \d \eta_1} G_{111}- 2\eta_5 \eta_2{\d\over \d \eta_2} G_{111}- 2\eta_5\eta_3 {\d\over \d \eta_3} G_{111}+\eta_4\eta_5 {\d\over \d \eta_4} G_{111}+  \eta_5^2{\d\over \d \eta_5} G_{111}\nn 
& +2(D-3)\eta_5 G_{111} - {K^2\over 2 s_0} \eta_5^2 G_{111}, ~~~\label{sunset-I234-m=0-n2}
\end{align}
These relations allow us to reduce any integral with indices $(n_1, n_2, n_3)$ to a linear combination of integrals with indices $(0, 0, N)$, where $N \leq n_1+n_2+n_3$, without increasing the complexity of the integral.

Since all four new reduction rules are TB-type, we do not perform the update step. This example makes the reason clear: once the T1B-type reduction rule for ${\d\over \d\eta_4}$ in \eref{sunset-I245plus345-m=0} is available, the previously derived degree-two operators ${\d\over \d\eta_4}{\d\over \d\eta_i}$ with $i=1,2,3$ all become descendants of that rule. To preserve the simpler and more convenient TA-type reduction rules, we therefore do not use TB-type rules in the global update.

\subsubsection*{Completeness check for the massless case}

At this stage we have seven new reduction rules: four degree-one rules, namely \eref{sunset-I245plus345-m=0}, \eref{sunset-I145minus345-m=0}, \eref{sunset-I234-m=0-n2}, and \eref{sunset-I135-m=0-n1}, together with three degree-two rules, namely \eref{sunset-I124-simpler}, \eref{sunset-I134-simpler}, and \eref{sunset-I235-simpler}. Applying the four degree-one rules to the remaining lattice points in \eref{unreduced-set-1}, we find that the points in the set
\bea {\cal U}_3=\{(0,0,n_3,0,0)| \forall n_3\geq 0\}~~~~\label{sunset-irr-2}\eea
cannot be reduced by these rules. 

Applying the three degree-two reduction rules does not improve the situation, since the same set ${\cal U}_3$ remains unreduced. The reason is clear: these rules are descendants of the operators ${\d\over \d\eta_i}$ with $i=1,2$. Nevertheless, they remain useful as simplification tools in the final iteration.

\subsubsection*{The third iteration: reducing the final sector}

Since only $n_3$ has not yet been fully reduced, we initiate a third round of computations by acting with $\partial_3$ on the four degree-one equations derived in the previous step, namely \eref{sunset-I245plus345-m=0}, \eref{sunset-I145minus345-m=0}, \eref{sunset-I234-m=0-n2}, and \eref{sunset-I135-m=0-n1}.

First, acting with $\partial_3$ on Eq.~\eqref{sunset-I245plus345-m=0} gives an equation containing three degree-two operators, namely $\partial_3 \partial_4$, $\eta_4 \partial_4\partial_3 \partial_4$, and $\eta_5 \partial_5\partial_3 \partial_4$, all of which are descendants of the operator $\partial_3 \partial_4$. After simplification, the new equation involves five degree-one operators: $\partial_3$, $\eta_3 \partial_3^2$, $\eta_1 \partial_1 \partial_3$, $\eta_2 \partial_2 \partial_3$, and $\eta_5 \partial_4 \partial_3$. These can be simplified using \eref{sunset-B1-2-5}, derived in the first iteration, together with \eref{sunset-I124-simpler}, \eref{sunset-I134-simpler}, and \eref{sunset-I235-simpler}, derived in the second iteration. The equation is then simplified to the following T1B-type equation:
\begin{align}
    \WH O_{0;II} {\d\over \d \eta_3} G_{111}
	= & -{(D-3)(3D-8)\over 2} G_{111} + {(5D-16)\over 2}\eta_3{\d\over \d \eta_3} G_{111}-\eta_3^2 {\d^2 \over \d \eta_3^2} G_{111}\nn
	& + (\eta_1,\,\eta_2,\,\eta_4,\,\eta_5)\,  \text{components},~~~\label{sunset-m=0-Rn3}
\end{align}
with
\begin{align}
    \WH O_{0;II}=  { (D-4) K^2\over 2 s_0}  -{ K^2\over s_0}\eta_3 {\d \over \d \eta_3}.~~~\label{sunset-m=0-Rn3-O}
\end{align}

Alternatively, applying $\partial_3$ to Eq.~\eqref{sunset-I145minus345-m=0} yields another T1B-type equation involving the same operator $\WH O_{0;II}$. In fact, when restricted to the irreducible set $\mathcal{U}_3$ (where $n_1=n_2=n_4=n_5=0$), this equation simplifies to the constraints of Eq.~\eqref{sunset-m=0-Rn3}, confirming their consistency.

Finally, applying $\partial_3$ to Eqs.~\eqref{sunset-I234-m=0-n2} and \eqref{sunset-I135-m=0-n1} initially yields degree-two equations containing operators ${\partial_3^2 }$ and $\eta_3{\partial_3^3 }$. However, after applying Gaussian elimination, these two equations can be combined to form a degree-one relation. After simplifying degree-one terms using existing rules, the resulting relation takes a revealing form:
\begin{align}
{(D-3) s_0\over K^2}\left\{\left((4-D) {\d\over \d \eta_2} G_{111}+2\eta_2{\d^2\over \d \eta_2^2} G_{111}\right)-\left((4-D) {\d\over \d \eta_1} G_{111}+2\eta_1{\d^2\over \d \eta_1^2} G_{111}\right)\right\},
\end{align}
which is a linear combination of Eqs.~\eqref{sunset-I234-m=0-n2} and \eqref{sunset-I135-m=0-n1}, and is therefore useless for the reduction of $\mathcal{U}_3$.

Employing one of these T1B-type equations yields a T1B-type reduction rule for ${\partial_3}$. The remaining propagator index $n_3$ can now be reduced to zero. Therefore, the only integral that cannot be reduced by any of our rules is
\begin{align}
    (0,0,0,0,0).
\end{align}
This single point constitutes the master-integral basis for the massless sunset topology. The massless branch therefore provides a complete example of how the iterative procedure closes: the first round isolates the unreduced subsets, the second round collapses them to a single line, and the third round removes the final propagator tower.

\subsection{Case study 2: Complete reduction in the massive case}
\label{sec:sunset-massive}
Having established the massless case, we now turn to the fully massive sunset diagram. Here the three masses and the external invariant $K^2$ restore the generic degree structure of the equations and enlarge the final master-integral basis. Even so, the logic of the algorithm is unchanged: first reduce the ISP directions, then constrain the propagator directions, and finally close the remaining unreduced set.

Starting from the six degree-two equations, the corresponding matrix is
\begin{align}
    \left( \begin{array}{c |c |c |c|c|c|l}{\partial_1^2} &  {\partial_2^2} & {\partial_3^2} &  {\partial_1} {\partial_2} &  {\partial_1} {\partial_3} & {\partial_2} {\partial_3} & \text{Source Eq.} \\  \hline 
	& {-2m_2^2 \over s_0} &  & 	{-f_{+++}\over s_0} &  {-2m_3^2 \over s_0} &  {-2m_3^2 \over s_0} &  {\eref{sunset-I124-simpler}}\\  \hline 
	& & {-2m_3^2 \over s_0} & {-2m_2^2 \over s_0}	& {-f_{+++}\over s_0}    & {-2m_2^2 \over s_0}  &  {\eref{sunset-I134-simpler}}\\   \hline 
	{2m_1^2 \over s_0}& &  & 	{f_{+++}\over s_0} & {2m_3^2 \over s_0}  & {2m_3^2 \over s_0}  &  {\eref{sunset-I125-simpler}}\\ \hline 
	& & {2m_3^2 \over s_0} & {2m_1^2 \over s_0}	& {2m_1^2 \over s_0}& {f_{+++}\over s_0}    &  {\eref{sunset-I235-simpler}}\\ \hline 
	& {-2m_2^2 \over s_0} & {2m_3^2 \over s_0} & & & & {\eref{sunset-I234-simpler}} \\ \hline 
	{2m_1^2 \over s_0} &  & {-2m_3^2 \over s_0} & & &  & {\eref{sunset-I135-simpler}}
\end{array}\right),~~~~~\label{sunset-massive-B-1}
\end{align}
where $f_{\pm\pm\pm}$ were defined previously in \eref{f-abc-def}. Through Gaussian elimination, the combination
\begin{equation}
({\eref{sunset-I124-simpler}} + {\eref{sunset-I125-simpler}}) - ({\eref{sunset-I234-simpler}} + {\eref{sunset-I135-simpler}})
\end{equation}
is zero, leaving five independent T2-type equations for six degree-two operators. To proceed, we choose an ordering scheme in which five operators (for example, $\partial_1^2$, $\partial_2^2$, and so on) are solved in terms of the sixth (for example, $\partial_3^2$). No update is needed for these five newly obtained reduction rules.

We next address the three degree-one equations. The corresponding matrix involves derivatives with respect to all five degree-one operators:
\begin{align}
    \left( \begin{array}{c| c |c |c| c| l} {\d\over \d\eta_1} & {\d\over \d\eta_2} &   {\d\over \d\eta_3} & {\d\over \d\eta_4} & {\d\over \d\eta_3} &  \text{Source Eq.} \\  \hline
	{2 m_1^2(m_2^2-m_3^2)\over s_0^2}  & {m_2^2 f_{++-}\over s_0^2} &  -{m_3^2 f_{+-+}\over s_0^2}   & -\WH O_{0;4,1}  &  \WH O_{0;5,2}  &  {\eref{sunset-I145-1}} \\  \hline
	{m_1^2 f_{++-}\over s_0^2}	&  {2 m_2^2(m_1^2-m_3^2)\over s_0^2}  & -{m_3^2 f_{-++}\over s_0^2}   & \WH O_{0;4,2}  &  -\WH O_{0;5,1}   &   {\eref{sunset-I245-1}} \\  \hline
	{m_1^2 f_{+-+}\over s_0^2}	  & -{m_2^2 f_{-++}\over s_0^2} & {2 m_3^2(m_1^2-m_2^2)\over s_0^2}  &  -\WH O_{0;4,1} &  \WH O_{0;5,1}  &  {\eref{sunset-I345-1}}\\  
\end{array}\right).~~~~~\label{sunset-massive-B-3}
\end{align}
After Gaussian elimination, the system simplifies to
\begin{align}
    \left( \begin{array}{c| c |c |c| c| l} {\d\over \d\eta_1} & {\d\over \d\eta_2} &   {\d\over \d\eta_3} & {\d\over \d\eta_4} & {\d\over \d\eta_3} &  \\  \hline
	  &  &      & -\WH O_{0;4,5}  &  \WH O_{0;5,4}  &  {\eref{sunset-I145-1}}- {\eref{sunset-I245-1}}+ {\eref{sunset-I345-1}}\\  \hline
h_1	  & g_{2,1,3}  & g_{3,1,2}  & \WH O_{0;I}  &    &   {\eref{sunset-I245-1}} + {\eref{sunset-I345-1}}\\  \hline
	g_{1,2,3} & h_2 & g_{3, 2, 1} &   &  \WH O_{0;I}  & {\eref{sunset-I145-1}} - {\eref{sunset-I345-1}}\\ 
\end{array}\right),
\end{align}
where
\begin{align}
    h_i = {2 m_i^2(m_i^2-K^2)\over s_0^2}, \quad
    g_{ijk} = {m_i^2(-m_i^2+ 3 m_j^2- 3m_k^2+K^2)\over s_0^2},~~~~~\label{extra_symbol}
\end{align}
\begin{align}
    \WH O_{0;i,j} =\eta_i{\d\over \d\eta_i}-2\eta_j{\d\over \d\eta_j}.~~~~~\label{extra_operator}
\end{align}

We identify the last two rows as T2-type equations, which are suitable for solving the ISP operators $\partial_4$ and $\partial_5$. Assuming the standard ordering in which ISP operators take precedence over propagator operators, we derive explicit reduction rules:
\begin{align}
    \WH O_{0;I}{\d \over \d \eta_4} G_{111}
=& -h_1 {\d \over \d \eta_1}G_{111} - g_{2,1,3} {\d \over \d \eta_2}G_{111} - g_{3,1,2} {\d \over \d \eta_3}G_{111}\nn  
&  +{f_{+--}\eta_5\over  s_0}{\d \over \d \eta_5}G_{111} -{(K^2-m_1^2)\eta_4+(K^2+m_2^2)\eta_5\over  s_0} {\d \over \d \eta_4}G_{111}\nn  
& -{K^2(D-2)-(D-3)(2m_1^2- m_2^2-m_3^2)\over s_0} G_{111} \nn 
& - ({K^2(f_{--+}- 2 m_2^2)\over 2 s_0^2}\eta_5+2{K^2 m_1^2\over s_0^2}\eta_4)G_{111}~~~~~~~~~\label{sunset-top-S1-n4-1}
\end{align}
\begin{align}
\WH O_{0;I} {\d \over \d \eta_5} G_{111}
=&-h_2 {\d \over \d \eta_2}G_{111} - g_{1,2,3} {\d \over \d \eta_1}G_{111} - g_{3,2,1} {\d \over \d \eta_3}G_{111}\nn 
&+ {f_{-+-}\eta_4\over  s_0}{\d \over \d \eta_4}G_{111} -{(K^2+m_1^2)\eta_4+(K^2-m_2^2)\eta_5\over  s_0} {\d \over \d \eta_5}G_{111}\nn 
& -{K^2(D-2)-(D-3)(2 m_2^2-m_1^2-m_3^2)\over s_0} G_{111}\nn 
&  -({K^2(f_{--+}-2m_1^2)\eta_4\over 2 s_0^2}+2{K^2 m_2^2\over s_0^2}\eta_5)G_{111}
~~~~~~\label{sunset-top-S1-n5-1}
\end{align}
Since these relations will be applied only to the set $(0,0,0,n_4,n_5)$, we retain only the terms that contribute nontrivially\footnote{In the corresponding recurrence relations, terms multiplied by $\eta_i$ with $i=1,2,3$ give zero contributions on this restricted set.}.

At this stage we have seven new reduction rules, two of which are degree-one rules for $n_4$ and $n_5$. Applying them to the irreducible set \eref{unreduced-set-1} reduces $n_4$ and $n_5$ to zero, so the remaining irreducible set becomes $(n_1, n_2, n_3, 0, 0)$. Applying the five degree-two propagator equations constrains the system further, leaving that set reducible except for ${\cal U}_3$ in \eref{sunset-irr-2} together with the two isolated points $(1,0,0,0,0)$ and $(0,1,0,0,0)$. The existence of infinitely many irreducible points along the $n_3$ direction necessitates a third iteration.

\subsubsection*{The third iteration: reducing the final set}

Since only $n_3$ remains unreduced, we proceed by acting with $\partial/\partial \eta_3$ on the existing equations. We focus on the descendant of the degree-one equation ${\eref{sunset-I245-1}} + {\eref{sunset-I345-1}}$.
Acting with $\partial_3$ on the full expression of this equation generates terms involving degree-two operators such as $\partial_3\partial_4$, $\partial_3\partial_2$, $\partial_3\partial_1$,  and $\partial_3^2$. Most of these can be simplified using the reduction rules found in the previous rounds. Crucially, the term $\partial_3^2$ does not cancel. 

After simplification, the new equation yields a relation involving the following degree-two propagator structure:
\begin{align}
    {g_{3,1,2}}{\d^2\over \d\eta_3^2}+ {g_{2,1,3}}{\d\over \d\eta_2} {\d\over \d\eta_3} + {h_1}{\d\over \d\eta_1} {\d\over \d\eta_3},~~~~~\label{sunset-massive-B-5b}
\end{align}
where $h_i$ and $g_{ijk}$ are defined in \eref{extra_symbol}. The new equation is linear independent with the five independent T2-type equations from the system \eref{sunset-massive-B-1}. Thus, we can uniquely solve for all six degree-two operators ${\partial_1^2}$, ${\partial_2^2}$, ${\partial_3^2}$, ${\partial_1}{\partial_2}$, ${\partial_1}{\partial_3}$, ${\partial_2}{\partial_3}$, giving six T2A-type reduction rules.

Consequently, any integral with $n_3 \ge 2$ can be reduced. Combined with the previous steps, the set of irreducible integrals is finite and consists of
\begin{equation}
{(0,0,0,0,0), (1,0,0,0,0), (0,1,0,0,0), (0,0,1,0,0)} ,.
\end{equation}
These four points constitute the master integrals for the generic massive sunset topology, in agreement with the standard expectation. The contrast with the massless branch makes the role of kinematics especially clear: the same algorithmic framework applies in both cases, but the degree structure of the descendant equations determines how quickly the irreducible lattice collapses and how many master integrals survive.

\subsection{Strategies for the application of reduction rules}
\label{sec:reduction-strategies}
Having completed both kinematic branches, it is useful to pause and discuss how the reduction rules may actually be applied in practice. The algorithm typically produces more valid relations than are strictly necessary for completeness, so one still has to choose an operational reduction strategy. For the sunset topology, two natural strategies already emerge.

\subsubsection*{Strategy I: The minimal basis}
The first approach prioritizes a minimal number of recurrence relations that remain valid across the full kinematic lattice. A convenient choice is the following five general relations:
\begin{itemize}
\item Eqs.~\eqref{sunset-I245plus345-m=0} and \eqref{sunset-I145minus345-m=0} for the ISP directions $\eta_4$ and $\eta_5$,
\item Eqs.~\eqref{sunset-I234-m=0-n2} and \eqref{sunset-I135-m=0-n1} for the propagator directions $\eta_2$ and $\eta_1$,
\item Eq.~\eqref{sunset-m=0-Rn3} for the final direction $\eta_3$.
\end{itemize}
The reduction flow is then straightforward. First, use the ISP relations to reduce any integral with $n_4>0$ or $n_5>0$ to the sector with $n_4=n_5=0$. Next, apply the propagator relations to reduce $n_2$ and $n_1$ to zero, at the cost of shifting weight into $n_3$. Finally, use the last relation to reduce the remaining $n_3$ tower to the master integral. This strategy is compact, but the individual equations retain their full algebraic complexity.

\subsubsection*{Strategy II: The hierarchical approach}
The second approach exploits the fact that boundary regions of the lattice satisfy much simpler equations. This strategy uses a larger collection of rules, but applies them in stages so that each step is algebraically lighter.
\begin{enumerate}
\item \textbf{General reduction.} Begin with the six fundamental degree-two rules in Eqs.~\eqref{sunset-B1-2-1}--\eqref{sunset-B1-2-6} to reduce a generic point $(n_1,\dots,n_5)$ to the boundary sets in Eq.~\eqref{unreduced-set-1}.
\item \textbf{ISP reduction on the boundary.} On the subset $(0,0,0,n_4,n_5)$, terms proportional to $n_1$, $n_2$, and $n_3$ vanish after expansion, so the general ISP relations simplify substantially. Setting $\eta_1=\eta_2=\eta_3=0$ in Eq.~\eqref{sunset-I245plus345-m=0} yields
\begin{align}
-\WH O_{0;I} \frac{\partial}{\partial\eta_4} G_{111} =& \frac{K^2(D-2)}{s_0} G_{111} + \frac{K^2}{s_0} \eta_5 \frac{\partial}{\partial \eta_5} G_{111} + \frac{K^2}{s_0} (\eta_4+\eta_5) \frac{\partial}{\partial \eta_4} G_{111}\nn
& -\frac{(K^2)^2\eta_5}{2s_0^2} G_{111},
\label{sunset-I245plus345-m=0-n4n5}
\end{align}
and setting the same variables to zero in Eq.~\eqref{sunset-I145minus345-m=0} gives
\begin{align}
-\WH O_{0;I} \frac{\partial}{\partial\eta_5} G_{111} =& \frac{K^2(D-2)}{s_0} G_{111}+ \frac{K^2}{s_0}\eta_4\frac{\partial}{\partial \eta_4}G_{111} + \frac{K^2}{s_0}(\eta_5+\eta_4) \frac{\partial}{\partial \eta_5}G_{111} \nn
& -\frac{(K^2)^2\eta_4}{2s_0^2} G_{111}.
\label{sunset-I145minus345-m=0-n4n5}
\end{align}
\item \textbf{Propagator reduction on the constrained line.} On the line $(n_1,n_2,n_3,0,0)$, setting $\eta_4=\eta_5=0$ simplifies Eqs.~\eqref{sunset-I234-m=0-n2} and \eqref{sunset-I135-m=0-n1} to
\begin{align}
(D-4)\frac{\partial}{\partial \eta_2} G_{111} - 2\eta_2 \frac{\partial^2}{\partial \eta_2^2} G_{111} &= (D-4) \frac{\partial}{\partial \eta_3} G_{111} - 2\eta_3 \frac{\partial^2}{\partial \eta_3^2} G_{111},
\label{sunset-I234-m=0-simpler}
\end{align}
\begin{align}
(D-4)\frac{\partial}{\partial \eta_1} G_{111} - 2\eta_1 \frac{\partial^2}{\partial \eta_1^2} G_{111} &= (D-4) \frac{\partial}{\partial \eta_3} G_{111} - 2\eta_3 \frac{\partial^2}{\partial \eta_3^2} G_{111}.
\label{sunset-I135-m=0-simpler}
\end{align}
Likewise, the specialized relation for the final subset $(0,0,n_3,0,0)$ reduces from Eq.~\eqref{sunset-m=0-Rn3} to
\begin{align}
\WH O_{0;II} \frac{\partial}{\partial \eta_3} G_{111} &= -\frac{(D-3)(3D-8)}{2} G_{111} + \frac{(5D-16)}{2}\eta_3\frac{\partial}{\partial \eta_3} G_{111} - \eta_3^2 \frac{\partial^2}{\partial \eta_3^2} G_{111}.
\label{sunset-m0-Rn3-constrain}
\end{align}
\end{enumerate}

Comparing the two strategies, Strategy II is often computationally superior. Although it requires a larger inventory of rules, the individual algebraic manipulations on the boundary are substantially lighter, which helps prevent intermediate expressions from swelling unnecessarily. A detailed optimization of reduction paths depends on the eventual software implementation, so we leave that issue for future work. The sunset example nevertheless already makes the main point: the framework naturally produces both globally valid rules and specialized boundary rules, and either type can be organized into a consistent reduction strategy.

\subsubsection{Reducing special lattice points}

The rule we found so far can be applied to reduce integrals with arbitrary propagator power and tensor rank. In practice, integrals could be special, for example, with power one for propagators. For the case of sunset, it corresponds to the lattice points with $n_1=n_2=n_3=0$. For these special points, we could use Eqs.~\eqref{sunset-I245plus345-m=0} and \eqref{sunset-I145minus345-m=0} to reduce the tensor part very efficiently. More explicitly, Eqs.~\eqref{sunset-I245plus345-m=0} and \eqref{sunset-I145minus345-m=0} will be simplified to 
\begin{align}
	& -\WH O_{0;I}{\d\over \d\eta_4} G_{111}\nn
	=& {K^2(D-2)\over s_0} G_{111}  +{K^2\over s_0} \eta_5 {\d \over \d \eta_5} G_{111}+{K^2\over s_0} (\eta_4+\eta_5) {\d \over \d \eta_4} G_{111}, 
	~~~\label{sunset-I245plus345-m=0-Special}
\end{align}
and
\begin{align}
	& -\WH O_{0;I}{\d\over \d\eta_5} G_{111}\nn
	=& {K^2\eta_4\over  s_0}{\d \over \d \eta_4}G_{111} +{K^2(\eta_5+\eta_4)\over  s_0} {\d \over \d \eta_5}G_{111}+{K^2(D-2)\over s_0} G_{111}, 
	~~~\label{sunset-I145minus345-m=0-Special}
\end{align}

In literature, using the syzyzy and moduler intersection, we can achieve similar thing, i.e., concentrated to integrals with propagators having power one. This idea has been realized in the program NeatIBP, which has been used in many cutting-edge computation. Here we see that our algorithm has provided another realization for such special case (see equations \eref{sunset-I245plus345-m=0-Special} and \eref{sunset-I145minus345-m=0-Special}). 

\section{Top sector of the two-loop planar massless double-box}
\label{sec:planar-top}

\begin{figure}[h]
\centering
\includegraphics[width= 6cm]{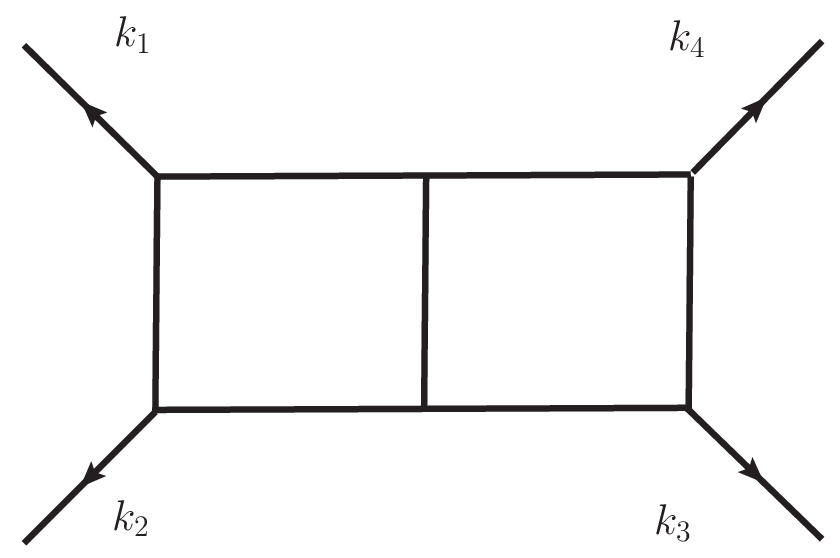}
\caption{The two-loop planar double-box diagram.}
\label{fig:doublebox}
\end{figure}

We next consider the two-loop planar massless double-box diagram shown in Fig.~\ref{fig:doublebox}. Relative to the sunset example, this topology probes a more demanding setting: it involves more propagators, two independent kinematic scales, and a substantially larger reduction lattice. It therefore serves as the first real test of whether the same algorithmic logic scales beyond the most economical two-loop family.

In this problem, there are four external momenta $k_1$, $k_2$, $k_3$, $k_4$ satisfying on-shell conditions $k_i^2=0$ and momentum conservation $\sum_{i=1}^4 k_i =0$, which leaves two independent scales $s=2 k_1 \cdot k_2$ and $t=2 k_2 \cdot k_3$. The complete set of Lorentz scalars is chosen as:

\begin{align}
	\begin{split}
		&\mathcal{D}_1=\ell_1^2,\ \mathcal{D}_2=(\ell_1+k_1)^2,\ \mathcal{D}_3=(\ell_1+k_1+k_2)^2,\\
		&\mathcal{D}_4= \ell_2^2,\ \mathcal{D}_5= (\ell_2-\ell_1)^2,\ \mathcal{D}_6=(\ell_2+k_1+k_2)^2,\\
		&\mathcal{D}_7=(\ell_2-k_4)^2,\ \mathcal{D}_8 = \ell_1 \cdot k_3,\ \mathcal{D}_9 = \ell_2 \cdot k_1,~~~\label{eq:lorentz-scalars}
	\end{split}
\end{align}
where $\mathcal{D}_{1},\dots,\mathcal{D}_{7}$ are the standard propagators and $\mathcal{D}_8, \mathcal{D}_9$ serve as ISPs. The generating function for the top sector is defined as:
\begin{equation}
G_{1111111}(\vec{\eta}) = \int\prod_{i=1}^{2}\frac{\md^{D}\ell_i}{\mi\pi^{D/2}} e^{\sum_{j=8}^9  \eta_j\mathcal{D}_j/s_0}
\prod_{i=1}^7 \frac{1}{\mathcal{D}_i-\eta_i s_0}, \label{eq:box-GF}
\end{equation}
For simplicity, we set $G = G_{1111111}$ throughout this section. To keep the presentation focused, we display only the highest-degree and next-to-highest-degree operator structures explicitly and suppress lower-degree terms with ellipses. Since our emphasis is the top sector, we also omit subsector contributions from the displayed equations.

\subsection{The first iteration from IBP relations}
\subsubsection{Generating and organizing the IBP relations}
For the double-box topology defined above, the standard derivation using total derivatives $\frac{\partial}{\partial \ell_i} \cdot \{k_j, \ell_k\}$ generates ten fundamental IBP relations. Acting with these on the generating function $G$ and extracting the corresponding differential operators, we obtain the primary system of equations:
\begin{align}
    0 &= \frac{s}{s_0}\frac{\partial}{\partial\eta_3}G +  (D-5)G-\sum_{i=1}^3(\eta_1+\eta_i){\d\over \d \eta_i} G+(\eta_4-\eta_1-\eta_5){\d\over \d\eta_5}G+ \eta_8 {\d\over \d\eta_8}G\,, \label{eq:box-IBP-1} 
\end{align}
\begin{align}
    0 &= 2 \frac{\partial}{\partial\eta_9} \frac{\partial}{\partial\eta_5}G - \frac{s}{s_0}\frac{\partial}{\partial\eta_3}G + {\cdots}\,, \label{eq:box-IBP-2}
\end{align}
\begin{align}
    0 &= 2 \frac{\partial}{\partial\eta_9}\frac{\partial}{\partial\eta_5} G - \frac{s}{s_0}\frac{\partial}{\partial \eta_1} G + {\cdots}\,, \label{eq:box-IBP-3}
\end{align}
\begin{align}
    0 &= 2\left(\frac{\partial}{\partial\eta_1} + \frac{\partial}{\partial\eta_2} + \frac{\partial}{\partial\eta_3} + \frac{\partial}{\partial\eta_5} \right)\frac{\partial}{\partial\eta_8}G - \frac{s+t}{s_0}\frac{\partial}{\partial\eta_2} G - \frac{s}{s_0}\left(\frac{\partial}{\partial\eta_3} + \frac{\partial}{\partial \eta_5}\right) G + {\cdots}\,, \label{eq:box-IBP-4}
\end{align}
\begin{align}
    0 &= 2 \frac{\partial}{\partial\eta_9}\frac{\partial}{\partial\eta_2} G - \frac{s}{s_0}\frac{\partial}{\partial \eta_3} G + {\cdots} \,,\label{eq:box-IBP-5}
\end{align}
\begin{align}
    0 &=  {s\over s_0}{\d\over \d \eta_6} G +  (D-5)G - \sum_{i=4}^7 (\eta_4+\eta_i){\d\over \d\eta_i} G+\eta_1{\d\over \d \eta_5} G+\eta_9{\d\over \d\eta_9} G\,,~~~~~\label{eq:box-IBP-6} 
\end{align}
\begin{align}
    0 &= 2 \frac{\partial}{\partial\eta_8}\frac{\partial}{\partial\eta_5} G - \frac{s}{s_0}\left(\frac{\partial}{\partial \eta_5} + \frac{\partial}{\partial \eta_6}\right) G + {\cdots}\,, \label{eq:box-IBP-7}
\end{align}
\begin{align}
    0 &= 2\left(\frac{\partial}{\partial\eta_4} + \frac{\partial}{\partial\eta_5} + \frac{\partial}{\partial\eta_6} + \frac{\partial}{\partial\eta_7}\right)\frac{\partial}{\partial\eta_9}G + \frac{s}{s_0}\frac{\partial}{\partial \eta_6} G - \frac{t}{s_0}\frac{\partial}{\partial \eta_7} G + {\cdots}\,, \label{eq:box-IBP-8}
\end{align}
\begin{align}
    0 &= 2\frac{\partial}{\partial\eta_8}\frac{\partial}{\partial\eta_5} G - \frac{s}{s_0}\left(\frac{\partial}{\partial \eta_4} + \frac{\partial}{\partial \eta_5}\right) G + {\cdots}\,, \label{eq:box-IBP-9}
\end{align}
\begin{align}
    0 &= 2 \frac{\partial}{\partial\eta_8}\frac{\partial}{\partial\eta_7} G - \frac{s}{s_0}\left(\frac{\partial}{\partial \eta_6} + \frac{\partial}{\partial \eta_7}\right) G + {\cdots}\,. \label{eq:box-IBP-10}
\end{align}
Thus, the first iteration begins with two degree-one equations and eight degree-two equations. The immediate task is to determine which lattice directions are already controlled by these relations and which irreducible subsets survive after the first pass.

\subsubsection{Solving the system of equations}
We begin with the degree-two equations, since they control the mixed propagator--ISP directions that dominate the first iteration. Gaussian elimination gives:
\begin{align}
    \left( \begin{array}{c| c| c| c| c| c| c |c| c | c |l}
	{\d_2\d_9} &	{\d_4\d_9} &  {\d_5\d_9} &  {\d_6\d_9} &  {\d_7\d_9} & 
	{\d_1\d_8} &  {\d_2\d_8} &  {\d_3\d_8} &  {\d_5\d_8} & {\d_7\d_8} & \text{Linear Comb.} \\  \hline
	2  & &  & & & & & & &  &		  \eqref{eq:box-IBP-5} \\  \hline
	& 2 &  &2 &2 & & & & &  &		  \eqref{eq:box-IBP-8} - \eqref{eq:box-IBP-2} \\  \hline
	& & 2 & & & & & & &  &		  \eqref{eq:box-IBP-2} \\  \hline
	& &  & & & 2 & 2 & 2 & &  & \eqref{eq:box-IBP-4} - \eqref{eq:box-IBP-9} \\  \hline
	& &  & & & & & & 2 &  &		  \eqref{eq:box-IBP-9} \\  \hline
	& &  & & & & & & & 2 &		  \eqref{eq:box-IBP-10} \\ \hline
	& &  & & & & & & &  &\eqref{eq:box-IBP-2}-\eqref{eq:box-IBP-3} \\  \hline
	& &  & & & & & &  &  &\eqref{eq:box-IBP-7}-\eqref{eq:box-IBP-9}
\end{array}\right).~~~~~\label{eq:box-matrix-deg2-reduced}
\end{align}
From this reduced system, we obtain four explicit T1A-type reduction rules for the composite operators $\d_2\d_9$, $\d_5\d_9$, $\d_5\d_8$, and $\d_7\d_8$. The two coupled rows also provide two T2A-type reduction rules, for example for $\d_2\d_8$ and $\d_7\d_9$. At this stage the operator structure is still simple enough that no additional update step is needed.

Next, we collect the degree-one primary equations (\eqref{eq:box-IBP-1} and \eqref{eq:box-IBP-6}) together with the degree-down combinations generated from the previous step ($\eqref{eq:box-IBP-2}-\eqref{eq:box-IBP-3}$ and ${\eqref{eq:box-IBP-7}} - {\eqref{eq:box-IBP-9}}$). The Gaussian elimination gives:
\begin{align}
\left( \begin{array}{c|c|c|c| l}
\partial_1 & \partial_3 & \partial_4 & \partial_6 & \text{Linear Comb.} \\  \hline
\frac{s}{s_0} &  &  &  & \eqref{eq:box-IBP-1} + \eqref{eq:box-IBP-2} - \eqref{eq:box-IBP-3} \\ \hline
 & \frac{s}{s_0} &  &  & \eqref{eq:box-IBP-1} \\ \hline
 &  & \frac{s}{s_0} &  & \eqref{eq:box-IBP-6} + \eqref{eq:box-IBP-7} - \eqref{eq:box-IBP-9} \\ \hline
  &  &  & \frac{s}{s_0} & \eqref{eq:box-IBP-6} \\
\end{array}\right). \label{eq:box-matrix-deg1-reduced}
\end{align}
Thus four T1A-type reduction rules are obtained for $\partial_1$, $\partial_3$, $\partial_4$, and $\partial_6$. These rules immediately remove four propagator directions from the irreducible lattice and sharply simplify the geometry of the remaining problem.

Furthermore, these four degree-one T1A-type reduction rules can be used to generate the degree-two operators ${\d_4\d_9}$, ${\d_6\d_9}$, ${\d_1\d_8}$, and ${\d_3\d_8}$. As a result, the two T2A-type reduction rules obtained in the previous step can be updated to fully resolved T1A-type reduction rules:
\begin{align}
    2 {\d\over \d \eta_8}{\d\over \d\eta_2}G
= & {s+t\over s_0}{\d\over \d\eta_2}G+{s\over s_0}{\d\over \d\eta_3}G+ {4(D-4)s_0\over s} {\d\over \d \eta_8}G+{4s_0\over s}\eta_8 {\d^2\over \d \eta_8^2}G\nn 
&-{2 s_0\over s}(\eta_1+\eta_3-\eta_4+2\eta_5-\eta_6) {\d\over \d \eta_8}{\d\over \d\eta_5}G -{2s_0\over s}(\eta_1+\eta_3){\d\over \d \eta_8}{\d\over \d\eta_3}G\nn 
&  -{2s_0\over s}(\eta_1+2\eta_2+\eta_3){\d\over \d \eta_8}{\d\over \d\eta_2}G-{2s_0\over s}(3\eta_1+\eta_3){\d\over \d \eta_8}{\d\over \d\eta_1}G+ {\cdots},~~~~~~\label{eq:box-rule-eta2eta8}
\end{align}
\begin{align}
    2{\d\over \d \eta_9} {\d\over \d \eta_7} G = & {s\over s_0}{\d\over \d\eta_6}G +{t\over s_0} {\d\over \d\eta_7}G   +{4(D-4)s_0\over s} {\d\over \d\eta_9}G +{4s_0\over s}\eta_9 {\d^2\over \d\eta_9^2}G \nn
&-{2s_0\over s} (3\eta_4+\eta_6){\d\over \d\eta_9} {\d\over \d\eta_4}G
+{2s_0\over s}(\eta_1+\eta_3-\eta_4-2\eta_5-\eta_6){\d\over \d\eta_9} {\d\over \d\eta_5}G  \nn
& -{2s_0\over s}(\eta_4+3\eta_6){\d\over \d\eta_9} {\d\over \d\eta_6}G  -{2s_0\over s}(\eta_4+\eta_6+2\eta_7){\d\over \d\eta_9}{\d\over \d\eta_7}G+ {\cdots}.~~~~~~\label{eq:box-rule-eta7eta9}
\end{align}
Although intermediate T2 choices are not unique, the updated equations correspond to the same underlying operator relations. In that sense, the ambiguity is organizational rather than mathematical.

\subsubsection{Completeness check}
To summarize the first round, we have obtained ten reduction rules for the following differential operators:
\begin{align}
\text{Degree one:}& \quad \left\{ \frac{\partial}{\partial \eta_1}, \frac{\partial}{\partial \eta_3}, \frac{\partial}{\partial \eta_4}, \frac{\partial}{\partial \eta_6} \right\}, \label{eq:box-10IBP-one} \\
\text{Degree two:}& \quad \left\{\frac{\partial}{\partial\eta_8}, \frac{\partial}{\partial\eta_9}\right\} \times \left\{\frac{\partial}{\partial\eta_2}, \frac{\partial}{\partial\eta_5}, \frac{\partial}{\partial\eta_7}\right\}. \label{eq:box-10IBP-two}
\end{align}

We can now interpret these first-round rules on the coefficient lattice. The degree-one relations in \eqref{eq:box-10IBP-one} reduce any integral with nonzero $n_1$, $n_3$, $n_4$, or $n_6$, so all unreduced points are confined to the five-parameter subspace $(0, n_2, 0, 0, n_5, 0, n_7, n_8, n_9)$. The degree-two rules in \eqref{eq:box-10IBP-two} then decouple the ISP indices from the remaining propagator indices whenever they appear together. In this sense, the first iteration reproduces, on a larger lattice, the same mixed-derivative reduction pattern already seen in the sunset example.
Applying these rules sequentially forces the indices to decouple into two distinct irreducible subsets\footnote{ One can see the irreducible subsets is exact the same one as the top-sector of sunset topology given in \eref{unreduced-set-1}.} :
\begin{align}
\mathcal{U}_1 &= { (0,n_2,0,0,n_5,0,n_7,0,0) } , \label{eq:box-remain-U1} \\
\mathcal{U}_2 &= { (0,0,0,0,0,0,0,n_8,n_9) } . \label{eq:box-remain-U2}
\end{align}
The appearance of this decoupling pattern shows that the reduction is already organizing itself into a small number of geometrically simple subsets. However, because the indices inside $\mathcal{U}_1$ and $\mathcal{U}_2$ are still unbounded, the first iteration is not complete. The second iteration must therefore probe the internal structure of these two residual subsets.

\subsection{The second iteration from descendant equations}
Before starting the second iteration, it is useful to note that the remaining irreducible subsets $\mathcal{U}_1$ and $\mathcal{U}_2$ impose strong kinematic restrictions on the operator content. In particular, any term whose operator index has a negative component in the $i$-th position ($i=1,3,4,6$) can be safely dropped, because its contribution vanishes on the reduced points belonging to $\mathcal{U}_1$ and $\mathcal{U}_2$.


\subsubsection{Generating the descendant equations}
Since the propagator indices $n_1$, $n_3$, $n_4$, and $n_6$ are already completely reduced, derivatives with respect to these variables produce only redundant equations. We therefore generate new equations by acting with $\partial_i$ for $i \in \{2,5,7,8,9\}$ on the first-round reduction rules.

When acting on the four degree-one reduction rules derived previously, all resulting degree-two equations are trivially generated. The nontrivial information instead comes from acting on the six degree-two rules and organizing the results through integrability conditions of the form $[\partial_a, \partial_b]\mathcal{O} = 0$. After simplification with the known descendant rules, we obtain nine linearly independent equations.

To write these cleanly, we introduce two sets of zero-index operators. For the ISP parameters ($i=8,9$), we define:
\begin{equation}
\WH O_{0;i}^{up} \equiv (D-4) + \eta_i \frac{\partial}{\partial\eta_i} ,\quad (i=8,9). \label{eq:box-O-isp}
\end{equation}
For the remaining propagator parameters ($i=2,5,7$), we define:
\begin{equation}
\WH O_{0;i}^{up} \equiv \frac{D-6}{2} - \eta_i \frac{\partial}{\partial\eta_i} ,\quad (i=2,5,7). \label{eq:box-O-prop}
\end{equation}

On the subset $\mathcal{U}_2$, the descendant equations generated for the ISP are:
\begin{enumerate}
\item from $[\partial_8,\partial_9]\partial_2 = 0$:
\begin{equation}
0 = \frac{2s_0}{s}\WH O_{0;8}^{up} \frac{\partial}{\partial\eta_8}\frac{\partial}{\partial\eta_9}G +\frac{1}{2}\WH O_{0;8}^{up} \frac{\partial}{\partial\eta_8}G - \left(\eta_8\frac{\partial}{\partial\eta_8} - \frac{1}{2}\eta_9 \frac{\partial}{\partial\eta_9}\right) \frac{\partial}{\partial\eta_9}G + \cdots, \label{eq:B-Int2-x82x92}
\end{equation}
\item from $[\partial_8,\partial_9]\partial_5 = 0$:
\begin{equation}
0 = \WH O_{0;8}^{up} \frac{\partial}{\partial\eta_8}G - \WH O_{0;9}^{up} \frac{\partial}{\partial\eta_9}G + \cdots, \label{eq:B-Int2-x85x95}
\end{equation}
\item from $[\partial_8,\partial_9]\partial_7 = 0$:
\begin{equation}
0 = \frac{2s_0}{s} \WH O_{0;9}^{up}\frac{\partial}{\partial\eta_8}\frac{\partial}{\partial\eta_9}G +\left(\WH O_{0;8}^{up}+\eta_9 \frac{\partial}{\partial\eta_9}-\frac{1}{2}\eta_8\frac{\partial}{\partial\eta_8} \right) \frac{\partial}{\partial\eta_8}G - \frac{1}{2}\WH O_{0;9}^{up}\frac{\partial}{\partial\eta_9}G + \cdots. \label{eq:B-Int2-x87x97}
\end{equation}
\end{enumerate}

On the subset $\mathcal{U}_1$, the descendant equations generated for the propagator are:
\begin{enumerate}
\item from $[\partial_2,\partial_5]\partial_8 = 0$:
\begin{align}
0 = & {t\over 2 s_0}{\d\over \d\eta_2}{\d\over \d\eta_5}G
 +{(D-5)\over 2}{\d\over \d\eta_2}G
+(D-{11\over 2}){\d\over \d\eta_5}G
-\eta_7 {\d\over \d\eta_7}{\d\over \d\eta_2}G\nn
& -\eta_7 {\d\over \d\eta_7}{\d\over \d\eta_5}G
-\eta_5{\d^2\over \d\eta_5^2}G
-{2t\eta_2+s(\eta_2+\eta_5+\eta_7)\over 2 s} {\d\over \d\eta_2}{\d\over \d\eta_5}G+ \cdots, \label{eq:B-Int2-x82x85}
\end{align}
\item from $[\partial_2,\partial_7]\partial_8 = 0$:
\begin{align}
    0 = & {t\over 2 s_0}{\d\over \d\eta_2}{\d\over \d\eta_7}G +{(D-5)\over 2}{\d\over \d\eta_2}G +(D-{11\over 2}){\d\over \d\eta_7}G 
-\eta_5 {\d\over \d\eta_5}{\d\over \d\eta_2}G \nn
& 
-\eta_5 {\d\over \d\eta_7}{\d\over \d\eta_5}G
-\eta_7{\d^2\over \d\eta_7^2}G
-{1\over 2}(\eta_2+\eta_5+\eta_7) {\d\over \d\eta_2}{\d\over \d\eta_7}G+ \cdots, \label{eq:B-Int2-x82x87}
\end{align}
\item from $[\partial_7,\partial_5]\partial_8 = 0$:
\begin{equation}
0 =  \WH O_{0;5}^{up}\frac{\partial}{\partial\eta_5}G - \WH O_{0;7}^{up}\frac{\partial}{\partial\eta_7}G  + \cdots, \label{eq:B-Int2-x85x87}
\end{equation}
\item from $[\partial_2,\partial_5]\partial_9 = 0$:
\begin{equation}
0 =  \WH O_{0;2}^{up}\frac{\partial}{\partial\eta_2}G - \WH O_{0;5}^{up}\frac{\partial}{\partial\eta_5}G  + \cdots, \label{eq:B-Int2-x92x95}
\end{equation}
\item from $[\partial_2,\partial_7]\partial_9 = 0$:
\begin{align}
0 = & {t\over 2 s_0}{\d\over \d\eta_2}{\d\over \d\eta_7}G
+{(D-5)\over 2}{\d\over \d\eta_7}G 
+(D-{11\over 2}){\d\over \d\eta_2}G
-\eta_5 {\d\over \d\eta_5}{\d\over \d\eta_7}G \nn 
& 
-\eta_5 {\d\over \d\eta_2}{\d\over \d\eta_5}G
-\eta_2{\d^2\over \d\eta_2^2}G
-{1\over 2}(\eta_2+\eta_5+\eta_7) {\d\over \d\eta_2}{\d\over \d\eta_7}G  + \cdots, \label{eq:B-Int2-x92x97}
\end{align}
\item from $[\partial_7,\partial_5]\partial_9 = 0$:
\begin{align}
= & {t\over 2 s_0}{\d\over \d\eta_5}{\d\over \d\eta_7}G
+{(D-5)\over 2}{\d\over \d\eta_7}G 
+(D-{11\over 2}){\d\over \d\eta_5}G 
-\eta_2 {\d\over \d\eta_2}{\d\over \d\eta_7}G\nn 
& 
-\eta_2{\d\over \d\eta_2}{\d\over \d\eta_5}G 
-\eta_5{\d^2\over \d\eta_5^2}G 
-{1\over 2}(\eta_2+\eta_5+\eta_7) {\d\over \d\eta_5}{\d\over \d\eta_7}G + \cdots. \label{eq:B-Int2-x95x97}
\end{align}
\end{enumerate}

\subsubsection{Solving the system of equations}
Among these nine equations, exactly six contain degree-two operators. After Gaussian elimination, we obtain
\begin{align}
    \left( \begin{array}{c | c | c |c |l}
	{\d_8\d_9} &   {\d_2\d_5} &  {\d_2\d_7} & {\d_5\d_7} & \text{Linear Comb.} \\  \hline
	{2s_0\over s}\WH O_{0;8}^{up}   &  & & &  {\eref{eq:B-Int2-x82x92}}\\    \hline
	& {t\over 2 s_0} & & &  {\eref{eq:B-Int2-x82x85}} \\    \hline
	&  &{t\over 2 s_0}  & &   {\eref{eq:B-Int2-x82x87}} \\     \hline
	&  &  & {t\over 2 s_0}   &  {\eref{eq:B-Int2-x95x97}} \\ \hline
	 &   & & & \WH O_{0;8}^{up} {\eref{eq:B-Int2-x87x97}}-  \WH O_{0;9}^{up}{\eref{eq:B-Int2-x82x92}}\\    \hline
	&  &  & &  {\eref{eq:B-Int2-x82x87}}-{\eref{eq:B-Int2-x92x97}}\\
\end{array}\right), \label{eq:box-solve-2-2-2}
\end{align}
which provides three T1A-type reduction rules for ${\d_2\d_5}$, ${\d_2\d_7}$, and ${\d_5\d_7}$, one T1B-type reduction rule for ${\d_8\d_9}$, and two residual degree-one equations:
\begin{equation}
0 = -\WH O_{0;2}^{up}\frac{\partial}{\partial\eta_2}G + \WH O_{0;7}^{up}\frac{\partial}{\partial\eta_7}G + \cdots, \label{eq:B-Int2-x92x97-x8287}
\end{equation}
\begin{equation}
0 = \WH O_{0;9,1}\frac{\partial}{\partial\eta_9} G - \WH O_{0;8,1}\frac{\partial}{\partial\eta_8} G + \cdots, \label{eq:B-Int2-x289-x789}
\end{equation}
where the zero-index operators are:
\begin{align}
\WH O_{0;9,1} &= \frac{1}{2}\WH O_{0;9}^{up} \left( \WH O_{0;9}^{up} - \eta_8\frac{\partial}{\partial\eta_8} \right), \\
\WH O_{0;8,1} &= \frac{1}{2}\WH O_{0;8}^{up} \left(\WH O_{0;9}^{up} + \eta_8\frac{\partial}{\partial\eta_8} \right) .
\end{align}

We now assemble the existing and newly generated degree-one equations. At this point an important subtlety appears. A naive elimination of $\partial_8$ and $\partial_9$ using Eqs.~\eqref{eq:B-Int2-x85x95} and \eqref{eq:B-Int2-x289-x789} would produce a leading coefficient proportional to
\begin{equation}
\WH O_{0;9,1} \WH O_{0;8}^{up} - \WH O_{0;8,1}\WH O_{0;9}^{up} = -\frac{1}{2}(\WH O_{0;9}^{up}+ \WH O_{0;8}^{up})\eta_8\frac{\partial}{\partial\eta_8} . \label{eq:box-bad-operator}
\end{equation}
Because this operator involves $\eta_8\frac{\partial}{\partial\eta_8}$, it vanishes on the boundary $n_8=0$. As a result, it cannot serve as a globally valid leading coefficient for reduction. To avoid this structural obstruction, we modify the elimination strategy and keep the rules in a form that remains safe on the relevant lattice boundaries:

\begin{align}
    \left( 
    \begin{array}{c | c  | c  |c | c |l}
	{\d_8}  &  {\d_9} &   {\d_2} &   {\d_5} &  {\d_7}  & \\
	\hline 
	(3-D)\eta_8{\d_8}- \eta_8^2{\d_8^2}&  &  &   &     & \left(  \WH O_{0;9}^{up}-\eta_8{\d_8}  \right) {\eref{eq:B-Int2-x85x95}  }+ {\eref{eq:B-Int2-x289-x789}  } \\  \hline 
	\WH O_{0;8}^{up} & -\WH O_{0;9}^{up}  &  &   &     &   {\eref{eq:B-Int2-x85x95}  } \\  \hline 
	& &\WH O_{0;2}^{up} &  -\WH O_{0;5}^{up} &   &  {\eref{eq:B-Int2-x92x95}} \\ \hline
	& & &\WH O_{0;5}^{up} & - \WH O_{0;7}^{up} &    {\eref{eq:B-Int2-x85x87}} \\  \hline
	&  & &   &    &  {\eref{eq:B-Int2-x92x97-x8287}  }+ {\eref{eq:B-Int2-x85x87}}+{\eref{eq:B-Int2-x92x95}}\\ 
\end{array}\right). \label{eq:box-solve-2-1-2-4}
\end{align}
The rules in the third and fourth rows reduce $(n_2, n_5, n_7)$ to $(0, N_5, 0)$ with $N_5 \leq n_2+n_5+n_7$. The second row then reduces $(n_8, n_9)$ to $(N_8, 0)$ with $N_8 \leq n_8+n_9$. Finally, the first row reduces every point with $n_8 \geq 2$ to the cases $n_8=0$ or $1$.

Since all four new degree-one reduction rules are TB-type, we do not use them in the global update step. This preserves the simpler degree-two TA-type rules that remain more convenient for practical reduction.

\subsubsection{Completeness check}
Substituting these newly derived rules back into the irreducible subsets reduces $n_2$, $n_7$, and $n_9$ completely to zero. In addition, the index $n_8$ is restricted to the values $0$ or $1$. The residual irreducible set descending from $\mathcal{U}_1$ and $\mathcal{U}_2$ is therefore restricted to
\begin{equation}
\mathcal{U}^{\text{2nd}} = { (0,0,0,0,n_5,0,0,0,0) } \cup { (0,0,0,0,0,0,0,1,0) }.
\end{equation}
The subset governed by $n_5$ still contains an infinite tower of points. A third iteration is therefore required to close the reduction completely.

\subsection{The third iteration for reducing the final set}

At this stage the residual set $\mathcal{U}^{\text{2nd}}$ is already so constrained that most newly generated equations are either trivial or redundant. It is therefore sufficient to focus on the essential operators $\partial_5$ and $\partial_8$ acting on selected degree-one relations from the previous iteration. We display only the new relations needed to eliminate the final irreducible subset.

First, we act with $\partial_5$ on the degree-one relation ${\eqref{eq:B-Int2-x85x87}}$. After using previously known reduction rules to simplify the result, we obtain a clean degree-two equation:
\begin{equation}
0 = \left[ \WH O_{0;5}^{up}-1\right] \frac{\partial^2}{\partial\eta_5^2} G+ \cdots, \label{eq:B-R3-eq-eta5-3}
\end{equation}
which is a degree-two T1B-type reduction rule for $\partial_5^2$. This rigorously allows us to reduce any integral with $n_5 \geq 2$ to $n_5 = 0,\,1$.

Next, we act with $\partial_8$ on the relation ${\eqref{eq:B-Int2-x92x95}}$. After simplification, the new equation naturally mixes $\partial_8$, $\partial_5$, and $\partial_2$. By combining the equation $-\frac{s+t}{2 s_0}{\eqref{eq:B-Int2-x92x95}}$, the $\partial_2$ component perfectly cancels, leaving a degree-one T1B-type equation between $\partial_8$ and $\partial_5$:
\begin{equation}
0 = \frac{s_0}{s}\left[ \WH O_{0;8}^{up}(\WH O_{0;8}^{up})-1\right] \frac{\partial}{\partial\eta_8} G + \frac{t}{2 s_0}\WH O_{0;5}^{up} \frac{\partial}{\partial\eta_5} G + \cdots, \label{eq:B-R3-eq-eta8-4}
\end{equation}
which gives a rule to fully express $\partial_8$ in terms of $\partial_5$. With this relation, we can systematically lower $n_8$ unconditionally down to $n_8 = 0$.

At the end of this third and final iteration, every index $n_i$ with $i \neq 5$ has been forced to zero, while the remaining propagator index $n_5$ is restricted to $\{0,1\}$. We therefore obtain two master integrals:
\begin{equation}
\mathcal{M} = \left\{ (0,0,0,0,0,0,0,0,0), \quad (0,0,0,0,1,0,0,0,0) \right\}, \label{eq:box-final-MI}
\end{equation}
which form the complete master-integral basis for the top sector. The planar double-box example therefore closes after three iterations and confirms that the same operator-based strategy remains effective well beyond the sunset family.


\section{Top sector of the two-loop non-planar massless double-box}
\label{sec:nonplanar-top}

\begin{figure}[h]
\centering
\includegraphics[width= 6cm]{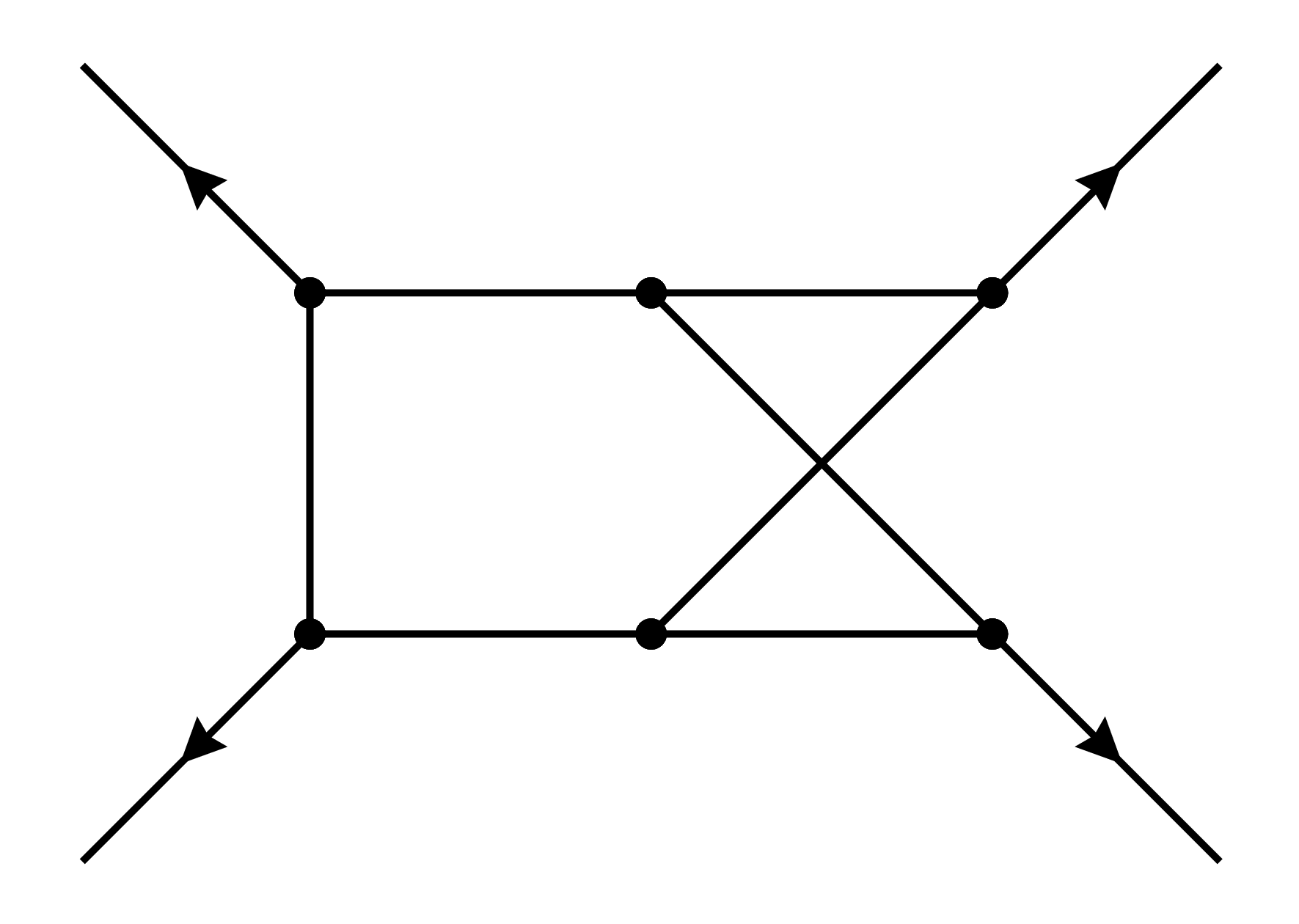}
\caption{The two-loop non-planar double-box diagram.}
\label{fig:npdoublebox}
\end{figure}

We now turn to the two-loop non-planar massless double-box diagram shown in Fig.~\ref{fig:npdoublebox}. This example is more intricate than the planar double box because the non-planar routing leads to a less transparent organization of the irreducible lattice. It therefore provides a useful robustness test: if the same iterative framework still closes cleanly here, then the method is not tied to especially simple routing structures.

Similar to the previous example, there are four external momenta $k_1$, $k_2$, $k_3$, $k_4$ satisfying on-shell conditions $k_i^2=0$ and momentum conservation $\sum_{i=1}^4 k_i =0$, which leaves two independent kinematic scales $s=2 k_1 \cdot k_2$ and $t=2 k_2 \cdot k_3$. The complete set of Lorentz scalars is explicitly chosen as:
\begin{align}
&\mathcal{D}_1={\ell_1}^2,\quad \mathcal{D}_2=(\ell_1-k_1)^2,\quad \mathcal{D}_3={(\ell_1-k_1-k_2)}^2,\nn
&\mathcal{D}_4= {\ell_2}^2,\quad \mathcal{D}_5={(\ell_2+k_4)}^2,\quad \mathcal{D}_6= {(\ell_2-\ell_1+k_1+k_2+k_4)}^2,\nn
&\mathcal{D}_7={(\ell_2-\ell_1)}^2,\quad \mathcal{D}_8 = \ell_1 \cdot k_4,\quad \mathcal{D}_9 = \ell_2 \cdot k_1,~~~\label{eq:npbox-scalars}
\end{align}
where $\mathcal{D}_1 \cdots \mathcal{D}_7$ are the standard propagators forming the topology, and $\mathcal{D}_8, \mathcal{D}_9$ serve as ISPs. The generating function for the top sector is defined as:
\begin{equation}
G_{1111111}(\vec{\eta}) = \int\prod_{i=1}^{2}\frac{\md^{D}\ell_i}{\mi\pi^{D/2}} e^{\sum_{j=8}^9  \eta_j\mathcal{D}_j/s_0}
\prod_{i=1}^7 \frac{1}{\mathcal{D}_i-\eta_i s_0}, \label{eq:npbox-GF}
\end{equation}
For simplicity, we set $G = G_{1111111}$ throughout the rest of this section.

Since the general workflow has already been explained in detail in the previous examples, we streamline the presentation here and display only the essential relations obtained at each iteration together with the auxiliary equations needed for the next step. As before, we keep only the leading and next-to-leading operator structures explicit and absorb lower-degree terms into ellipses. The emphasis in this section is therefore not on reproducing every intermediate manipulation, but on showing how the same iterative logic survives in a more entangled lattice geometry.

\subsection{The first iteration from IBP relations}
Following the same workflow as before, we begin with the ten fundamental IBP relations generated by acting with total derivatives on the generating function. Their degree-two operator content can be summarized by the matrix
\begin{align}
    \left(\begin{array}{c |c | c | c | c | c | c |c |c | c | c |c |c |c }
	{\d_1\d_8} & {\d_2\d_8}  & {\d_3\d_8} & {\d_4\d_8} & {\d_5\d_8} & {\d_6\d_8} &  {\d_7\d_8}  & 
	{\d_1\d_9} & {\d_2\d_9}  & {\d_3\d_9} & {\d_4\d_9} & {\d_5\d_9} & {\d_6\d_9} &  {\d_7\d_9} \\
    \hline 
	&  & & & & 2 &  &  	&  &  & & & & \\  \hline 
	&  & & & &  &  &  	&  &  & & &2  & 2 \\   \hline 
	&  &  & & &2  & 2 & 	&  &  & & & -2 & -2 \\   \hline 
	-2	& -2 & -2 & & & -2 &  -2 & &  &  & & & & \\   \hline 
	&  & 2 & & & 2 &  & & 2 &  & & & &  \\    \hline 
	&  &  & & & -2 &  & &  &  & & & & \\    \hline 
	&  &  & & & &  & &  &  & -2 &-2 &-2  & -2\\    \hline 
	&  &  & 2 & 2 & &  & 	&  &  & & & & \\    \hline 
	&  &  & & &2 & 2 & &  &  & & & & \\    \hline 
	&  &  & &-2  &-2 &  & &  &  & & & & \\
\end{array}\right),
\end{align}

After Gaussian elimination, we first extract the degree-one core of the system. This gives two T1A-type reduction rules for the operators
\begin{align}
    {\d\over \d\eta_3},\quad {\d\over \d\eta_1}, \label{eq:npbox-IBP1}
\end{align}
which allow us to reduce the propagator indices $n_1$ and $n_3$ to zero.

We then return to the degree-two IBP relations. After updating the operator content with the newly solved descendants of $\partial_1$ and $\partial_3$, we obtain a set of T1A-type and T2A-type reduction rules. Six T1A-type rules are obtained for the operators
\begin{align}
    {\d\over \d\eta_7}{\d\over \d\eta_8},\quad {\d\over \d\eta_6}{\d\over \d\eta_8},\quad {\d\over \d\eta_5}{\d\over \d\eta_8},\quad
    {\d\over \d\eta_4}{\d\over \d\eta_8},\quad {\d\over \d\eta_2}{\d\over \d\eta_8},\quad {\d\over \d\eta_2}{\d\over \d\eta_9},\label{eq:npbox-IBP2-1}
\end{align}
and two T2A-type rules for the operators
\begin{align}
    {\d\over \d\eta_6}{\d\over \d\eta_9} \rightarrow   {\d\over \d\eta_7}{\d\over \d\eta_9},\quad {\d\over \d\eta_4}{\d\over \d\eta_9} \rightarrow {\d\over \d\eta_5}{\d\over \d\eta_9}.\label{eq:npbox-IBP2-2}
\end{align}
Here $\rightarrow$ indicates that the reduction rule for ${\d\over \d\eta_6}{\d\over \d\eta_9}$ is expressed in terms of ${\d\over \d\eta_7}{\d\over \d\eta_9}$.

With these first-round rules in hand, the lattice reduction can be organized as follows:
\begin{enumerate}
\item Applying the degree-one relations for operators \eqref{eq:npbox-IBP1}, all integrals with $n_1 \geq 1$ or $n_3 \geq 1$ are reduced, giving irreducible subset as $(0,n_2, 0,n_4, n_5, n_6, n_7, n_8, n_9)$.

\item Utilizing the degree-two relations for operators \eqref{eq:npbox-IBP2-1} and \eqref{eq:npbox-IBP2-2}, we completely decouple $n_8$ from $n_2$, $n_4$, $n_5$, $n_6$, $n_7$, and $n_9$ from $n_2$, $n_4$, $n_6$, giving $3$ irreducible subsets:
\begin{align}
  \mathcal{U}_{1}^{1st} &= (0,n_2, 0,n_4, n_5, n_6, n_7, 0, 0), \\
  \mathcal{U}_{2}^{1st} &= (0, 0, 0, 0, n_5, 0, n_7, 0, n_9), \\
  \mathcal{U}_{3}^{1st} &= (0, 0, 0, 0, 0, 0, 0, n_8, n_9).
\end{align}
\end{enumerate}
The first iteration therefore reduces the problem to three residual subsets with a much simpler structure than the full nine-dimensional lattice. Since each subset still contains infinitely many points, the reduction is not yet complete, but the remaining geometry is now simple enough to guide the next round.

\subsection{The second iteration from descendant equations}
Since $n_1$ and $n_3$ are already fully reduced, the second iteration only needs to probe the remaining directions. We therefore act with $\partial/\partial\eta_i$ for $i\neq 1,3$ on the degree-two rules established in the first iteration\footnote{One can check that acting on the degree-one reduction rules does not produce any non-trivially generated equation.}. After removing trivial descendants, the integrability conditions produce sixteen new equations, fourteen of which still carry degree-two operators. Their leading structure is summarized by
\begin{align}
    \left( \begin{array}{c| c |c|c|c|c|c|c|c|c   |c|c|c|c }
	\d_2\d_4 & \d_2\d_5 & \d_2\d_6 & \d_2\d_7  & \d_4\d_5 & \d_4\d_6 & \d_4\d_7 & \d_5\d_6 & \d_5\d_7 & \d_5\d_9 &  \d_6\d_7 & \d_7\d_9 & \d_8\d_9 & \d_8\d_8 \\ \hline  
	 -{s+t\over 2 s_0}&    &   &  &  &  &   &  &  &  &  &  &  &  \\  \hline  
	&  -{t\over 2 s_0}  &   &  &  &  &   &  &  &  &  &  &  &  \\  \hline 
	&    & -{s+t\over 2 s_0}   &  &  &  &   &  &  &  &  &  &  & \\  \hline  
	&    &   & -{t\over 2 s_0} &  &  &   &  &  &  &  &  &  &  \\  \hline  
	&    &   &  & &  &   &  &  &  &  &  & -2\WH O_{0;89} & \WH O_{0;89} \\  \hline  
	&    &   &  & {s\over 2 s_0} &  &   &  &  &  &  &  &  &  \\  \hline  
	&    &   &  &  &    &  {s\over 2 s_0}   &  &  &  &  &  &  & \\  \hline  
	&    &   &  &  &  &   &  &  & {-s\over 2 s_0} &  &  &  &  \\  \hline  
		&    &   &  &  &  &   &{-s\over 2 s_0}  &  &  &  &  &  & \\  \hline  
	&    &   &  &  &  &   &  &  &  & {s\over 2 s_0} &  &  & \\  \hline  
	&    &   &  &  &  &   &  &  &  &  &  {-s\over 2 s_0} &  & \\  \hline  
	&  {-t\over 2 s_0}  &   &  &  &  &   &  &  &  &  &  &  & \\  \hline  
	&    &  {-t-s\over 2 s_0}  &  &  &  &   &  &  &  &  &  &  & \\  \hline  
	&    &   &  &  &{-s\over 2 s_0}  &   & {-s\over 2 s_0} & {t\over 2 s_0} &  &  &  &  & \\
	\end{array}\right),
\end{align}
where we introduce the zero-index operator
\begin{align}
	 \WH O_{0;89} = &{2(D-4)s_0\over s}  +{s_0\over s}\eta_8 {\d\over \d\eta_8} +{s_0\over s}\eta_9  {\d\over \d\eta_9}.
\end{align}

Gaussian elimination now produces ten degree-two T1A-type equations, two degree-two T2A-type equations, and two degree-one equations. From them we derive ten T1A-type reduction rules for the operators
 ${\d_2}{\d_4}$, ${\d_2}{\d_5}$, ${\d_2}{\d_6}$,   ${\d_2}{\d_7}$, ${\d_4}{\d_5}$,  ${\d_4}{\d_7}$,  ${\d_5}{\d_9}$,
  ${\d_5}{\d_6}$, ${\d_6}{\d_7}$, and ${\d_7}{\d_9}$\footnote{It is worth pointing out that the T1A-type reduction rules for $\d_5\d_9$ and $\d_7\d_9$ turn the original T2A-type reduction rules for $\d_4\d_9$ and $\d_6\d_9$ into T1A-type reduction rules.}, together with one T2A-type reduction rule $\d_5\d_7\rightarrow \d_4\d_6$ and one T2B-type reduction rule $\d_8\d_9\rightarrow \d_8\d_8$. After using these new reduction rules to simplify the two degree-one equations, along with the two original degree-one equations, the coefficient matrix of degree-one operators is:
\begin{align}
    \left( \begin{array}{ c| c |c|c|c}
	{\d_2} & {\d_4} & {\d_5}  & {\d_6} & {\d_7} \\   \hline  
    & \WH O_{0;4}^{up} &  & -\WH O_{0;6}^{up} & \\ \hline
    &  & -\WH O_{0;5}^{up} &  & \WH O_{0;7}^{up} \\ \hline
   \WH O_{0;2}^{up} & -\WH O_{0;4}^{up} & -\WH O_{0;5}^{up} &  & \\ \hline
    -\WH O_{0;2}^{up} &  &  &  \WH O_{0;6}^{up} &  \WH O_{0;7}^{up} \\ 
\end{array}\right),
\end{align}
where the zero-index operators are defined as
\begin{equation}
    \widehat{O}_{0;i}^{up} = \frac{D-6}{2}-\eta_i \frac{\partial}{\partial\eta_i}.\label{eq:up-operator}
\end{equation}
One linear combination drops to degree zero. The remaining structure yields three degree-one T2B-type rules for the operators
\begin{align}
    \widehat{O}_{0;2}^{up}{\d\over \d\eta_2}\rightarrow\left(\widehat{O}_{0;4}^{up}{\d\over \d\eta_4},\widehat{O}_{0;5}^{up}{\d\over \d\eta_5}\right),\quad \widehat{O}_{0;6}^{up}{\d\over \d\eta_6}\rightarrow\widehat{O}_{0;4}^{up}{\d\over \d\eta_4},\quad\widehat{O}_{0;7}^{up}{\d\over\d\eta_7} \rightarrow \widehat{O}_{0;5}^{up}{\d\over \d\eta_5}, \label{eq:npbox-2-1}
\end{align}

These newly extracted rules further fragment the irreducible sets as follows:
\begin{enumerate}
    \item Applying T2B-type rules for operator $\widehat{O}_{0;2}^{up}{\d\over \d\eta_2}$, $\widehat{O}_{0;6}^{up}{\d\over \d\eta_6}$ and $\widehat{O}_{0;7}^{up}{\d\over \d\eta_7}$, the set $\mathcal{U}_{1}^{1st}$ is reduced to
    \begin{align}
        (0,0,0,n_4,n_5,0,0,0,0).
    \end{align}
    Then, using T1A-type rule for operator ${\d\over \d\eta_4}{\d\over \d\eta_5}$, it can be fragmented to
    \begin{align}
        (0,0,0,n_4, 0,0,0,0,0) \cup(0,0,0,0,n_5,0,0,0,0).
    \end{align}
    \item Applying the T1A-type rule for operator ${\d\over \d\eta_5}{\d\over \d\eta_9}$ and the T2B-type rule for $\widehat{O}_{0;7}^{up}{\d\over \d\eta_7}$, the set $\mathcal{U}_{2}^{1st}$ fragments to
    \begin{align}
        (0,0,0,0,n_5,0,0,0,0) \cup (0,0,0,0,0,0,0,0,n_9).
    \end{align}
    \item Applying the T2B-type rule for reducing operator $2\WH O_{0;89}  {\d\over \d\eta_8} {\d\over \d\eta_9}$ to $\WH O_{0;89}  {\d^2\over \d\eta_8^2}$, the set $\mathcal{U}_{3}^{1st}$ fragments to
    \begin{align}
        (0,0,0,0,0,0,0,n_8,0) \cup (0,0,0,0,0,0,0,0,n_9).
    \end{align}
\end{enumerate}
After the second iteration, every surviving irreducible subset is supported on a single nonzero index:
\begin{align}
  \mathcal{U}_{1}^{2nd} &= (0,0,0,n_4, 0,0,0,0,0), \\
  \mathcal{U}_{2}^{2nd} &= (0,0,0,0,n_5,0,0,0,0), \\
  \mathcal{U}_{3}^{2nd} &= (0,0,0,0,0,0,0,n_8,0), \\
  \mathcal{U}_{4}^{2nd} &= (0,0,0,0,0,0,0,0,n_9).
\end{align}

This is already a substantial simplification: the coupled subsets from the first round have collapsed to one-dimensional towers. A third iteration is nevertheless required, because each of those towers remains infinite.

\subsection{The third iteration for reducing the final sets}

Continuing the same procedure, and using the fact that $n_2$, $n_6$, and $n_7$ have already been fully reduced, we now act only with ${\d\over \d\eta_i}$ for $i=4,5,8,9$ on the three degree-one relations established in the second iteration. This produces four degree-two equations involving three degree-two operators:
\begin{align}
    \left( \begin{array}{ c| c |c}
	{\d\over \d\eta_4}  {\d\over \d\eta_6} & {\d^2\over \d\eta_4^2}  & {\d^2\over \d\eta_5^2} \\   \hline  
-\WH O_{0;6}^{up}	& \WH O_{0;4}^{up}-1 & \\ \hline
{(D-6)(s+t)\over 2 t}	&   & 1- \WH O_{0;5}^{up} \\ \hline
	 & 1-\WH O_{0;4}^{up} &  \\ \hline
 &  &    1- \WH O_{0;5}^{up} \\ 
\end{array}\right),
\end{align}
and four legitimate degree-one equations involving four degree-one operators:
\begin{align}
    \left( \begin{array}{ c| c |c|c}
	{\d\over \d\eta_4} &  {\d\over \d\eta_5} & {\d\over \d\eta_8}  & {\d\over \d\eta_9}  \\   \hline  
{(s+t)\over 2 s_0} \WH O_{0;4}^{up}	& & \WH O_{0; 89,1} & \\ \hline
	& {t\over 2 s_0} \WH O_{0;5}^{up}  & \WH O_{0; 89,1} & \\ \hline
{s+t\over 2 s_0}\WH O_{0;4}^{up}	&{t\over 2 s_0}\WH O_{0;5}^{up} & \WH O_{0; 89,2}  &  \\ \hline
 &{t\over 2 s_0}\WH O_{0;5}^{up}	& \WH O_{0;89,3} &    \WH O_{0;89,1} \\ 
\end{array}\right),
\end{align}

After removing trivially generated dependencies, simplifying with descendant reduction rules, and carrying out Gaussian elimination, we first obtain three degree-one T2B-type rules for the operators $\frac{\partial}{\partial\eta_4}$, $\frac{\partial}{\partial\eta_8}$, and $\frac{\partial}{\partial\eta_9}$:
\begin{align}
    \WH O_{0;5}^{up} {\d\over \d\eta_5} \rightarrow \WH O_{0;4}^{up}{\d\over \d\eta_4}, \quad
    \WH O_{0; 89,1}{\d\over \d \eta_8} \rightarrow \WH O_{0;4}^{up}{\d\over \d \eta_4}, \quad
    \WH O_{0; 89,2} {\d\over \d \eta_9} \rightarrow \left( \WH O_{0; 89,3} {\d\over \d \eta_8}, \WH O_{0;4}^{up} {\d\over \d\eta_4} \right), \label{eq:npbox-3-1}
\end{align}
where zero-index operators $\WH O_{0; 89,1}$, $\WH O_{0; 89,2}$ and $\WH O_{0; 89,3}$ defined by:
\begin{align}
    \WH O_{0; 89,1} = & {(D-5)(2D-9)s_0\over s}+ {(3D-13)\eta_9 s_0\over s}    {\d\over \d\eta_9} + {\eta_9^2s_0\over s}   {\d^2\over \d\eta_9^2}, \nn
    &  + {(D-5)\eta_8 s_0\over s} {\d\over \d\eta_8}   +{\eta_8\eta_9s_0\over s} {\d\over \d\eta_8} {\d\over \d\eta_9},\\
    \WH O_{0; 89,2} = &  {(D-4)(2D-9)s_0\over s}   + {3(D-4)\eta_9 s_0\over s}  {\d\over \d\eta_9}  + {\eta_9^2 s_0\over s} {\d^2\over \d\eta_9^2} G   \nn 
    &  + {6(D-4) \eta_8 s_0\over 2s}  {\d\over \d\eta_8}   + {2 \eta_8\eta_9 s_0\over s} {\d\over \d\eta_8} {\d\over \d\eta_9} G + {\eta_8^2 s_0\over s}  {\d^2\over \d\eta_8^2},\\
    \WH O_{0; 89,3} = & {(D-6)s_0\over 2s}\left( (2D-9)+\eta_8 {\d\over \d\eta_8} +\eta_9 {\d\over \d\eta_9}\right),
\end{align}
and $\WH O_{0;4}^{up},\WH O_{0;5}^{up}$ have been defined in \eqref{eq:up-operator}.

In addition, we obtain two crucial T1B-type reduction rules for the second-order derivatives associated with $\eta_4$ and $\eta_5$:
\begin{align}
    (1-\WH O_{0;4}^{up}) {\d^2\over \d\eta_4^2} G\,, \quad (1-\WH O_{0;5}^{up}) {\d^2\over \d\eta_5^2} G\,.\label{eq:npbox-3-2}
\end{align}

With these final rules, the remaining irreducible sets collapse as follows:
\begin{enumerate}
    \item Applying the T2B-type rules for the operators $\frac{\partial}{\partial\eta_5}$, $\frac{\partial}{\partial\eta_8}$, and $\frac{\partial}{\partial\eta_9}$, the indices $n_5, n_8,$ and $n_9$ are unconditionally reduced to zero. Thus, the infinite lines corresponding to the sets $\mathcal{U}_{2}^{2nd}$, $\mathcal{U}_{3}^{2nd}$, and $\mathcal{U}_{4}^{2nd}$ are entirely reduced to $\mathcal{U}_{1}^{2nd}$.
  
    \item Applying the T1B-type rule for the reducing operator ${\d^2\over \d\eta_4^2}$, the remaining subset $\mathcal{U}_{1}^{2nd}$ is finally restricted to $n_4 \in \{0, 1\}$.
\end{enumerate}

At this stage the infinite towers have been fully eliminated, and only a finite set of master integrals remains:
\begin{align}
    I_1 = (0,0,0,0, 0,0,0,0,0), \quad
    I_2 = (0,0,0,1, 0,0,0,0,0),
\end{align}
This completes the reduction for the non-planar double box. Compared with the planar case, the intermediate operator structure is more intricate and the decoupling pattern is less transparent, but the same iterative logic still drives the irreducible lattice to a finite master-integral basis.

\section{The subsector of the sunset diagram}
\label{sec:sunset-sub}
The previous examples focused on top sectors. To show that the formalism is not restricted to that setting, we now consider the subsector $G_{110}$ of the sunset family. This example is conceptually important because it shows how the same operator-based strategy works after part of the propagator structure has already been removed. Using the kinematics introduced in Section~\ref{sec:sunset-top}, we again extract the reduction rules directly and track their action on the remaining lattice.

\subsection{Degree-one rules from fundamental IBP relations}
From the six fundamental IBP relations, we obtain the coefficient matrix for four degree-two equations:
\begin{align}
    \left( \begin{array}{c | c | c | c |  c | c }
	{\d_3\d_1} & {\d_4\d_1} & {\d_5\d_1} & 
	{\d_3\d_2} & {\d_4\d_2} & {\d_5\d_2}\\  \hline
	1 & 2  & 2 &  &  & \\  \hline
	 & 2  &  &  &  & \\  \hline
	  &   &  & 1 & 2 & 2 \\  \hline
	   &   &  &  &  & 2 \\
\end{array}\right),
\end{align}
and the coefficient matrix for two degree-one equations:
\begin{align}
    \left( \begin{array}{c | c }
	{\d_1} & 	{\d_2} \\  \hline
	{{2m_1^2}/s_0} & \\  \hline
	&  {{2m_2^2}/s_0} \\  
\end{array}\right).
\end{align}
Diagonalizing the degree-one system immediately yields two explicit T1A-type rules for the operators $\partial/\partial\eta_1$ and $\partial/\partial\eta_2$:
\begin{align}
    \frac{2m_1^2}{s_0}{\d\over \d\eta_1}G_{110} &\rightarrow \left((D-2) -2\eta_1 {\d\over \d\eta_1} + \eta_3 {\d\over \d\eta_3} + \eta_4{\d\over \d\eta_4} + 2 \eta_3 {\d\over \d\eta_5} \right) G_{110} + \dots \,, \label{eq:sunset-G110-eta1} \\
    \frac{2m_2^2}{s_0}{\d\over \d\eta_2}G_{110} &\rightarrow \left((D-2) -2\eta_2 {\d\over \d\eta_2} + \eta_3 {\d\over \d\eta_3} + \eta_5{\d\over \d\eta_5} + 2 \eta_3 {\d\over \d\eta_4} \right) G_{110} + \dots \,. \label{eq:sunset-G110-eta2}
\end{align}
Applying these T1-type rules reduces $n_1$ and $n_2$ unconditionally to zero, leaving the irreducible set
\begin{align}
    \mathcal{U}_1 = (0,0,n_3,n_4,n_5).
\end{align}

\subsection{Updating degree-two operators and final reduction}
At this point, it is more efficient to update the remaining degree-two relations from the original IBP system than to generate a large family of new descendants. Substituting the newly solved degree-one rules, Eqs.~\eqref{eq:sunset-G110-eta1} and \eqref{eq:sunset-G110-eta2}, turns the problem into a degree-one system with coefficient matrix
\begin{align}
    \left( \begin{array}{c | c  |c | c |c |c}
	{\d_3} & 	{\d_4}  &  	{\d_5} & \eta_3 {\d_5^2} & \eta_3 {\d_5 \d_4} & \eta_3 {\d_4^2} 
	\\  \hline
	\WH O_{s;34,1} &  & 2\WH O_{s;34,1}+ {s_0\over m_1^2}\eta_3 {\d_3} & {2 s_0 \over m_1^2} & & \\  \hline
	 & \WH O_{s;34,1}   &   & &  { s_0 \over m_1^2}  & \\  \hline
	 \WH O_{s;35,2} & 2\WH O_{s;35,2}+{s_0\over m_2^2}\eta_3 {\d_3} & & &  & {2 s_0 \over m_2^2}  \\  \hline
	 &  &\WH O_{s;35,2}   &   &  { s_0 \over m_2^2}  & \\  
\end{array}\right).
\end{align}

Gaussian elimination on this updated system yields four new degree-one equations. From them we extract three key reduction rules for the operators $\frac{\partial}{\partial\eta_3}$, $\frac{\partial}{\partial\eta_4}$, and $\frac{\partial}{\partial\eta_5}$:
\begin{align}
    \WH O_{s;34,1} {\d\over \d\eta_3} &\rightarrow \left((2\WH O_{s;34}+ {s_0\over m_1^2}\eta_3 {\d\over \d\eta_3}) {\d\over \d\eta_5}, \eta_3 {\d^2\over \d\eta_5^2} \right), \label{eq:sunset-G110-3} \\
    \WH O_{s;34,1} {\d\over \d\eta_4} &\rightarrow \eta_3 {\d\over \d\eta_5 }{\d\over\d\eta_4}, \label{eq:sunset-G110-4} \\
    \WH O_{s;35,2} {\d\over \d\eta_5} &\rightarrow \eta_3 {\d\over \d\eta_5}{\d\over \d\eta_4}, \label{eq:sunset-G110-5}
\end{align}
where we have introduced the respective zero-index coefficient operators:
\begin{align}
    \WH O_{s;ij,k} &= {s_0\over 2 m_k^2}\left({(D-1)}+ \eta_i {\d\over \d\eta_i} +\eta_j {\d\over \d\eta_j}\right).
\end{align}

With these updated equations, the remaining irreducible set is resolved in a simple hierarchical way:
\begin{enumerate}
    \item We first apply the rule for $\WH O_{s;34,1} {\d\over \d\eta_3}$ (Eq.~\eqref{eq:sunset-G110-3}), which has the highest priority. Although the right-hand side contains an operator with a negative index component, namely $\eta_3 {\d^2\over \d\eta_5^2}$, this causes no difficulty for the reduction of $n_3$. The factor of $\eta_3$ means that the corresponding contribution starts only at $n_3=1$, so it does not obstruct the reduction from $n_3=1$ to $n_3=0$. The set $\mathcal{U}_1$ is therefore reduced to
    \begin{align}
        \mathcal{U}_2 = (0,0,0,n_4,n_5).
    \end{align}

    \item Once $n_3 = 0$, every term carrying an explicit factor of $\eta_3$ vanishes identically. As a result, the right-hand sides of Eq.~\eqref{eq:sunset-G110-4} and Eq.~\eqref{eq:sunset-G110-5} collapse, and these two relations degenerate into pure T1A-type rules for $\frac{\partial}{\partial\eta_4}$ and $\frac{\partial}{\partial\eta_5}$. Applying them reduces the residual set to a single master integral with all zero indices:
    \begin{align}
        I &= (0,0,0,0,0).
    \end{align}
\end{enumerate}

In summary, the sunset subsector closes without requiring a separate descendant-equation analysis beyond the updated IBP system. This example complements the top-sector discussion by showing that the same framework also handles lower sectors naturally and still leads to a finite master-integral basis, in this case a single master integral.

\section{The subsector of two-loop massless non-planar box}
\label{sec:nonplanar-sub}

We close the sequence of worked examples with a nontrivial subsector of the two-loop massless non-planar box. Using the kinematics introduced in Section~\ref{sec:nonplanar-top}, we consider the subsector in which $\mathcal{D}_2$ is treated as an ISP. This example is not redundant with the top non-planar sector: it shows that the same method still works when the sector structure changes while the non-planar geometry is retained. To keep the presentation focused, we suppress long intermediate matrices whenever they do not add conceptual value and concentrate on the key reduction rules.

\subsection{The first iteration from fundamental IBP relations}
The original ten fundamental IBP relations give the coefficient matrix of degree-two equations as:
\begin{align}
    \left( \begin{array}{c| c| c| c |  c| c| c| c| c | c  |  c| c| c| c }
	{\d_2\d_1} & {\d_2\d_3} & {\d_2\d_6} & {\d_2\d_7}%
	& {\d_1\d_8} & 	{\d_3\d_8} & 	{\d_4\d_8} & 	{\d_5\d_8} & 	{\d_6\d_8} & 	{\d_7\d_8}%
	& 	{\d_4\d_9} & 	{\d_5\d_9} & 	{\d_6\d_9} & 	{\d_7\d_9} \\  \hline 
	&  &  &    & &  &   &  & 2 &   &  &  &  &  \\   \hline 
	&  &  &    &  & 2 &   &  & 2 &   & &  &  &  \\   \hline 
	1& 1 & 1 &  1  &  &  &   &  &  &   &  &  & 2 & 2 \\    \hline 
	-1& -1 & -1 &  -1  &  &  &   &  &  2&  2 & &  & -2 & -2 \\    \hline 
	&  &  &    & 2	&2  &   &  &  &   &  &  &  &  \\    \hline 
	&  &  &    &  &  &   & -2 & -2 &   &  &  &  &  \\    \hline 
	&  &  &    &  &  &   &  & -2 &   &  &  &  &  \\    \hline 
	&  & -1 &  -1  & &  &   &  &  &   & -2& -2 & -2 & -2 \\    \hline 
	&  & 1 &  1  &  &  &  -2 & -2 & -2 & -2  &  2& 2 & 2 & 2 \\   \hline 
	&  &  &    &  &  &  2 & 2 &  &   &  &  &  &  \\
\end{array}\right).
\end{align}

After Gaussian elimination, we extract four degree-two T1A-type reduction rules for the operators $\d_i\d_8$ with $i=4,5,6,7$, together with two explicit degree-one T1A-type scaling rules for the operators
\begin{align}
{\d\over \d\eta_1},\quad {\d\over \d\eta_3}. \label{eq:NP-sub-1-1}
\end{align}
Applying these T1-type rules reduces $n_1$ and $n_3$ completely to zero.

These first-round rules already have an important strategic consequence: once $\partial/\partial\eta_1$ and $\partial/\partial\eta_3$ are under control, the remaining degree-two relations can be updated to expose a new degree-one reduction rule. In particular, updating the equations containing $\partial/\partial\eta_1\,\partial/\partial\eta_8$ and $\partial/\partial\eta_3\,\partial/\partial\eta_8$ and then performing Gaussian elimination yields the key T1B-type rule
\begin{align}
\WH O_{0;8}^{a}  {\d\over \d\eta_8}, \label{eq:NP-sub-eta8}
\end{align}
where the zero-index operator is given by:
\begin{align}
\WH O_{0;8}^{a} = {s_0\over s} \left({2(D-4)}+ \eta_9 {\d\over \d\eta_9}+ \eta_8 {\d\over \d\eta_8}+ \eta_2 {\d\over \d\eta_2}\right).
\end{align}
Applying this rule reduces the index $n_8$ to zero.

In addition, we record two T2A-type reduction rules for the degree-two operators $\frac{\partial}{\partial\eta_2}\frac{\partial}{\partial\eta_6}$ and $\frac{\partial}{\partial\eta_4}\frac{\partial}{\partial\eta_9}$, namely
\begin{align}
    \frac{\partial}{\partial\eta_2}\frac{\partial}{\partial\eta_6}&\rightarrow \left(\frac{\partial}{\partial\eta_2}\frac{\partial}{\partial\eta_7}, \frac{\partial}{\partial\eta_6}\frac{\partial}{\partial\eta_9}, \frac{\partial}{\partial\eta_7}\frac{\partial}{\partial\eta_9} \right),\\
    \frac{\partial}{\partial\eta_4}\frac{\partial}{\partial\eta_9}&\rightarrow \frac{\partial}{\partial\eta_5}\frac{\partial}{\partial\eta_9}.
\end{align}
Applying these rules decouples the index $n_2$ from $n_6$ and $n_4$ from $n_9$.

These rules fragment the irreducible set in the following way:
\begin{enumerate}
    \item Applying degree-one rules for operators ${\d\over \d\eta_1}$, ${\d\over \d\eta_3}$ and $\WH O_{0;a,8}  {\d\over \d\eta_8}$, we have irreducible set:
    \begin{align}
        (0, \eta_2, 0, \eta_4, \eta_5, \eta_6, \eta_7, 0, \eta_9).
    \end{align}
    \item Utilizing the degree-two rules for operators $\frac{\partial}{\partial\eta_2}\frac{\partial}{\partial\eta_6}$ and $\frac{\partial}{\partial\eta_4}\frac{\partial}{\partial\eta_9}$, we decompose the irreducible set to 4 irreducible subsets:
    \begin{align}
        \mathcal{U}^{1st}_1 &= (0, n_2, 0, n_4, n_5, 0, n_7, 0, 0),\\
        \mathcal{U}^{1st}_2 &= (0, 0, 0, n_4, n_5, n_6, n_7, 0, 0),\\
        \mathcal{U}^{1st}_3 &= (0, 0, 0, n_4, n_5, 0, n_7, 0, n_9),\\
        \mathcal{U}^{1st}_4 &= (0, 0, 0, 0, n_5, n_6, n_7, 0, n_9).
    \end{align}
\end{enumerate}

Since infinitely many lattice points remain unreduced, a second iteration is still required.

\subsection{The second iteration from descendant equations}

After the first iteration, the remaining unreduced set is spanned by $n_2$, $n_4$, $n_5$, $n_6$, $n_7$, and $n_9$. The second iteration therefore probes only these directions. Following the general algorithm, we act with the remaining derivatives $\partial/\partial\eta_i$ for $i=2,4,5,6,7,9$ on the equations established in the first step, paying particular attention to the degree-zero relation because it can generate structurally new degree-one constraints.

After updating the newly generated equations with the reduction rules already in hand, we obtain twelve degree-one equations and seven degree-two equations. For the full reduction, the degree-one equations already suffice, and their coefficient matrix is:
\begin{align}
{
\left( \begin{array}{c | c |c |c |c |c |c |c | c| c|c|c}
		{\d_2}  &  {\d_4} &  {\d_5} &  {\d_6} & {\d_7} & {\d_8} & {\d_9} & 
		 {\eta_2\d_6\d_9} &  {\eta_2\d_7\d_9} &  {\eta_2\d_8\d_9}  &  {\eta_2\d_9\d_9} &
		{\eta_9\d_2\d_8}  \\  \hline
		 \WH O_{0;2}^{b} &  &  &  &  & -\WH O_{0;2}^{c} & 2\WH O_{0;2}^{c} &  %
		  &  &  &  & -{t \over 2s}  \\  \hline 
		    & \WH O_{0;4}^{b} &  &  &  & &  &  %
		  &  &  &  & \\  \hline 
		  &  & \WH O_{0;5}^{b} &  &  & &  &  %
		  &  &  &  &  \\  \hline 
		  &  &  & \WH O_{0;6}^{b} &  & &  &  %
		 2 &  &  &  & \\  \hline 
		  &  &  &  &  \WH O_{0;7}^{b} & &  &  %
		  & 2 &  &  & \\  \hline 
		  &  &  &  &  &-{t \over 2s}\WH O_{0;9}^{c} & \WH O_{0;9}^{b} &  %
		  &  & -1 & 2 & \\  \hline 
		  & \WH O_{0;4}^{e} &  &  &  & &  &  %
		  &  &  &  & \\  \hline 
		  &  & \WH O_{0;5}^{e} &  &  & &  &  %
		  &  &  &  & \\  \hline 
		  &  &  & \WH O_{0;6}^{e}&  & &  &  %
		-1  &  &  &  & \\  \hline 
		  &  &  &  & \WH O_{0;7}^{e} & &  &  %
		  & -1 &  &  & \\  \hline 
		  & \WH O_{0;4}^{up} &  &\WH O_{0;6}^{up}  &  & &  &  %
		  &  &  &  & \\  \hline 
		  &  & \WH O_{0;5}^{up} &  & \WH O_{0;7}^{up} & &  &  %
		  &  &  &  & \\   
	\end{array}\right).}
\end{align}
where we have defined zero-index operators as
\begin{align}
\WH O_{0;4}^{b} = \WH O_{0;5}^{b}=& 1-{\eta_9\over 2} {\d\over \d\eta_9} -\eta_7 {\d\over \d\eta_7} -\eta_6 {\d\over \d\eta_6}  +\eta_5 {\d\over \d\eta_5}  + \eta_4 {\d\over \d\eta_4}  +{\eta_2\over 2} {\d\over \d\eta_2},\\
\WH O_{0;6}^{b}=\WH O_{0;7}^{b}=&-1-{\eta_9\over 2} {\d\over \d\eta_9} -\eta_7 {\d\over \d\eta_7} -\eta_6 {\d\over \d\eta_6}  +\eta_5 {\d\over \d\eta_5}  + \eta_4 {\d\over \d\eta_4}  +{\eta_2\over 2} {\d\over \d\eta_2},\\
\WH O_{0;4}^{e}=&{4-D\over 2} +\eta_6 {\d\over \d\eta_6}  -\eta_5 {\d\over \d\eta_5}  -{\eta_2\over 2} {\d\over \d\eta_2},\\
\WH O_{0;5}^{e}=&{D-6\over 2} +\eta_6 {\d\over \d\eta_6}  -\eta_5 {\d\over \d\eta_5}  +{\eta_9\over 2} {\d\over \d\eta_9},\\
\WH O_{0;6}^{e}=&{-D+6\over 2} +\eta_6 {\d\over \d\eta_6}  -\eta_5 {\d\over \d\eta_5}  -{\eta_2\over 2} {\d\over \d\eta_2},\\
\WH O_{0;7}^{e}=&{-4+D\over 2} +\eta_6 {\d\over \d\eta_6}  -\eta_5 {\d\over \d\eta_5}  +{\eta_9\over 2} {\d\over \d\eta_9},\\
\WH O_{0;2}^{b} = & {1\over 2}-{\eta_9\over 2} {\d\over \d\eta_9} -\eta_7 {\d\over \d\eta_7} -\eta_6 {\d\over \d\eta_6}  +\eta_5 {\d\over \d\eta_5}  + \eta_4 {\d\over \d\eta_4}  +{\eta_2\over 2} {\d\over \d\eta_2},\\
\WH O_{0;9}^{b} = & \WH O_{0;2}^{b}-1,\\
\WH O_{0;i}^{c} = &  1+\eta_i{\d\over \d\eta_i},
\end{align}
and the operators $\WH O_{0;i}^{up}$ have been defined in \eqref{eq:up-operator}.

Through Gaussian elimination, together with simplification by the existing rules, we extract four T1B-type rules for the degree-one operators:
\begin{align}
    \WH O_{0;4}^{b}\frac{\partial}{\partial\eta_4},\quad 
    \WH O_{0;5}^{b}\frac{\partial}{\partial\eta_5},\quad 
    \left(\WH O_{0;6}^{b}+2\WH O_{0;6}^{e}\right)\frac{\partial}{\partial\eta_6},\quad 
    \left(\WH O_{0;7}^{b}+2\WH O_{0;7}^{e}\right)\frac{\partial}{\partial\eta_7}.
\end{align}
We also extract two T2B-type rules, namely
\begin{align}
    \WH O_{0;8}^{a} \WH O_{0;2}^{b}  \frac{\partial}{\partial\eta_2} &\rightarrow \left(\WH O_{0;8}^{a} \WH O_{0;2}^{c}\frac{\partial}{\partial\eta_8}, \WH O_{0;8}^{a} \WH O_{0;2}^{c}\frac{\partial}{\partial\eta_9}\right),\\
    \WH O_{0;9}^{b} \frac{\partial}{\partial\eta_9}&\rightarrow \left(\WH O_{0;9}^{c}\frac{\partial}{\partial\eta_8}, \eta_2 \frac{\partial}{\partial\eta_8}\frac{\partial}{\partial\eta_9},\eta_2 \frac{\partial^2}{\partial\eta_9}\right),
\end{align}
where the operator $\WH O_{0;8}^{a}$ defined previously is applied because we use the T1B-type reduction rule \eqref{eq:NP-sub-eta8} for updating the operator $\eta_9 {\d\over\d \eta_2} {\d\over\d \eta_8}$.

These newly extracted rules reduce the irreducible sets as follows:
\begin{enumerate}
    \item Utilizing the four T1B-type rules, we reduce $\mathcal{U}^{1st}_1$ to $(0, n_2, 0,0,0,0,0,0,0)$, $\mathcal{U}^{1st}_2$ to $(0, 0, 0,0,0,0,0,0,0)$, and both $\mathcal{U}^{1st}_3$ and $\mathcal{U}^{1st}_4$ to $(0, 0, 0,0,0,0,0,0,n_9)$.
    \item Applying the reduction rule for $\WH O_{0;8}^{a} \WH O_{0;2}^{b}  \frac{\partial}{\partial\eta_2}$ to $(0, n_2, 0,0,0,0,0,0,0)$) moves this set to $(0, 0, 0,0,0,0,0, n_8, n_9)$, which already covers the subset $(0, 0, 0,0,0,0,0,0,n_9)$.
    \item At the bound $n_2 = 0$, the reduction rule for $\WH O_{0;9}^{b} \frac{\partial}{\partial\eta_9}$ simplifies as
    \begin{align}
        \WH O_{0;9}^{b} \frac{\partial}{\partial\eta_9}&\rightarrow \WH O_{0;9}^{c}\frac{\partial}{\partial\eta_8}.
    \end{align}
    Thus, along with the reduction rule \eqref{eq:NP-sub-eta8}, the set $(0, 0, 0,0,0,0,0, n_8, n_9)$ can be reduced to $(0, 0, 0,0,0,0,0,0,0)$.
\end{enumerate}

After the second iteration, we have established a complete set of reduction rules corresponding to every $\frac{\partial}{\partial\eta_i}$. Consequently, any integral in this subsector reduces to the unique master integral,
\begin{align}
    I &= (0,0,0,0,0,0,0,0,0).
\end{align}

\section{Degenerate topology: elimination of the top sector}
\label{sec:degenerate-top}

In this section, we study a qualitatively different situation: a family whose top sector is degenerate, meaning that it contains no master integrals and reduces completely to subsectors. This example is important because it shows that the formalism can do more than construct finite master-integral bases. It can also detect when the top sector itself is absent as an independent component of the reduction problem. We consider the topology defined by the integral family
\begin{align}
    I(a_i) = \int \prod_{j=1}^{2} \frac{\md^{D}\ell_j}{\mi\pi^{D/2}} \frac{1}{\prod_{i=1}^5 \mathcal{D}_i^{a_i}}\,, \label{Hu-0-1}
\end{align}
with the conventional 2-loop configuration:
\begin{align}
    \mathcal{D}_1=\ell_1^2\,,\quad \mathcal{D}_2=(\ell_1-K)^2\,,\quad \mathcal{D}_3=\ell_2^2\,,\quad \mathcal{D}_4=(\ell_2-K)^2\,,\quad \mathcal{D}_5=(\ell_1-\ell_2)^2\,. \label{Hu-0-2}
\end{align}
Since $L=2, E=1$, the number of complete basis invariants is exactly $N=5$. Consequently, there are no Irreducible Numerator Parts (ISPs). The generating function of the top sector is explicitly given by:
\begin{align}
    G = \int \prod_{j=1}^{2}\frac{\md^{D}\ell_j}{\mi\pi^{D/2}} \frac{1}{\prod_{i=1}^5 (\mathcal{D}_i-s_0\eta_i)}\,. \label{Hu-0-3}
\end{align}
To describe the reduction to lower topologies economically, we introduce a compact notation for the subsector generating functions. For instance, the boundary function where the first propagator is contracted ($\eta_1=0$) is denoted as:
\begin{align}
    B_1 \equiv G(\eta_1=0,\eta_2,\eta_3,\eta_4,\eta_5)\,, \label{Hu-0-4}
\end{align}
with analogous definitions for $B_2$, $B_3$, $B_4$, and $B_5$.

\subsection{IBP generation and degree-zero constraints}

Applying the standard IBP differential operators, we generate six fundamental relations. Since there are no ISPs, this initial system contains only degree-one operators acting on $G$, balanced against subsector terms:
\begin{align}
    0 ={}& \frac{s}{s_0} {\d\over \d\eta_2} G + \left[ (D-4) + (-\eta_1+\eta_3-\eta_5){\d\over \d\eta_5}  -(\eta_1+\eta_2){\d\over \d\eta_2}  -2\eta_1 {\d\over \d\eta_1} \right] G \nn
    & -\left[{\d\over \d\eta_5}+{\d\over \d\eta_2}\right] B_1+ {\d\over \d\eta_5} B_3, \label{Hu-R1-Eq-1-1}
\end{align}
\begin{align}
    0 ={}& \frac{s}{s_0} {\d\over \d\eta_2} G - \left[ (\eta_1-\eta_3+\eta_5){\d\over \d\eta_5}  +(\eta_1+\eta_4-\eta_5){\d\over \d\eta_2}  +(\eta_1+\eta_3+\eta_5){\d\over \d\eta_1} \right] G \nn
    & - \left[{\d\over \d\eta_5} + {\d\over \d\eta_2}\right] B_1 + \left[{\d\over \d\eta_5}-  {\d\over \d\eta_1}\right] B_3 - {\d\over \d\eta_2} B_4+ \left[{\d\over \d\eta_2} +  {\d\over \d\eta_1}\right] B_5, \label{Hu-R1-Eq-1-2}
\end{align}
\begin{align}
    0 ={}& \frac{s}{s_0} \left[ {\d\over \d\eta_2} -{\d\over \d\eta_1} \right] G + \left[ (-\eta_1+\eta_2)\left[{\d\over \d\eta_5}+{\d\over \d\eta_2}+{\d\over \d\eta_1}\right]+(\eta_3-\eta_4){\d\over \d\eta_5}\right] G \nn
    & -\left[{\d\over \d\eta_5}+ {\d\over \d\eta_2}\right]B_1 + \left[{\d\over \d\eta_5}+  {\d\over \d\eta_1}\right] B_2+ {\d\over \d\eta_5} B_3- {\d\over \d\eta_5} B_4, \label{Hu-R1-Eq-1-3}
\end{align}
\begin{align}
    0 ={}& \frac{s}{s_0} {\d\over \d\eta_4} G + \left[ (\eta_1-\eta_3+\eta_5){\d\over \d\eta_5} -(\eta_2+\eta_3-\eta_5){\d\over \d\eta_4}  -(\eta_1+\eta_3-\eta_5){\d\over \d\eta_3} \right] G \nn
    & + \left[{\d\over \d\eta_5}-{\d\over \d\eta_3}\right] B_1- \left[{\d\over \d\eta_5}+{\d\over \d\eta_4}\right] B_3 -{\d\over \d\eta_4} B_2+\left[{\d\over \d\eta_4} +{\d\over \d\eta_3} \right]B_5, \label{Hu-R1-Eq-1-4}
\end{align}
\begin{align}
    0 ={}& \frac{s}{s_0} {\d\over \d\eta_4} G + \left[ (D-4) +(\eta_1-\eta_3-\eta_5){\d\over \d\eta_5} -(\eta_3+\eta_4){\d\over \d\eta_4} -2\eta_3 {\d\over \d\eta_3} \right] G \nn
    & + {\d\over \d\eta_5} B_1 -\left[{\d\over \d\eta_5} + {\d\over \d\eta_4}\right] B_3, \label{Hu-R1-Eq-1-5}
\end{align}
\begin{align}
    0 ={}& \frac{s}{s_0} \left[ {\d\over \d\eta_4} - {\d\over \d\eta_3} \right] G + \left[ (\eta_1-\eta_2){\d\over \d\eta_5} + (-\eta_3+\eta_4)\left[{\d\over \d\eta_5} +{\d\over \d\eta_4} +{\d\over \d\eta_3} \right]\right] G \nn
    & + {\d\over \d\eta_5} B_1- {\d\over \d\eta_5} B_2- \left[{\d\over \d\eta_5}+{\d\over \d\eta_4}\right] B_3+ \left[{\d\over \d\eta_5} + {\d\over \d\eta_3}\right] B_4. \label{Hu-R1-Eq-1-6}
\end{align}

We therefore have six equations but only four independent degree-one operators controlling the top-sector system, namely $\{\frac{\d}{\d\eta_1}, \frac{\d}{\d\eta_2}, \frac{\d}{\d\eta_3}, \frac{\d}{\d\eta_4}\}$. Diagonalizing the corresponding coefficient matrix yields four T1A-type reduction rules, which reduce any top-sector integral to the subset $(0,0,0,0,n_5)$. More importantly, the remaining two independent linear combinations lose all degree-one content and become pure degree-zero constraints. These degree-zero equations provide the key signal that the top sector may be degenerate.

Specifically, the combination \eqref{Hu-R1-Eq-1-1}-\eqref{Hu-R1-Eq-1-2} evaluates to:
\begin{align}
    0 ={}& \left[(D-4) -2\eta_5{\d\over \d\eta_5} + (-\eta_2+\eta_4-\eta_5){\d\over \d\eta_2}  +(-\eta_1+\eta_3-\eta_5) {\d\over \d\eta_1} \right] G \nn
    & + \left[ {\d\over \d\eta_2} B_4 - {\d\over \d\eta_2} B_5+ {\d\over \d\eta_1} B_3-  {\d\over \d\eta_1} B_5\right]\,, \label{Hu-R1-Eq-2-1}
\end{align}
while the combination \eqref{Hu-R1-Eq-1-5}-\eqref{Hu-R1-Eq-1-4} generates:
\begin{align}
    0 ={}& \left[ (D-4) -2\eta_5{\d\over \d\eta_5} +(\eta_2-\eta_4-\eta_5){\d\over \d\eta_4}  +(\eta_1-\eta_3-\eta_5) {\d\over \d\eta_3} \right] G \nn
    & + \left[{\d\over \d\eta_4} B_2- {\d\over \d\eta_4} B_5 +{\d\over \d\eta_3} B_1-{\d\over \d\eta_3} B_5\right]\,. \label{Hu-R1-Eq-2-2}
\end{align}
These degree-zero constraints are the defining feature of the present degenerate topology.

\subsection{Update the degree-zero equations}
To exploit these constraints, we update the degree-zero equations with the four T1A-type reduction rules. On the restricted subset $(0, 0, 0, 0, n_5)$, Eq.~\eqref{Hu-R1-Eq-2-1} simplifies to
\begin{align}
    0 = &  \left[(D-4) -2\eta_5{\d\over \d\eta_5} \right]G+{2s_0\over s} \eta_5\left[ {(D-4) } -{\eta_5} {\d\over \d\eta_5}\right] G \nn
& + \left[ {\d\over \d\eta_2} B_4 - {\d\over \d\eta_2} B_5+ {\d\over \d\eta_1} B_3-  {\d\over \d\eta_1} B_5\right]\nn
& +{s_0\over s}\eta_5\left[\left[{\d\over \d\eta_2}+{\d\over \d\eta_5}\right] B_1 -\left[{\d\over \d\eta_5} + {\d\over \d\eta_1}\right] B_2- {\d\over \d\eta_5} B_3+ {\d\over \d\eta_5} B_4  \right], ~~~~\label{Hu-R1-Eq-3-1}
\end{align}
and Eq.~\eqref{Hu-R1-Eq-2-2} simplifies to
\begin{align}
    0 = & \left[(D-4) -2\eta_5{\d\over \d\eta_5} \right]G+{2s_0\over s} \eta_5\left[ {(D-4) } +{\eta_5} {\d\over \d\eta_5}\right] G \nn
&  +\left[{\d\over \d\eta_4} B_2- {\d\over \d\eta_4} B_5 +{\d\over \d\eta_3} B_1-{\d\over \d\eta_3} B_5\right]\nn
& + {s_0\over s}\eta_5\left[ {\d\over \d\eta_5} B_1 +{\d\over \d\eta_5} B_2-\left[{\d\over \d\eta_5} + {\d\over \d\eta_4}\right] B_3- \left[{\d\over \d\eta_5} + {\d\over \d\eta_3}\right] B_4 \right].~~~~\label{Hu-R1-Eq-3-2}
\end{align}

We now find that both \eref{Hu-R1-Eq-3-1} and \eref{Hu-R1-Eq-3-2} are T1-type equations for the zero-index operator
\begin{align}
    \WH O_{0;1}\equiv (D-4)  -2\eta_5{\d\over \d\eta_5}.
\end{align}
Using this operator, we can reduce all remaining top-sector integrals to subsector contributions. The conclusion is therefore stronger than ordinary completeness: the top sector itself contains no master integral. 

An additional observation is that subtracting \eref{Hu-R1-Eq-3-1} from \eref{Hu-R1-Eq-3-2} yields a nontrivial relation among subsectors, indicating that some subsectors are not independent. We do not pursue that point further here. This example highlights an additional strength of the framework: the same operator-based analysis can certify the elimination of a top sector, not merely reduce it to a finite basis.

\section{Conclusion}

In this paper, we reformulated the reduction of multi-loop Feynman integrals in terms of generating functions and differential operators. Within this framework, integration-by-parts identities become differential equations, while symbolic reduction rules appear as operator identities acting on the lattice of expansion coefficients. This viewpoint makes it possible to discuss reduction not only as a computational task but also as a structured algebraic problem.

Building on this formalism, we proposed an iterative algorithm that generates differential equations, simplifies them with previously derived rules, extracts new reduction rules by solving linear equations, and checks completeness through the geometry of irreducible lattice subsets. The worked examples show that the same logic applies across qualitatively different situations: simple top sectors, higher-point two-loop topologies, subsectors, and a degenerate case in which the top sector is eliminated altogether.

Several aspects of the present work remain intentionally conservative. We have focused on the structural formulation of the method and on explicit examples rather than on implementation-level optimization or on an exhaustive comparison with other reduction strategies. These topics are natural directions for future work. In particular, it would be useful to study automated operator-selection strategies, a more systematic incorporation of symmetries, and applications of the framework to broader classes of multi-loop families.

Overall, the generating-function viewpoint provides a unified language in which symbolic reduction rules, descendant equations, and completeness conditions can be discussed on equal footing. We hope that this formulation will be useful both for conceptual studies of IBP reduction and for the development of practical symbolic-reduction algorithms.
\section{Acknowledgment}
We thank Mingxing Luo, Antonela Matijašić and Stefan Weinzierl for many helpfull discussions. Bo Feng is supported by the National Natural Science Foundation of China (NSFC) through Grants No. 12535003, No.11935013, No.11947301, No.12047502.  Xiang Li and Yan-Qing Ma are supported by NSFC through Grant No. 12325503.  Yuanche Liu is  supported by NSFC through Grant No. 124B1014, and Yang Zhang is supported by NSFC through Grant No. 12575078 and  12247103, and would like to thank the Erwin Schr\"odinger International Institute
for Mathematics and Physics (ESI), University of Vienna (Austria), for the opportunity to participate
in the Thematic Programme “Amplitudes and Algebraic Geometry” in 2026 where a significant part
of this work has been accomplished and for the support given.

\bibliographystyle{unsrturl}
\bibliography{references}

\end{document}